\documentclass[fleqn,usenatbib]{mnras}
\usepackage[T1]{fontenc}
\usepackage{ae,aecompl}
\usepackage{graphicx}
\usepackage{amsmath}
\usepackage{amssymb}
\usepackage{natbib}
\usepackage{epstopdf}

\usepackage{float}

\usepackage{times}
\usepackage{amsmath}
\usepackage{amssymb}
\usepackage{amsfonts}
\usepackage{mathrsfs}
\title [Rotation and lithium depletion in M35]{The effects of rotation on the lithium depletion of G- and K-dwarfs in Messier 35}
\author[R. D. Jeffries, R. J. Jackson, Q. Sun, C. P. Deliyannis]
  {R. D.~Jeffries$^1$, R. J.~Jackson$^1$, Qinghui Sun$^2$ and Constantine
    P.~Deliyannis$^2$\\
   $^1$ Astrophysics Group, Keele University, Keele, 
      Staffordshire ST5 5BG\\
     $^2$Department of Astronomy, Indiana University.
727 E 3rd Street,
Bloomington, IN 47405-7105, USA\\
}

\date{Accepted 7th October 2020}

\pagerange{\pageref{firstpage}--\pageref{lastpage}} \pubyear{2020}

\def\LaTeX{L\kern-.36em\raise.3ex\hbox{a}\kern-.15em
    T\kern-.1667em\lower.7ex\hbox{E}\kern-.125emX}

\begin{document}
\label{firstpage}
\maketitle

\begin{abstract}
New fibre spectroscopy and radial velocities from the WIYN telescope are used to measure photospheric lithium in 242 high-probability, zero-age-main-sequence (ZAMS) F- to K-type members of the rich cluster M35. Combining these with published rotation periods, the connection between lithium depletion and rotation is studied in unprecedented detail. At $T_{\rm eff}<5500$\,K there is a strong relationship between faster rotation and less Li depletion, although with a dispersion larger than measurement uncertainties. Components of photometrically identified binary systems follow the same relationship. A correlation is also established between faster rotation rate (or smaller Rossby number), decreased Li depletion and larger stellar radius at a given $T_{\rm eff}$. These results support models where starspots and interior magnetic fields lead to inflated radii and reduced Li depletion during the pre main sequence (PMS) phase for the fastest rotators. However, the data are also consistent with the idea that all stars suffered lower levels of Li depletion than predicted by standard PMS models, perhaps because of deficiencies in those models or because saturated levels of magnetic activity suppress Li depletion equally in PMS stars of similar $T_{\rm eff}$ regardless of rotation rate, and that slower rotators subsequently experience more mixing and post-PMS Li depletion.
 \end{abstract}


\section{Introduction}

For several decades, observations of lithium abundances in young low-mass stars have indicated that standard models of pre-main-sequence (PMS) stellar evolution have missing ingredients. 

Lithium is present in the gas from which stars are formed, but is destroyed at relatively low temperatures ($\sim 2.5 \times 10^{6}$ K) in stellar interiors. As low-mass PMS stars contract towards the zero age main sequence (ZAMS) their cores become hot enough to "burn" Li in p, $\alpha$ reactions \citep{Bodenheimer1965a, Deliyannis1990a, Bildsten1997a}. This Li destruction will be observed at the photosphere if standard, convective mixing reaches down as far as the Li-burning regions. In a low-mass star ($< 0.35 M_{\odot}$) that remains fully convective all the way to the ZAMS, complete Li depletion is expected. Higher mass PMS stars develop a radiative core that hinders any further mixing of depleted material to the surface once the convection zone base falls short of the Li-burning temperature. Standard PMS models \citep[e.g.][]{Dantona1997a, Baraffe1998a, Siess2000a, Piau2002a} predict that the Li abundance of PMS stars should be a smooth, single-valued function of mass and age (and also metallicity) among G-, K- and early M-type stars, with increasing Li depletion at lower masses and older ages (and higher metallicity). 

Establishing the extent and time-dependence of PMS Li depletion and identifying the parameters that control it are of course important in understanding the physics of stellar interiors. It is also a pre-requisite for understanding how much depletion takes place subsequently on the main sequence and hence for using Li abundances as a means of estimating the ages of low-mass main sequence stars, whose structure and position in the Hertzsprung-Russell diagram change relatively little over billions of years \citep[e.g][]{Randich2009a, Soderblom2010a}.

The predictions of basic PMS models have been contradicted by many determinations of Li abundance in young, coeval clusters of stars \citep[e.g. see reviews by][]{Jeffries2000a, Jeffries2006a}. Whilst the general shape and progression of the Li depletion pattern (Li abundance versus effective temperature, $T_{\rm eff}$) with age is as expected as far as the ZAMS, 
there is continuing depletion in G-dwarfs whilst on the main sequence and
a significant scatter in Li abundance at a given $T_{\rm eff}$ among late-G and K-dwarfs in the same cluster, that presumably share a similar age and overall chemical composition. Since
standard theory predicts no post-PMS photospheric Li depletion for G-dwarfs and no
scatter at a given $T_{\rm eff}$ for dwarfs of any spectral type, these phenomena
betray the action of physical processes not included in standard models. A small scatter may appear first among very young ($<10$ Myr) cool PMS stars \citep{Bouvier2016a, Lim2016a}, increases amongst clusters with age 20--40\,Myr \citep{Randich2001a, Messina2016a} and reaches 2 orders of magnitude for ZAMS K-dwarfs at $\sim 100$ Myr \citep[e.g.][]{Duncan1983a, Butler1987a, Balachandran1988a, Soderblom1993a, Jeffries1998a, Randich1998a}. 

Important clues to the origin of the dispersion are that it is much smaller among hotter G-dwarfs and that fast rotation is correlated with higher Li abundances. Early studies used spectroscopically measured projected equatorial velocities as a rotation proxy; the uncertain inclination angle leaving room for debate about the strength of the correlation. Recent studies of stars in the Pleiades (age $\simeq 120$ Myr), using rotation periods determined from starspot modulation, have demonstrated that the correlation is very strong \citep[][hereafter B18]{Barrado2016a, Bouvier2018a}.

The connection between fast rotation and Li abundance is still uncertain.
One hypothesis links this to another puzzle in low-mass stellar astrophysics - that the components of magnetically active, close, tidally locked eclipsing binaries are often $\sim 10\%$ larger than models predict \citep[e.g.][]{Morales2009a, Torres2013a}. If convective heat
transfer is inhibited, either by dynamo-generated magnetic fields in the convection zone or by the blocking of photospheric flux by dark, magnetic starspots, then an inflated radius is expected \citep[e.g.][]{Spruit1986a, Ventura1998a, Feiden2013a, MacDonald2013a, Jackson2014a}. This leads to
cooler interior temperatures, slower Li destruction and for stars with radiative cores, shallower convection zones and less photospheric Li depletion. Hence the suggestion that the fastest rotating young stars, with the strongest magnetic dynamos and most spotted surfaces, may be more inflated and suffer less PMS Li depletion than their more slowly rotating siblings
\citep{Somers2015a, Somers2015b, Jeffries2017a, Somers2017a}. 

Others have interpreted the spread as due to additional mixing at the base of the convection zone. The rotation dependence may then be ascribed to greater early angular momentum loss and consequent differential rotation and mixing in those PMS stars that remained locked to their accretion discs for longer durations \citep{Bouvier2008a, Eggenberger2012a} or
to less efficient convective penetration ("overshooting") into the radiative zone for faster rotators \citep{Montalban2000a, Baraffe2017a}. It is also possible that the dispersion, or at least some fraction of it, could be attributed to the formation conditions of the main Li~{\sc i} line diagnostic. Starspots, chromospheric activity or intense magnetic fields might lead to some amplification of the line equivalent width that is indirectly related to rotation rate but that does not require a genuine spread in abundance \citep[e.g.][]{Stuik1997a, King2004a, Leone2007a}.

The purpose of this paper is to investigate the lithium-rotation connection in solar-type and lower-mass stars of the open cluster M35 (NGC~2168). With an age of $\sim 150$ Myr and at a distance of $\sim 800$ pc \citep{Sung1999a, vonHippel2002a}, it is a much richer analogue of the well-studied Pleiades cluster, which should enable a more detailed picture of the lithium depletion pattern in ZAMS stars. Previous spectroscopic investigations of lithium in the cluster have been limited to relatively small numbers of targets and focused more on the hotter (and brighter) F- and G-stars. These studies do show evidence for some Li depletion among the F-stars and that at least some of the Li depletion dispersion observed in the Pleiades cool stars is also present in M35 \citep[][hereafter AT18]{Barrado2001a, Steinhauer2004a, Anthony-Twarog2018a}. 

Section \ref{s2} describes how a sample of targets were selected for spectroscopic observation at the WIYN 3.5-m telescope; most of these have published rotation periods. The section goes on to explain the observations, data reduction and analysis of the spectra. Section \ref{s3} discusses cluster membership and combines radial velocities with astrometry from the Gaia DR2 catalogue \citep{Gaia2018a} to provide individual membership probabilities and how multiwavelength photometry is used to estimate the luminosity, $T_{\rm eff}$ and hence radius of the M35 members. Section \ref{s4} presents results for the cluster members and investigates the lithium depletion pattern as a function of $T_{\rm eff}$, rotation and binarity. Section \ref{s5} discusses these results in terms of both standard evolutionary models and those that include magnetic fields, starspots and radius inflation. The conclusions are presented in section \ref{s6}.

\section{Targets and Spectroscopic Measurements}

\label{s2}

The young open cluster M35 has a mean parallax of $1.1301 \pm 0.0013$\,mas \citep{Gaia2018b} giving a cluster distance $d_C=885$\ pc ($(M-m)_0=9.73$ mag), with a conservative uncertainty of $<80$\,pc caused by remaining systematics in the Gaia data \citep{Lindegren2018a}. Other cluster parameters are reviewed extensively by AT18 and we adopt their choice of reddening ($E(B-V)=0.20$) and an age of about 120-160 Myr (all our targeted stars have reached the ZAMS). From the reddening value, we adopt extinctions of $A_V=0.62$ and $A_K=0.07$ \citep{Rieke1985a}. Our observations cover stars in the colour range $1.0 < (V-K)_0 < 3.5$, which, using a 120\,Myr solar-metallicity isochrone from \cite{Baraffe2015a}, is equivalent to a temperature range of approximately $4000 < T_{\rm eff} < 6600$\,K and a mass range of  $0.6 < M/M_{\odot} <1.35$. The metallicity of M35 is likely to be slightly sub-solar; previous spectroscopic work by \cite{Barrado2001a} and \cite{Steinhauer2004a} indicates [Fe/H]$=-0.21\pm 0.10$ and $-0.143\pm 0.014$ respectively. However,  AT18 present some spectroscopic evidence that the metallicity may be closer to solar. The precise metallicity has little influence on our main results and conclusions; we adopt a solar metallicity and discuss the effects of a slightly lower metallicity where necessary. 
 
\begin{figure*}
	\centering
	\begin{minipage}[t]{0.93\textwidth}
	\centering
	\includegraphics[width = 170mm]{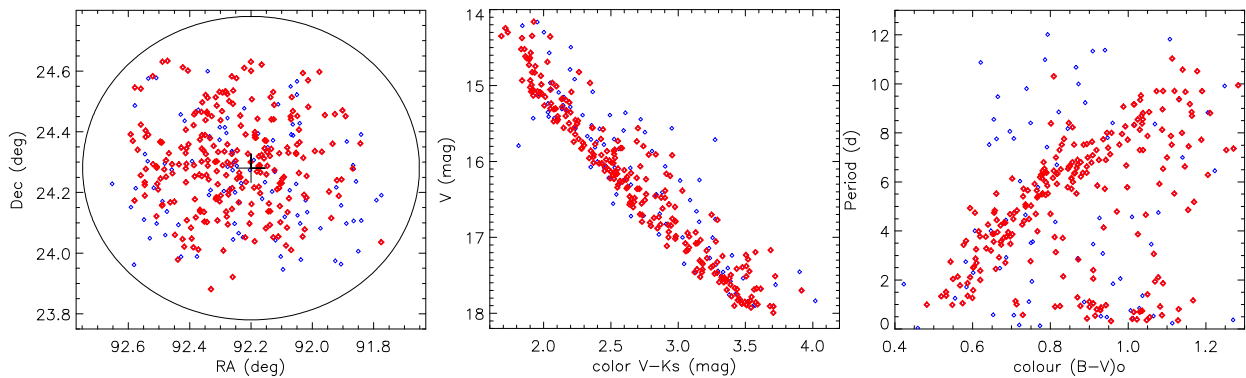}
	\end{minipage}
	\caption{Targets observed in open cluster M35. Plot (a) shows the spatial distribution of  the 342 targets,  plot (b) shows their $V$ versus $V-K_S$ colour magnitude diagram and plot (c) shows the period versus $(B- V)_0$ colour for the 324 targets with measured periods. The larger, red symbols are targets that were later identified as ($P_{\rm mem}>0.95$) probable cluster members from their measured radial velocities and Gaia DR2 proper motions (see Section~\ref{s3.1}).}
	\label{fig1}
\end{figure*}

\subsection{Target selection}
\label{s2.1}

\begin{figure}
	\centering
	\includegraphics[width = 72mm]{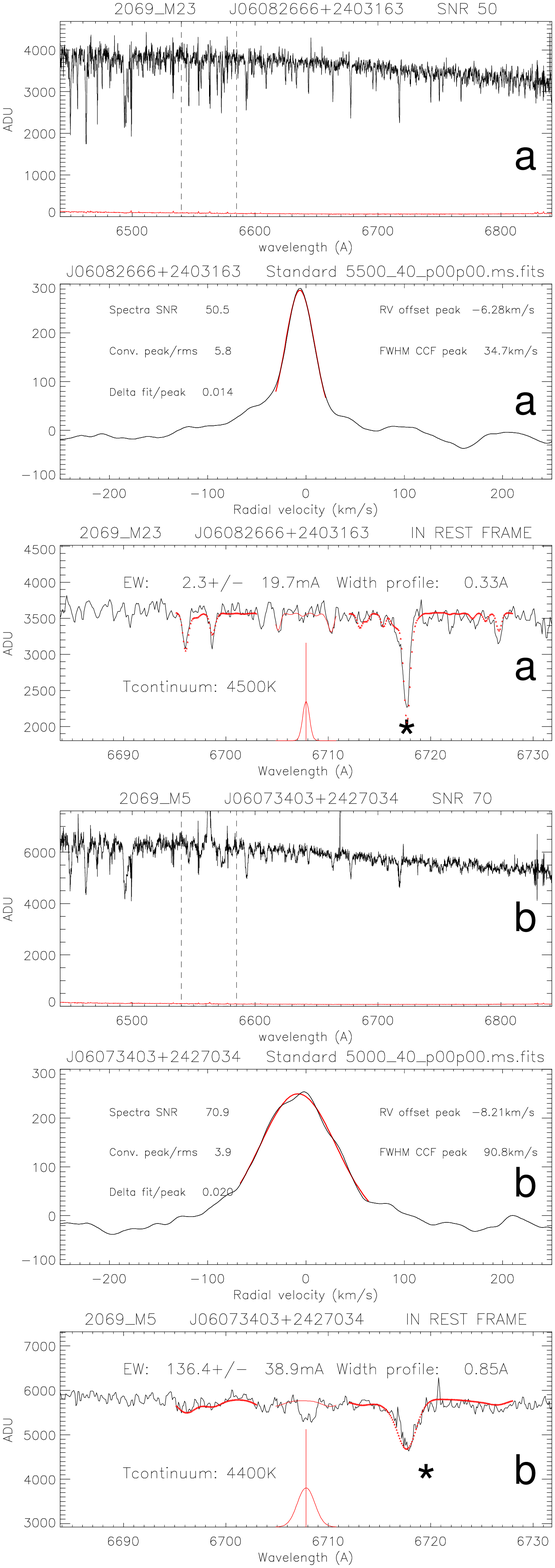}
	\caption{Typical spectra observed in M35. Plots marked (a) show results for a slowly rotating star. The upper plot shows the reduced spectrum, the second plot shows the cross-correlation function (CCF) of the spectrum convolved with a synthetic spectrum, and the lower plot shows the spectrum around the 6708\,\AA\ Li~{\sc i} line. The solid red line shows the weighting profile used to measure EW$_{\rm Li}$ and a dotted red line shows where the template spectrum was fitted (see Section~\ref{s2.4}). Plots marked (b) show similar results for a rapidly rotating star, where the CCF and weighting profile are wider due to rotational broadening. The asterisks mark the Ca~{\sc i} line at 6717.7\,\AA.}
	\label{fig2}
\end{figure}

Targets for fibre spectroscopy were assembled from three sources. A total of 310 stars with $14<V<18$ were identified that had periods measured as part of the Kepler K2 campaign \citep{Libralato2016a} and a M35 membership probability $>0.2$ from the DANCe proper motion study of \cite{Bouy2015a}. Many of these also had ground-based rotation periods recorded in \cite{Meibom2009a}. A further 28 stars that only had rotation periods in \cite{Meibom2009a} were added, with similar $V$ magnitudes and proper motion membership probabilities. Finally, a set of lower priority targets were selected; these had no measured periods but were likely proper motion and photometric members of M35, with $15.5<g<18.2$ \citep[from the][catalogue]{Bouy2015a} and adopting $V \simeq 0.911g + 0.91$ as a transformation for the purposes of target selection. For sky subtraction, a set of "blank sky" targets were also identified that were $>20$ arcsec away from any source in the 2MASS catalogue \citep{Skrutskie2006a}.

From these lists a total of 342 targets were observed; 327 with a measured rotation period; 301 from \cite{Libralato2016a}; 172 from \cite{Meibom2009a} (147 are in both catalogues) and 15 with no period data. Where rotation periods appear in both sources, the value from \cite{Meibom2009a} was adopted (see section~\ref{s4.1}). Figure~\ref{fig1}a shows the spatial distribution of the observed targets and Fig.~\ref{fig1}b their $V$ versus $V-K_s$ photometry. Figure~\ref{fig1}c shows a rotation period vs colour plot for observed targets with measured rotation periods. Optical photometry comes from \cite{Nardiello2015a} for objects with periods from K2 or from \cite{Meibom2009a} otherwise\footnote{The photometry used to select targets, listed in Table~\ref{targets} and shown in Fig.~\ref{fig1}, was superseded by new photometry as described in Section~\ref{s3.2}.}. 
The $K_s$ magnitudes were taken from 2MASS \citep{Skrutskie2006a}. The observed targets are listed in Table~\ref{targets}.

\begin{table*}
\caption{Targets in M35, giving the information from which the targets were selected (see Section~\ref{s2.1}), measurements from the spectra (see Sections~\ref{s2.3} and~\ref{s2.4}) and kinematic membership probability (see Section~\ref{s3.1}).  Target names are from \protect\cite{Bouy2015a}, but the coordinates are those used at the telescope, which come from the Gaia DR1 catalogue \protect\citep{Lindegren2016a}. The table has 342 rows and a sample is shown here. The full table is available electronically.}
\begin{tabular}{l@{\hspace{0.5\tabcolsep}}r@{\hspace{0.5\tabcolsep}}r@{\hspace{0.8\tabcolsep}}r@{\hspace{0.5\tabcolsep}}r@{\hspace{0.5\tabcolsep}}r@{\hspace{0.5\tabcolsep}}r@{\hspace{0.8\tabcolsep}}r@{\hspace{0.8\tabcolsep}}r@{\hspace{0.8\tabcolsep}}r@{\hspace{0.8\tabcolsep}}r@{\hspace{0.8\tabcolsep}}r@{\hspace{0.9\tabcolsep}}r} 
\hline
OBJECT	          &	RA (ICRS)	&	Dec (ICRS)&	$V\ \ $	    &	$K_s\ \ $	& $B-V$  & 	Period$^{\dagger}$   &	SNR	& RV $ \ \ \ \ \ \ $&	FWHM	&	EW(Li) & EW(Ca) & $p_{\rm mem}^{\ddagger}$	\\
	                &	deg$\ \ $	  &	deg$\ \ $	  &	mag$\ $&	mag$\ $&	mag$\ $& d$\ \ $	&   & km\,s$^{-1}\ \ \ \ $ & km\,s$^{-1}$ &	m\AA$\ \ $  &m\AA$\ \ $&	\\\hline
J06070601+2411272   &   91.77509     &   24.19086     &   17.468   &  14.340   &  1.196  & 16.569(2)      & 60    & $ -11.73\pm 0.47$ 
        &   31.3    &   $3 \pm 21$ &$225 \pm 21$ & 0.002 \\
J06070616+2402101	&	91.77568	&	24.03614	&	16.999	&	14.072	&	1.109	&	7.087(2)	&	61	&	$-6.48	\pm	0.54$	&	36.1	&	$148\pm	17$	& $223 \pm 17$	&0.999	\\
J06070982+2410280   & 91.79092       &    24.17442    &   17.539  & 14.057    &   1.278
&   4.750(2) &   52  &   $+31.23\pm 0.79$    &   45.6    &   $31 \pm 53$ & $226 \pm 53$ & 0 \\
... & ... & ...&... & ... & ...&... & ... & ...&... & ... & ...& ... \\
\hline
\multicolumn{13}{l}{$\dagger$ Source of the rotation period is noted in brackets: (1) \cite{Meibom2009a}, (2) \cite{Libralato2016a}.}\\ 
\multicolumn{13}{l}{$\ddagger$ Kinematic membership probability. $-1$ indicates missing information or $|RV|>50$~km\,s$^{-1}$.}
  \end{tabular}
  \label{targets}
\end{table*}

\begin{figure}
	\centering
	\includegraphics[width = 75mm]{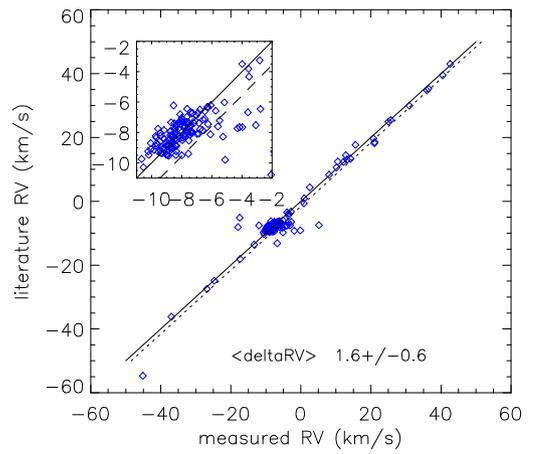}
	\caption{Comparison of radial velocities from Geller et al. (2010) to radial velocities measured here after correction for a mean offset (see Section~\ref{s2.3}). The solid line shows a one to one correspondence, the dashed line shows the line prior to applying the 2$\sigma$ clipped, mean offset between the two data sets.}
	\label{fig3}
\end{figure}

\begin{table}
\caption{Hydra Configurations for targets observed in open cluster M35.}
\begin{tabular}{c@{\hspace{0.6\tabcolsep}}ccccc} \hline
Config.	&	No. of	&	No. of	&	Date first	&	No. of	&	Exposure	\\
number	&	targets	&	sky fibres	&	exposure	&	exposures	&	time (s)	\\\hline
1a	&	61	&	14	&	22:11:2017	&	6	&	21600	\\
1b	&	60	&	12	&	23:11:2017	&	5	&	18000	\\
2a	&	60	&	15	&	21:11:2017	&	9	&	32410	\\
2b	&	60	&	14	&	22:11:2017	&	8	&	28800	\\
3a	&	60	&	15	&	25:11:2017	&	5	&	12000	\\
3b	&	43	&	20	&	25:11:2017	&	5	&	12000	\\
\hline
    \end{tabular}
  \label{configurations}
\end{table}

\subsection{Observations}
\label{s2.2}

Observations were made at the WIYN 3.5-m telescope 
using the Hydra multi-object fibre spectrograph \citep{Bershady2008a} over 5 nights, beginning on 21 November 2017. 

Six fibre configurations with a similar nominal centre of RA $=92.20$ deg, Dec $=+24.28$ deg. were observed. The "blue" Hydra fibres were used giving a resolving power of $\sim 14\,000$. Spectra were recorded over a $\sim 400$\AA\ interval, centred at $\sim 6640$\AA. The FWHM of a resolution element was sampled by $\sim 2.3$ (binned) CCD pixels of size 0.2\AA. 

Details of the fibre configurations and exposure times
are given in Table~\ref{configurations}. Two configurations (1a and 1b) comprised fainter stars with $1.7 < (V-K_s)_0 < 3.5$, (recall that $E(V-K_s)=0.55$) the remaining four comprised of brighter targets with   $1.0 < (V-K_s)_0 < 2.5 $. Each configuration included $\sim 15$ fibres placed on blank sky. Observations were built up from exposures of 30--60 minutes, which were repeated until the stacked spectra measured for each field showed a signal-to-noise ratio (SNR) of $\sim 25$ per pixel for the faintest targets.

\subsection{Data reduction}
\label{s2.3}

The observational data were reduced using the pipeline described in \cite{Jackson2018a}. De-biased science frames were normalised with day-time tungsten lamp flat-field exposures. Spectra were extracted from the normalised images using an optimal extraction algorithm \citep{Horne1986a}. Day-time Th-Ar lamp exposures were
used to define polynomial relations between pixels and wavelength in the extracted spectra. Small corrections were applied for any drift in the calibration using prominent emission lines from the median sky spectrum obtained in each exposure. The spectra were rebinned to a common wavelength range of 6441.5--6841.5\AA\ in 0.1\AA~steps and sky-subtracted using median sky spectra, weighted according to fibre transmission efficiencies estimated from the flat-field. After heliocentric correction, spectra from repeat exposures were summed. 

Radial velocities (RVs) were measured by cross correlation against synthetic spectra with solar metallicity from  \cite{Coelho2005a}, which were broadened to match the resolution of the target spectra. Spectra  in 500\,K steps were selected to match the target temperature, which was estimated from the $(V-K_s)_0$ colour using a \cite{Baraffe2015a} solar-metallicity 120\,Myr isochrone. Representative spectra for slow- and fast-rotating stars and cross correlation functions (CCFs) are shown in Fig.~\ref{fig2}. Figure~\ref{fig3} compares the RV for 182 targets in common with \cite{Geller2010a}, some of which will be binary stars. This was used to determine the offset between the measured RV (relative to the synthetic spectra) and the absolute RV. The 2$\sigma$ clipped mean offset between the two data sets is ($1.6 \pm 0.1$)\ km\,s$^{-1}$. This offset was applied to the data in this paper to give the absolute RV values shown in Table~\ref{targets}. 

\cite{Geller2010a} estimated RV uncertainties of 0.6\,km\,s$^{-1}$. The 2$\sigma$ clipped standard deviation of the difference between the two data sets is 0.8\,km\,s$^{-1}$, consistent with a similar precision for our measurements. The precision of individual RVs was estimated more directly by comparing RVs from subsets of the summed spectra as a function of the FWHM (where FWHM refers here to the width of the CCF) and SNR, although the comparison was hampered by the difficulty in measuring the sky line correction at low cumulative exposure times. This gave an estimated precision of $\sigma_{\rm RV}=0.90\times $\,FWHM$/$SNR in km\,s$^{-1}$ (where FWHM is $\sim 30$ km\,s$^{-1}$ for a slowly rotating star). As a cross-check the precision was estimated using the empirical formula derived in \cite{Jackson2018a} for RV measurements in the Pleiades/Praesepe clusters using a similar but {\it not} identical WIYN/Hydra set up. The results were, for practical purposes, the same.   

\subsection{Equivalent width of the 6708\AA~Lithium line}
\label{s2.4}

The equivalent width (EW) of the Li~{\sc i}~6707.8\AA~line (hereafter, EW(Li)) was measured by comparing the target spectrum (corrected to a rest wavelength scale) to a template spectrum, with no lithium, matched to the target $T_{\rm eff}$ in 100\,K steps. The synthetic spectra were generated using the {\sc moog} software \citep{Sneden2012a} and solar-metallicity Kurucz model atmospheres \citep{Kurucz1992a}.  The template spectrum was broadened to match the measured FWHM and scaled to match the target spectrum either side of the Li line as shown in Fig.~\ref{fig2}. This latter step ensures (and confirmed with simulations) that EW(Li) is correctly estimated, without systematic bias, for rapid rotators. A weighted profile $P(\lambda$) was used to measure EW(Li) from the difference between the target ($S_{T}(\lambda)$) and template ($S_{C}(\lambda)$) spectra;
\begin{equation}
	{\rm EW(Li)} = \int [S_{C}(\lambda)-S_{T}(\lambda)]P(\lambda)\, d\lambda \ / \int
	P(\lambda)^2\, d\lambda\, ,
\end{equation}
where $P(\lambda)$ is a Gaussian profile with the FWHM of the CCF (see Fig.~2). 
There is a weak (10-20\,m\AA) Fe~{\sc i} line at 6707.4\AA\ that is blended with the Li line in all our spectra. The template subtraction accounts for this blend (and any others), but EW(Li) may have been underestimated by 3-6\,m\AA\ if M35 has a subsolar metallicity (see Section~\ref{s2.1}). 
The uncertainty in EW(Li) was estimated as the RMS value of the EWs measured using the same procedure with $P(\lambda)$ centred at five wavelengths either side of the Li line.  These error bars were validated by comparing EW(Li) measured from individual spectra from different nights prior to any summation. EW(Li) and its error bar are listed in Table~\ref{targets}.

\section{Cluster Membership, Stellar Parameters and Lithium Abundances}
\label{s3}

\subsection{Membership Probabilities}
\label{s3.1}

Target RA and Dec were cross matched with Gaia DR2 data \citep{Gaia2018a} to give proper motions (pmRA and pmDec) and parallax data for 337 objects in our sample. RVs and proper motion velocities  ($V_{\rm RA}=4.74\, d_C\,{\rm pmRA}$~and~$V_{\rm Dec}=4.74\, d_C\,{\rm pmDec}$ were used to determine the three dimensional (3D) velocity of the observed stars, where $d_C = 885$\,pc. The kinematic distribution of 331 stars with absolute values of $V_{\rm RA},V_{\rm RA}$~and $|$RV$|<50$\,km\,s$^{-1}$ was modelled with a pair of 3D Gaussians, one narrow component representing the cluster and a broader component to represent any residual contamination. A maximum likelihood method was used to find the best-fit cluster velocities and intrinsic dispersions, taking into account the uncertainties in each measurement, and to estimate membership probabilities  \citep[see Appendix~\ref{appa} and][]{Jackson2020a}. The estimated RV of the cluster centre $-8.10\pm 0.07$\,km\,s$^{-1}$ compares well with values of $-8.16 \pm 0.05$\,km\,s$^{-1}$ from \cite{Geller2010a}\footnote{Though note that our RV values were offset to agree with Geller et al. (2010).} and $-7.70 \pm 0.27$\,km\,s$^{-1}$ determined by \cite{Gaia2018b}. 

Membership probabilities for individual targets are shown in Table~\ref{targets}. Cluster members for subsequent analysis were defined as having $p_{\rm mem}>0.95$ {\it and} a measured rotation period, giving 244 cluster members, and an expected number of contaminants (from the sum of $1-p_{\rm mem}$ for these 244 targets) of just 0.4.

Figure~\ref{fig1} shows how cleaning the sample of less probable members sharpens up the cluster sequence in the colour-magnitude diagram and also more clearly delineates the slow-rotating "I-sequence" and fast-rotating "C-sequence" for M35 in a plot of rotation period versus colour \citep[e.g.][]{Barnes2003a}.

\subsection{SED fitting}
\label{s3.2}
\begin{table*}
\caption{The properties of cluster members (with $p_{\rm mem}>0.95)$ for which rotation periods and SED fits are available. The columns give effective temperatures and luminosities from the SED fits (Section~\ref{s3.2}), NLTE lithium abundances (Section~\ref{s3.3}, error bars of zero indicate an upper limit), the relative over-radius (Section~\ref{s3.4}), the deviation of EW(Li) from the trend defined by slowly rotating stars (positive means a higher EW(Li)), the rotation period (repeated from Table~\ref{targets}, the Rossby number and a flag indicating binary status (Section~\ref{s3.4}, $1 =$ likely binary).}
\begin{tabular}{lrrrrrrrr}
\hline
Object  &	$T_{\rm eff}$ & $\log L/L_{\odot}\ $	& $A$(Li)$\ \ $ & $\rho$	&$\Delta$EW$_{\rm Li}$ & Period & $\log N_R$ & Binary \\
 & K$\ \ $ &	&	& &	m\AA$\ \ \ $& d$\ \  \ $ &	& \\\hline
J06070616+2402101 & 4900 & $-0.597\pm 0.004$ & $1.91^{+0.10}_{-0.10}$ & 0.971 & $31\pm 17$ &   7.087 & $-0.492$ & 0 \\
J06072249+2421401 & 4900 & $-0.418\pm 0.006$ & $2.45^{+0.21}_{-0.19}$ & 1.194 & $147\pm 39$ & 0.911 & $-1.335$ & 0 \\
J06072843+2416426 & 5700 & $-0.030\pm 0.003$ & $2.66^{+0.07}_{-0.07}$ & 1.134 & $-13\pm 11$ & 2.001 & $-0.697$ & 0 \\
... & ... & ... & ... & ... & ... & ... & ... & ...	\\\hline
\end{tabular}
\label{resultstable}
\end{table*}
The spectral energy distributions (SEDs) of targets were modelled for the purposes of estimating luminosities, $T_{\rm eff}$ and hence radii. 

The available photometry is summarised in Appendix~\ref{appb}. The observed SEDs were analysed using the Virtual Observatory SED Analyser \citep[VOSA - version 6.0;][]{Bayo2008a}. Observed SEDs were built assuming a fixed cluster distance of 885\,pc and de-reddened using a fixed $A_V=0.62$ (i.e. assuming all targets are members of M35). These were then compared with synthetic SEDs derived using BT-NextGen-GNS93 model atmosphere \citep{Allard2012a}, assuming $\log g = 4.5$ and [Fe/H]$=0$, to determine the best fit luminosity and $T_{\rm eff}$ using chi-squared minimisation. The uncertainty in $\log L/L_{\odot}$ quoted in Table~\ref{resultstable} is estimated from the chi-squared minimisation, is usually $<0.01$ dex and is likely comparable with uncertainty due to distance spread for stars within the cluster  (and ignores the systematic uncertainty associated with error in the mean cluster distance, which could be as large as 0.08~dex). The statistical uncertainty in $T_{\rm eff}$ is usually much less than the 100\,K grid spacing of the atmosphere models; the 1-sigma $T_{\rm eff}$ uncertainty is set to $\pm 50$\,K. Two members exhibiting a very poor fit to $K_s$ in the SED were cut from the sample, leaving a total of 242 $P_{\rm mem}>0.95$ members with $P$, $\log L/L_{\odot}$ and $T_{\rm eff}$ values. The results for these objects are shown in Table~\ref{resultstable}.

 \subsection{Lithium abundance}

\label{s3.3}

Armed with EW(Li) and $T_{\rm eff}$ for each star, the abundance of lithium, expressed as $A$(Li)$ = 12 + \log N({\rm Li})/N({\rm H})$ was estimated 
    using a spline interpolation of the curves of growth given by \cite{Soderblom1993a}. These 1D LTE abundances were adjusted using the 3D NLTE corrections provided by the {\sc breidablik} code\footnote{https://github.com/ellawang44/Breidablik} (E. Wang, private communication), from interpolating synthetic spectra from the {\sc stagger} 3D model atmosphere grid \citep{Magic2013a}. The 1D LTE to 3D NLTE additive corrections range from $+0.1$~dex for the coolest stars in the sample to $-0.3$ dex for the most Li-rich stars at $T_{\rm eff} \simeq 5200$\,K. The overall effect of the correction is to slightly decrease the inferred spread of Li abundance at a given $T_{\rm eff}$. The uncertainties in the abundances are calculated by propagating the error bars in EW(Li) and $T_{\rm eff}$ as independent sources of uncertainty. The error bars are asymmetric because the relationship between EW(Li) and $A$(Li) is non-linear. Note also that the uncertainties in $A$(Li) and $T_{\rm eff}$ are strongly correlated, with an over-estimated $T_{\rm eff}$ leading to an over-estimated abundance. The effects of EW(Li) and $T_{\rm eff}$ uncertainties have a comparable size for most objects. For a few cool stars where the error bar in EW(Li) makes the EW compatible with zero, we use $2\,\Delta$EW(Li) to define an upper limit to $A$(Li), which also includes the uncertainty due to $T_{\rm eff}$.

\subsection{Radius inflation and defining a sample of probable binary systems}
\label{s3.4}
\begin{figure}
	\centering
	\includegraphics[width = 85mm]{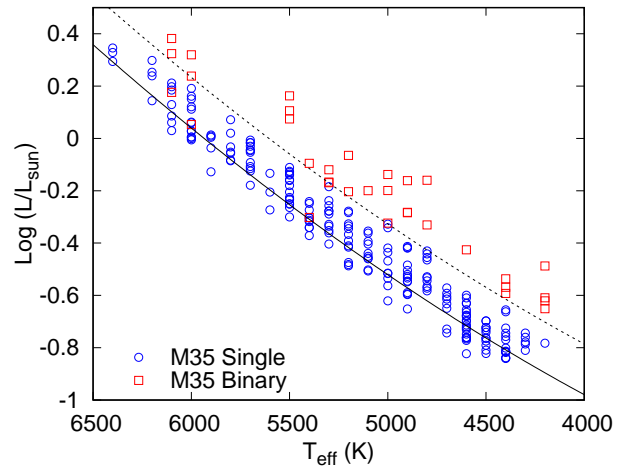}
	\caption{The Hertzsprung-Russell diagram for members of M35 with measured rotation periods. $\log L/L_{\odot}$ and $T_{\rm eff}$ were determined by SED fitting (see Section~\ref{s3.2}). The solid reference line is an empirical fit used to define the over-radius. Red squares denote objects likely to be binary systems (see Section~\ref{s3.4}).}
\label{fig4}
\end{figure}
Figure~\ref{fig4} shows a Hertzsprung--Russell (HR) diagram for the $p_{\rm mem}>0.95$ members of M35 that have measured rotation periods. The solid line shows a second order polynomial defining the lower quartile value of $\log L/L_{\odot}$ versus $T_{\rm eff}$. This is {\it not} an empirical isochrone, it is a reference line from which the relative increase in luminosity, $\Delta \log L/L_{\odot}$ of individual stars can be measured at a given $T_{\rm eff}$. This was used to define an {\it apparent} "over-radius" $\rho = \sqrt{10^{\Delta \log L}}$, that corresponds to the factor by which the stellar radius needs to increase to produce the observed $\Delta \log L/L_{\odot}$, assuming no contribution from a binary companion. 

In practice it is not possible to separate the effects of over-radius from the effects of binarity using the HR diagram alone, so $\rho$ is an upper limit to the true over-radius. Our target list will include a fraction of near equal mass binaries which will show $\Delta \log L/L_{\odot} \sim 0.3$. The dotted line in Fig.~\ref{fig4} separates out 28 stars with $\rho >1.25$ ($\Delta \log L/L_{\odot} > 0.194$), as objects that are more likely to be binaries than other cluster members, and which can be examined separately in subsequent analyses. Stars below this cut will still be a mixture of single and (lower mass ratio) binary stars. Note also that the membership probability calculation in Section~\ref{s3.1}, may have filtered out a small number of (short period) binaries with RV measurements that are discrepant from the cluster mean. A further three objects with $\rho <1.25$, identified as spectroscopic binaries\footnote{J060816660+2400372, J06083296+2408164, J06083644+2404530.} by Geller et al. (2010) because they exhibited small, but significant, RV variations in that paper, are also flagged as probable binary members.

\section{Results}
\label{s4}
\begin{figure*}
	\centering
\begin{minipage}[b]{0.48\textwidth}
    \includegraphics[width=\textwidth]{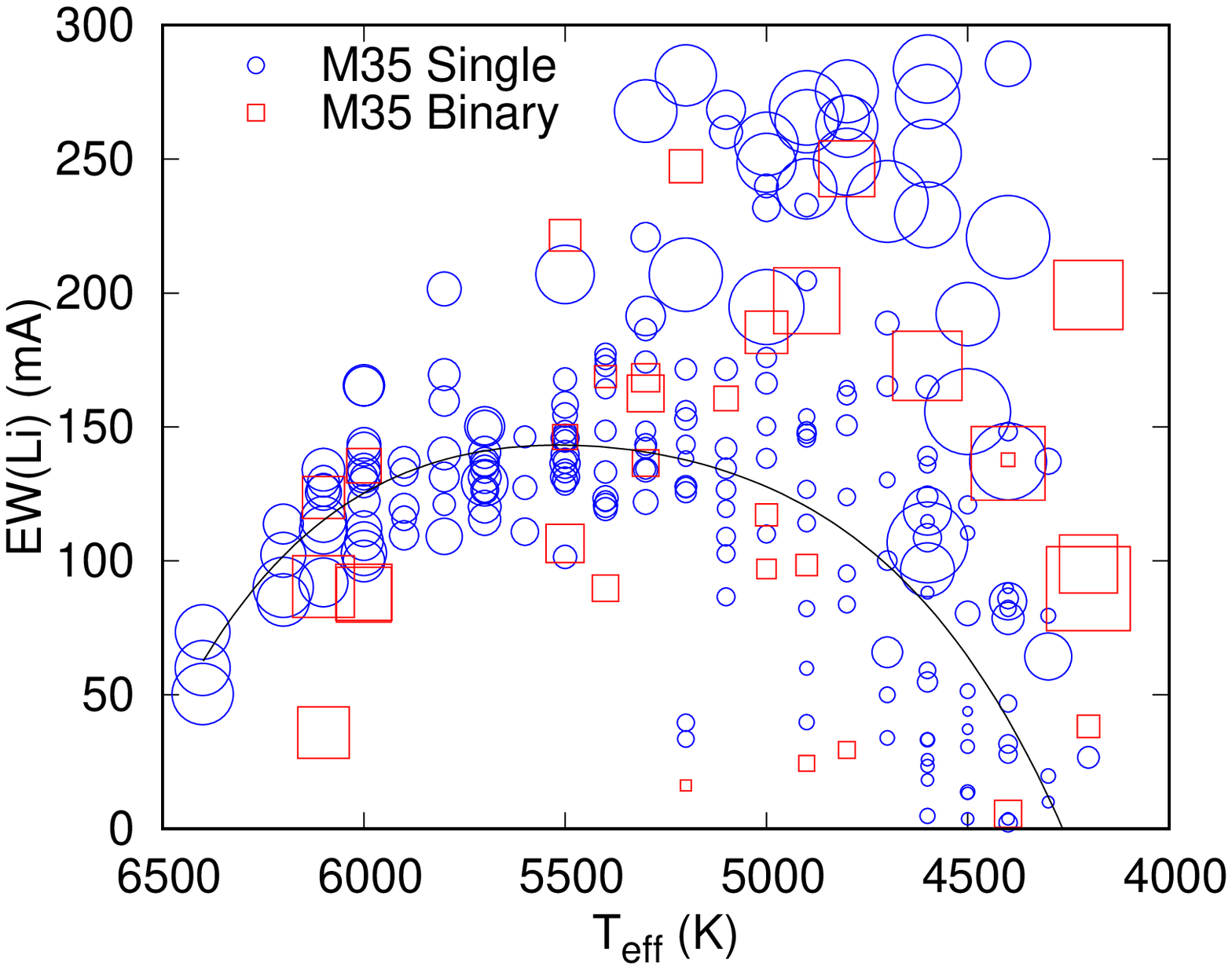}
  \end{minipage}
  \hfill
  \begin{minipage}[b]{0.48\textwidth}
    \includegraphics[width=\textwidth]{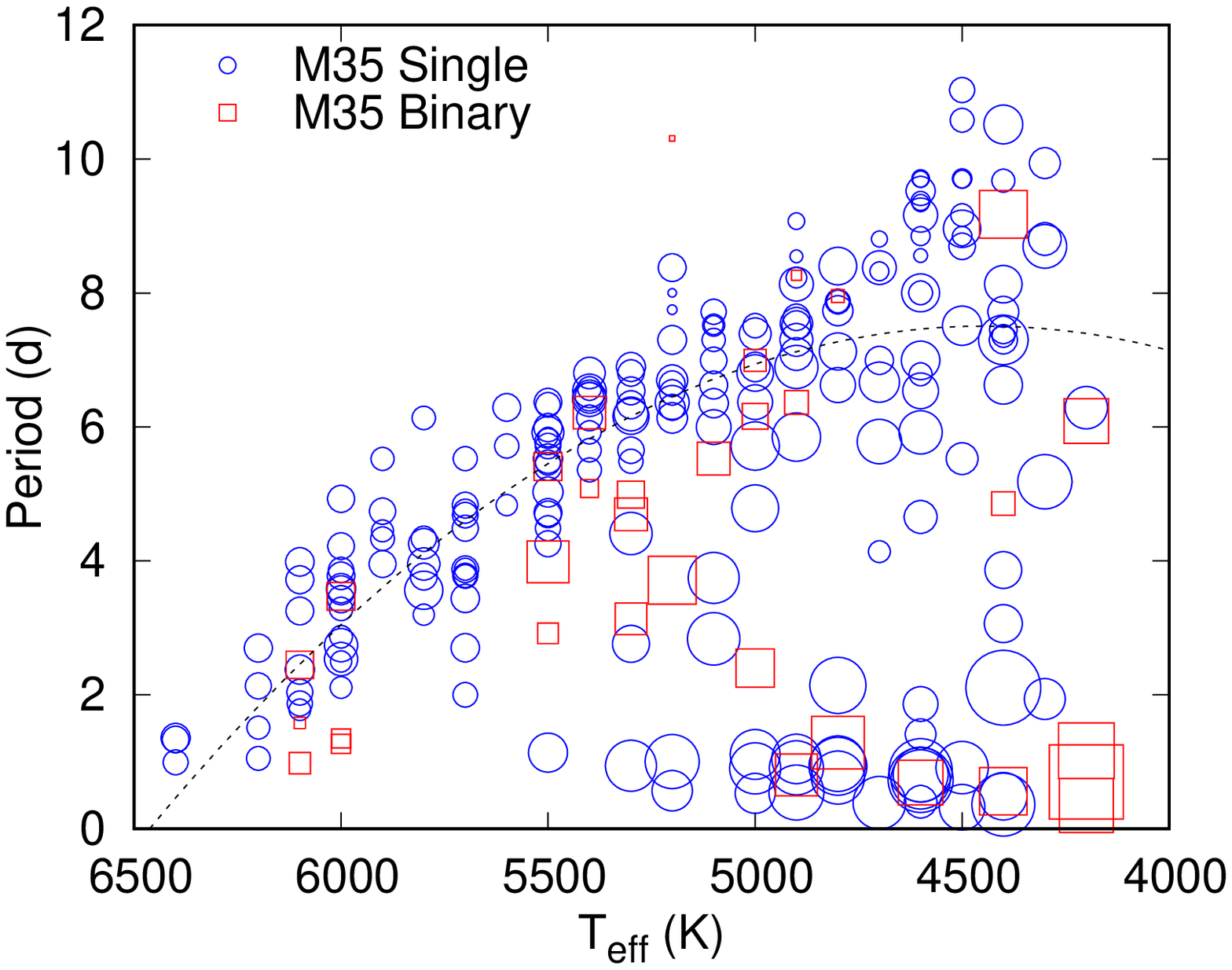}
  \end{minipage}
	\caption{The equivalent width of the Li~{\sc i} line and rotation period versus $T_{\rm eff}$ for the sample of $p_{\rm mem}>0.95$ members with rotation periods. (a) The symbol size is proportional to the logarithm of the angular velocity. The solid line is a fourth-order polynomial fitted to the slowest rotating half of the sample in 250\,K bins. (b) The symbol size is proportional to $\Delta$EW$_{\rm Li}$, the difference between EW(Li) and the solid curve in (a). The dashed line shows a quadratic fit to the median rotation period in 250\,K bins.} 
\label{fig5}
\end{figure*} 
Table~\ref{targets} lists the positions $V$, $K_s$ and $B-V$ photometry used in the selection of all observed targets, along with a rotation period \citep[if available from either][in that precedence order]{Meibom2009a, Libralato2016a}), along with the SNR of the combined Hydra spectra for that target, its measured $RV$, the FWHM of its CCF, EW(Li), EW(Ca) (see Section~\ref{s4.1})  and calculated membership probability. For the 242 $p_{\rm mem}>0.95$ members with rotation periods and a valid SED fit, Table~\ref{resultstable} gives the $T_{\rm eff}$, luminosity, $A$(Li), relative over-radius and a flag indicating whether the object is a potential binary system (Section~\ref{s3.4}). 

\subsection{Trends of lithium with effective temperature and rotation}
\label{s4.1}

\begin{figure}
	\centering
	\includegraphics[width = 85mm]{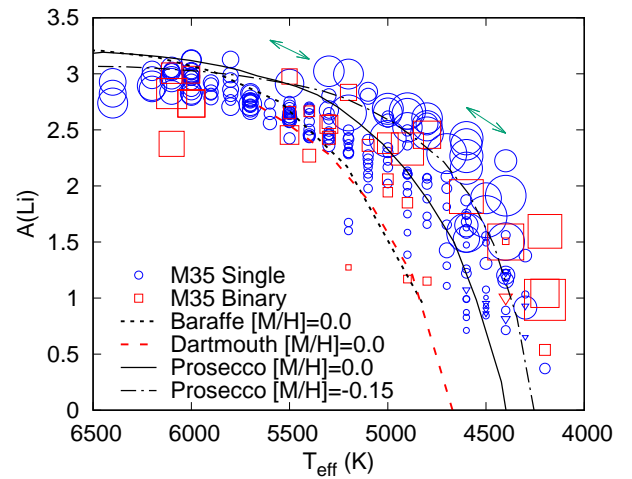}
	\caption{The NLTE lithium abundance of M35 members versus $T_{\rm eff}$. Symbols sizes are proportional to the logarithm of angular velocity. Triangles denote upper limits. The lines represent the predictions of various "standard" stellar evolutionary models for an age of 120 Myr, assuming an initial Li abundance of $3.26+$[Fe/H] (see Section~\ref{s4.4}). The double-headed arrows show the effects of a $\pm 100$ K uncertainty in $T_{\rm eff}$ on the inferred $A$(Li) at two different temperatures.} 
	\label{fig6}
\end{figure}

Figure~\ref{fig5} shows the basic observational results of our investigation.
Figure~\ref{fig5}a plots EW(Li) as a function of $T_{\rm eff}$, with  a symbol size proportional to $\log (1/{\rm Period})$.  A fiducial fourth order polynomial was fitted in 250\,K bins to the median of the slowest rotating half of the sample and is shown as a solid line\footnote{${\rm EW(Li)}= -31530 +22.89T_{\rm eff} - 0.006261T_{\rm eff}^2 +7.6828\times 10^{-7}
T_{\rm eff}^3 -3.56675\times 10^{-11}T_{\rm eff}^4$}. Likely binary stars are identified. Figure~\ref{fig5}b shows how rotation period varies with $T_{\rm eff}$. The dashed line shows a quadratic fit to the median rotation period in 250\,K bins\footnote{$P = -28.87 +0.016371T_{\rm eff}^2 -1.84186\times 10^{-6}T_{\rm eff}^2$}. Stars above this line comprise the sample used to define the locus in Fig.~\ref{fig5}a and the symbol size now varies (linearly) with by how much EW(Li) differs from that locus. This quantity is referred to as $\Delta$EW$_{\rm Li}$. Larger symbols mean a star has a larger EW$_{\rm Li}$ at a given $T_{\rm eff}$. Figure~\ref{fig6} shows the same stars with EW(Li) replaced by $A$(Li).

The most obvious results from Figs.~\ref{fig5} and ~\ref{fig6} are that for $T_{\rm eff} <5500$\,K there is a clear trend that faster rotating stars have larger EW(Li) and larger Li abundance at a given $T_{\rm eff}$. The total spread in EW(Li) reaches $\sim 300$m\AA\ at $T_{\rm eff} \sim 4500\,K$, corresponding to $\sim 2$ orders of magnitude in $A$(Li). The median uncertainty in EW(Li) is 13\,m\AA\ and the median uncertainty in $A$(Li) is 0.08 dex, so the dispersion is much larger than any plausible star-to-star measurement uncertainties. There is also a hint that for stars with $T_{\rm eff}>5900$\,K that the opposite trend may be true, though the range of rotation rates and EW(Li)/$A$(Li) is much smaller.

\begin{figure}
	\centering
	\includegraphics[width = 85mm]{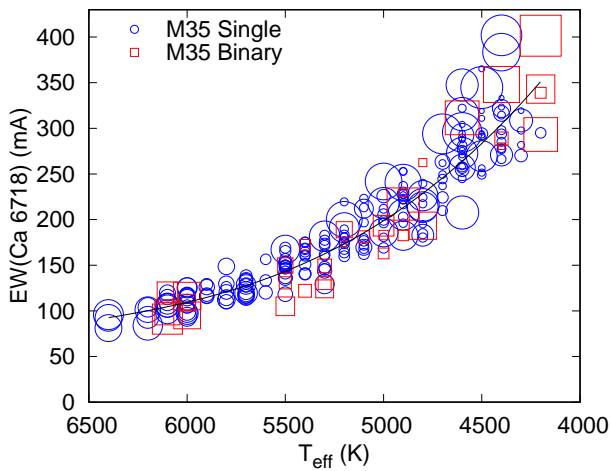}
	\caption{EW of the Ca~{\sc i}~6717.7\AA\ line versus  $T_{\rm eff}$. The meanings of symbol types, sizes and colours are the same as in Fig.~\ref{fig5}a. The solid line is a cubic fit to the data.} 
	\label{fig7}
\end{figure}

We considered whether systematic measurement error of EWs in rapidly rotating stars might play some role in these relationships. There are several other lines (mainly Fe~{\sc i}) close to the Li~{\sc i} line which are blended-in for fast rotating stars. However, our measurement technique is differential in that the target spectrum is compared with the fiducial spectrum of a similar star and should be relatively immune to such error. To test this, the EW of the neighbouring Ca~{\sc i} line at 6717.7\AA\ line (EW(Ca)) was measured in the same way. The results are shown in Fig.~\ref{fig7}. This line has a similar strength to the Li~{\sc i} line and is equally affected by blending at fast rotation rates. There is no indication that fast rotating stars have larger EWs; the RMS dispersion around a cubic fit to the mean relation is 20\,m\AA\ for slow and intermediate rotators, increasing to 40\,m\AA\ for the fastest rotating quartile, with no significant systematic offset. These dispersions are consistent with (actually, slightly smaller than) the RMS measurement uncertainties of 22\,m\AA\ and 47\,m\AA\ for the same stars, giving further confidence in the robustness of our EW uncertainty estimates. Note, we choose $T_{\rm eff}$ as the ordinate for these relationships rather than colour. The dispersion in EW(Li) and EW(Ca) would appear larger if plotted versus (e.g.) $B-V$ and the dispersion in EW(Ca) would also show some rotation dependence. The colours of active stars appear to be changed by activity and starspot coverage \citep{Stauffer2003a} and this is explored further in Sections~\ref{s4.3} and \ref{s5.2}.    
 
\subsection{A more detailed look at the Li-rotation correlation}

\label{s4.2}

\begin{figure*}

	\begin{minipage}[t]{0.44\textwidth}
	\includegraphics[width=75mm]{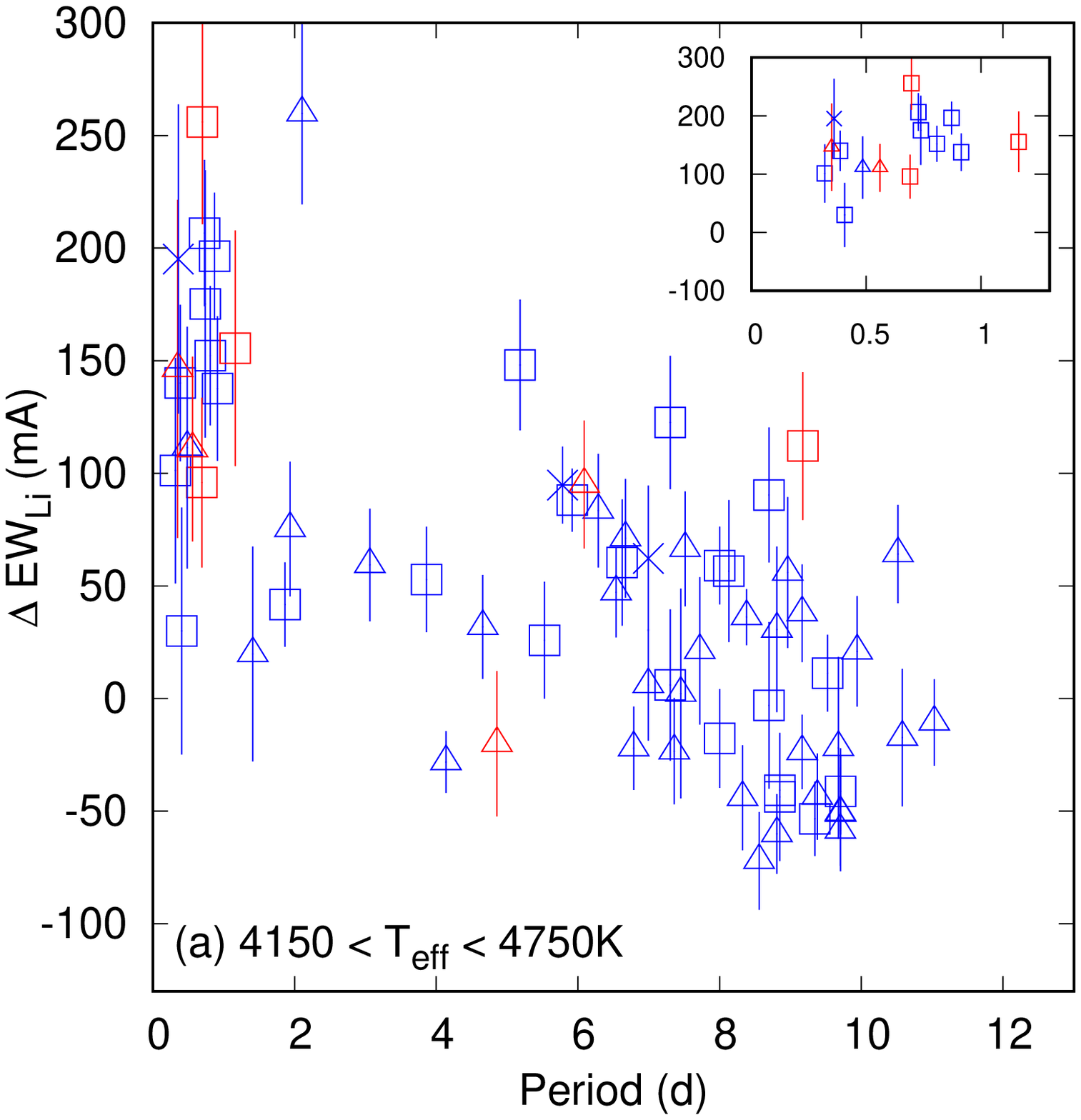}
	\end{minipage}
	\hfill
	\begin{minipage}[t]{0.5\textwidth}
	\includegraphics[width=75mm]{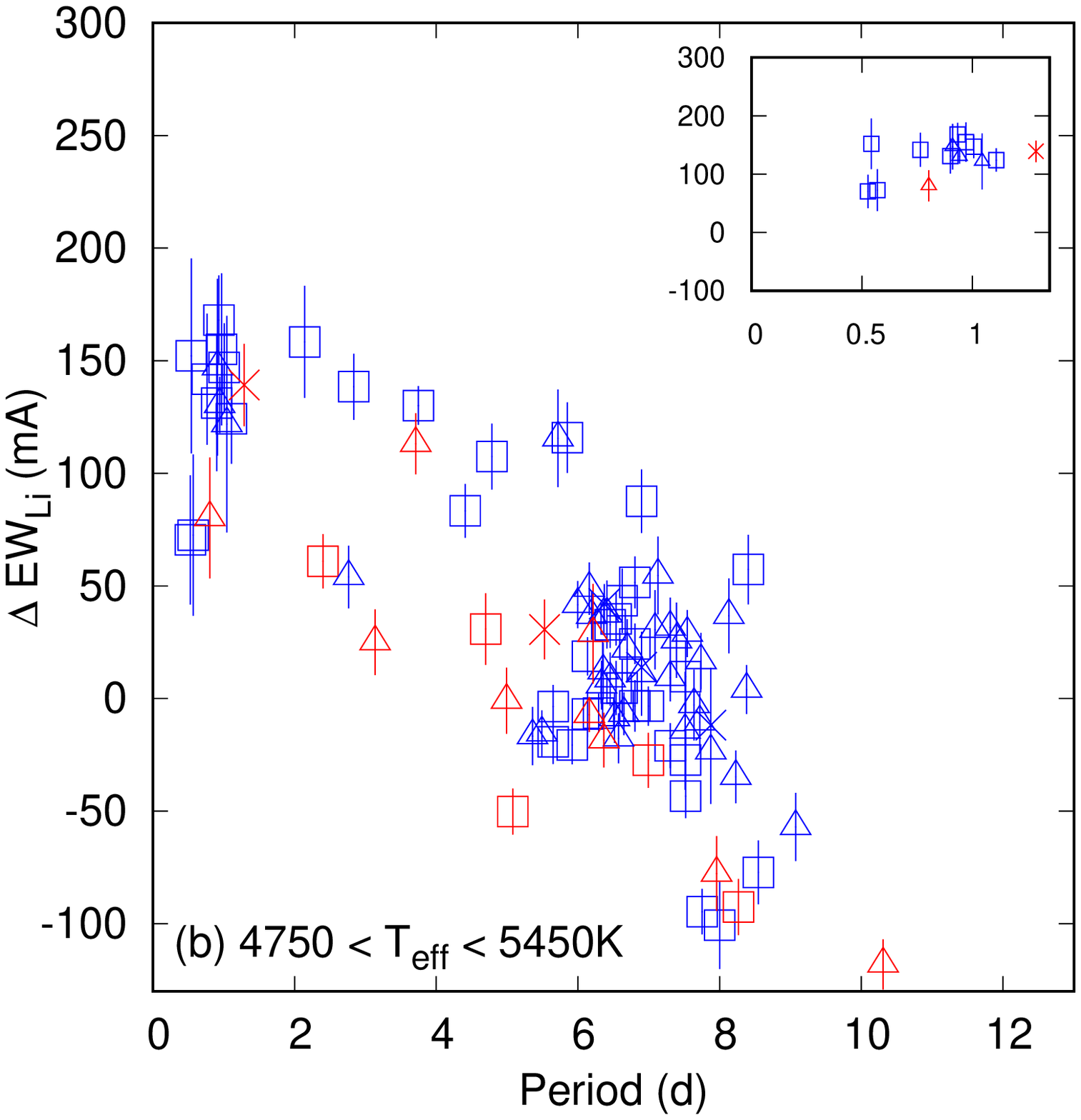}
	\end{minipage}
	\hfill
	\newline
	\begin{minipage}[t]{0.44\textwidth}
	\includegraphics[width=75mm]{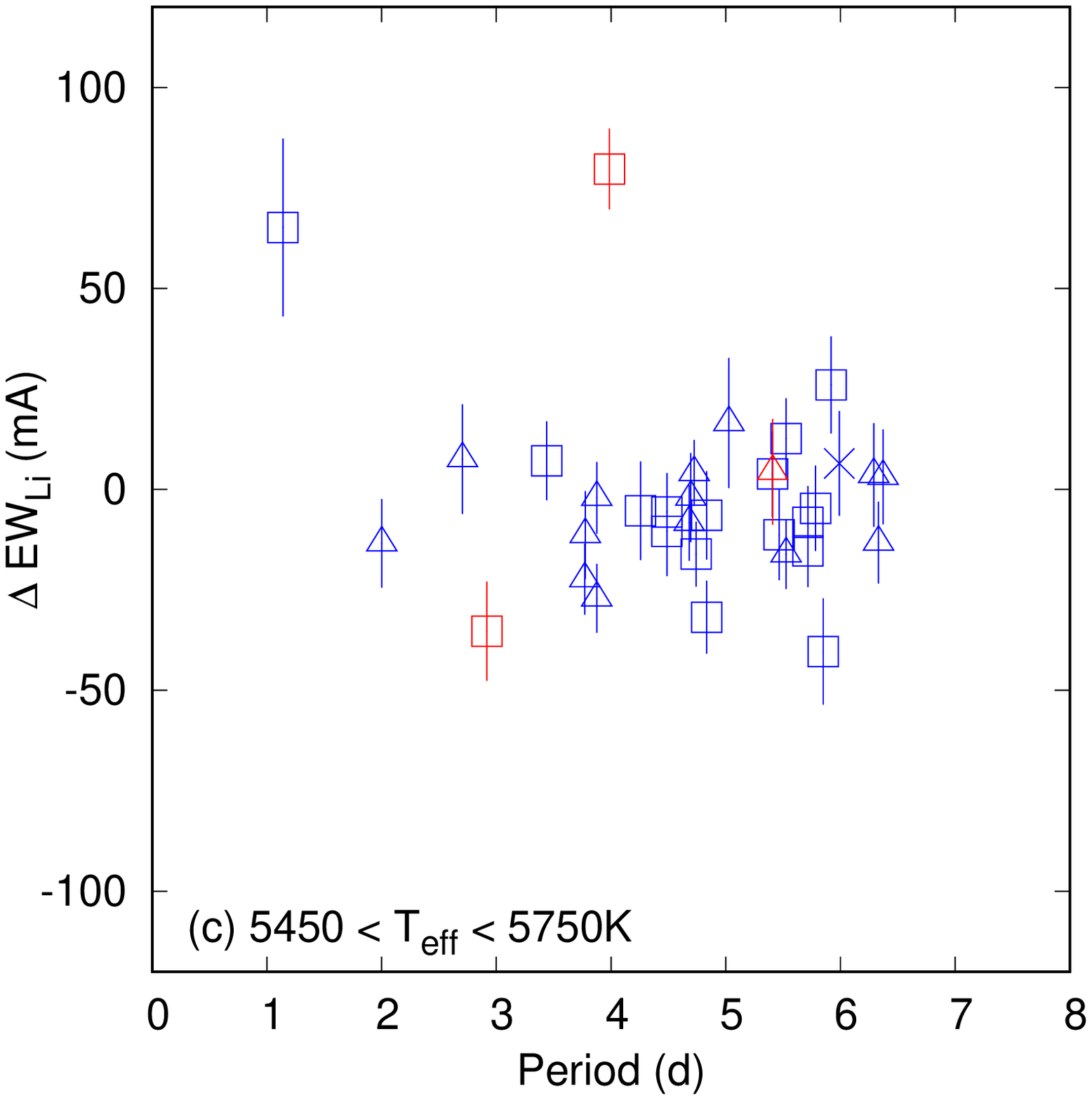}
	\end{minipage}
	\hfill
	\begin{minipage}[t]{0.44\textwidth}
	\includegraphics[width=75mm]{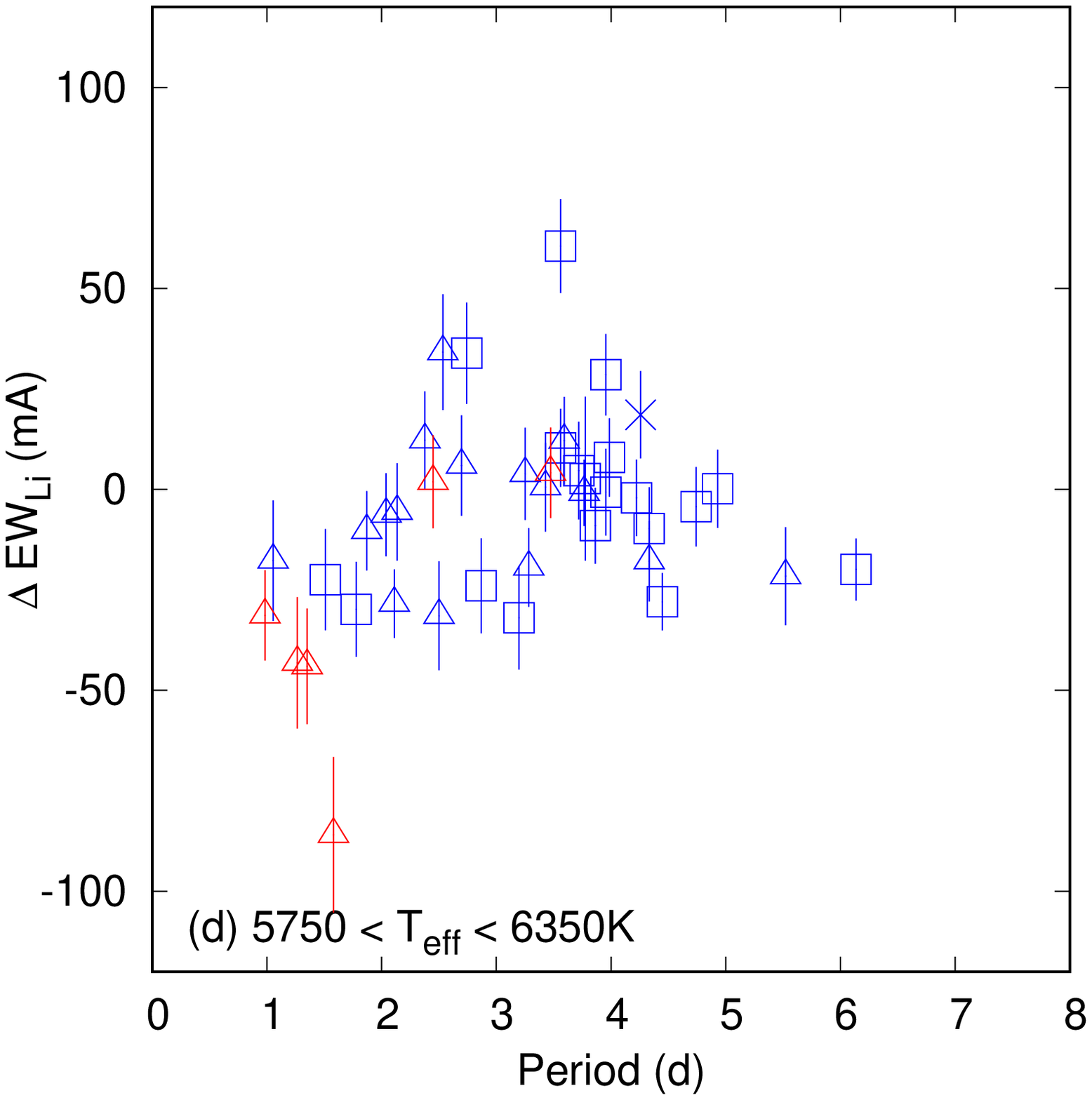}
	\end{minipage}
	\hfill	
	\caption{$\Delta$EW$_{\rm Li}$ versus rotation period for stars in four temperature ranges. The symbols encode whether targets have consistently measured periods from two sources (squares), period from just one source (triangles), or periods from two sources which are inconsistent (crosses). Blue and red symbols denote probable single and probably binary stars respectively. The inserts in plots (a) and (b)  show the rapidly rotating stars in more detail. Note the different axis ranges in the plots on the upper and lower row.}
	\label{fig8}
\end{figure*}

To investigate the correlation of EW(Li) with rotation in more detail, Fig~\ref{fig8} shows how $\Delta$EW$_{\rm Li}$ depends on rotation period for stars in four temperature ranges, illustrating a diversity of behaviour.

Figures~\ref{fig8}a and~\ref{fig8}b, containing the coolest stars in our sample, show the first important result, that there is a clear dependence of $\Delta$EW$_{\rm Li}$ on rotation period, reflecting the impression gained from Fig.~~\ref{fig5}a that this behaviour is confined to stars cooler than 5500\,K. The correlation appears cleaner for stars with $4750 < T_{\rm eff} < 5450$\,K, though this is mostly, if not entirely, explained by the larger measurement errors for the cooler, fainter stars in the sample.

A second important result is that these correlations are not perfect or single-valued. The scatter in Fig.~~\ref{fig8}b is larger than the measurement uncertainties. In particular, there is a significant dispersion for periods greater than 2 days, but perhaps not for faster rotating objects, where the observed scatter is consistent with the error bars. The same dispersion appears to be present in Fig.~~\ref{fig8}a, but the scatter due to uncertainties is larger. Note that uncertainties in $T_{\rm eff}$ inject some scatter into this diagram via the definition of the baseline locus for slow rotators in Fig.~\ref{fig5}a. For $\pm 50$\,K $T_{\rm eff}$ uncertainties, this additional error is about $\pm$10--20 m\AA\ in the coolest stars of the sample, negligible for those in Figs.~\ref{fig8}b and c, increasing again to $\pm 10$ m\AA\ for the hottest stars, and has been included in quadrature with the $\Delta$EW(Li) uncertainties in the error bars shown in Fig.~\ref{fig8} (but not in the values listed in Table~\ref{resultstable}). The picture would not change much even if the $T_{\rm eff}$ uncertainties were doubled.

To investigate whether there is any possibility that rotation period unreliability  plays a role in this dispersion (the formal uncertainties are very small), a comparison of periods was made for 120 $p_{\rm mem}>0.95$ objects with independent measurements available in both \cite{Meibom2009a} and \cite{Libralato2016a}. 

For 110 of the objects there is good or reasonable (less than 20 per cent difference) agreement on the period (marked as squares in Fig.~\ref{fig8}). For 9 of the 10 objects with a larger disagreement (marked with crosses in Fig.~\ref{fig8}) the K2 period is much shorter than the period found by \cite{Meibom2009a} and in 6 of these cases the K2 period is close to half that of Meibom et al. This suggests that \cite{Libralato2016a} may have identified a false period associated with an antisymmetric pair of spot groups. The K2 dataset of Libralato et al. covers a shorter observing window than does Meibom et al.'s data, so may be vulnerable to this type of period misidentification. This is why Meibom et al.'s period was adopted here where it is available.  Of the 122 objects with just one independent measurement of the period (103 of which were measured by K2), then we might expect another handful of spurious (probably underestimated) periods.
Overall then, the main results and trends noted above appear robust; only one or two, but not all, of the discrepant objects that define a dispersion in the Li-rotation correlations might be explained as due to erroneous identification of rotation periods. 

A third result is that the $\Delta$EW$_{\rm Li}$-rotation correlation either vanishes or even reverses at $T_{\rm eff}>5500$\,K. The interval $5450 < T_{\rm eff} <5750$\,K shown in Fig.~\ref{fig8}c is characterised by very little spread around the mean relationship for both period and  $\Delta$EW$_{\rm Li}$. In Fig.~\ref{fig8}d there is a broader dispersion in rotation period and some indication that slower rotators have larger $\Delta$EW$_{\rm Li}$ than faster rotators. However, the significance of this result is low because the steepness of the $T_{\rm eff}$-dependence of both rotation rate and EW(Li) in this temperature range (see Fig.~\ref{fig5}), combined with $T_{\rm eff}$ uncertainties, introduces correlated errors that would lead to such a correlation: a positive $T_{\rm eff}$ error leads to the inference that an object is a slow rotator for its $T_{\rm eff}$ and also upwardly biases $\Delta$EW$_{\rm Li}$ (and vice versa). This is much less of an issue at lower temperatures.

A fourth important result emerging from Fig.~\ref{fig8}, and made possible by the large sample size, is that stars that are probable components of binaries (shown with red symbols) appear to behave in the same way to the rest of the sample. Thus although there may be unrecognised binaries (with small mass ratios) in the "single" star sample, the mere fact that they are in binary systems does not appear to drive directly the lithium-rotation correlation or the scatter that has been identified at a given rotation rate. Note that most of these binaries are likely to have widely separated components. By using RV as part of the membership selection criteria it is possible that some very close binary systems have been excluded from the sample and it may be that in these cases the presence of tidally interacting components could directly influence Li depletion \citep[e.g.][]{Thorburn1993a}.

\subsection{M35 and the Pleiades}

\label{s4.3}
\begin{figure*}
	\centering
\begin{minipage}[b]{0.32\textwidth}
    \includegraphics[width=\textwidth]{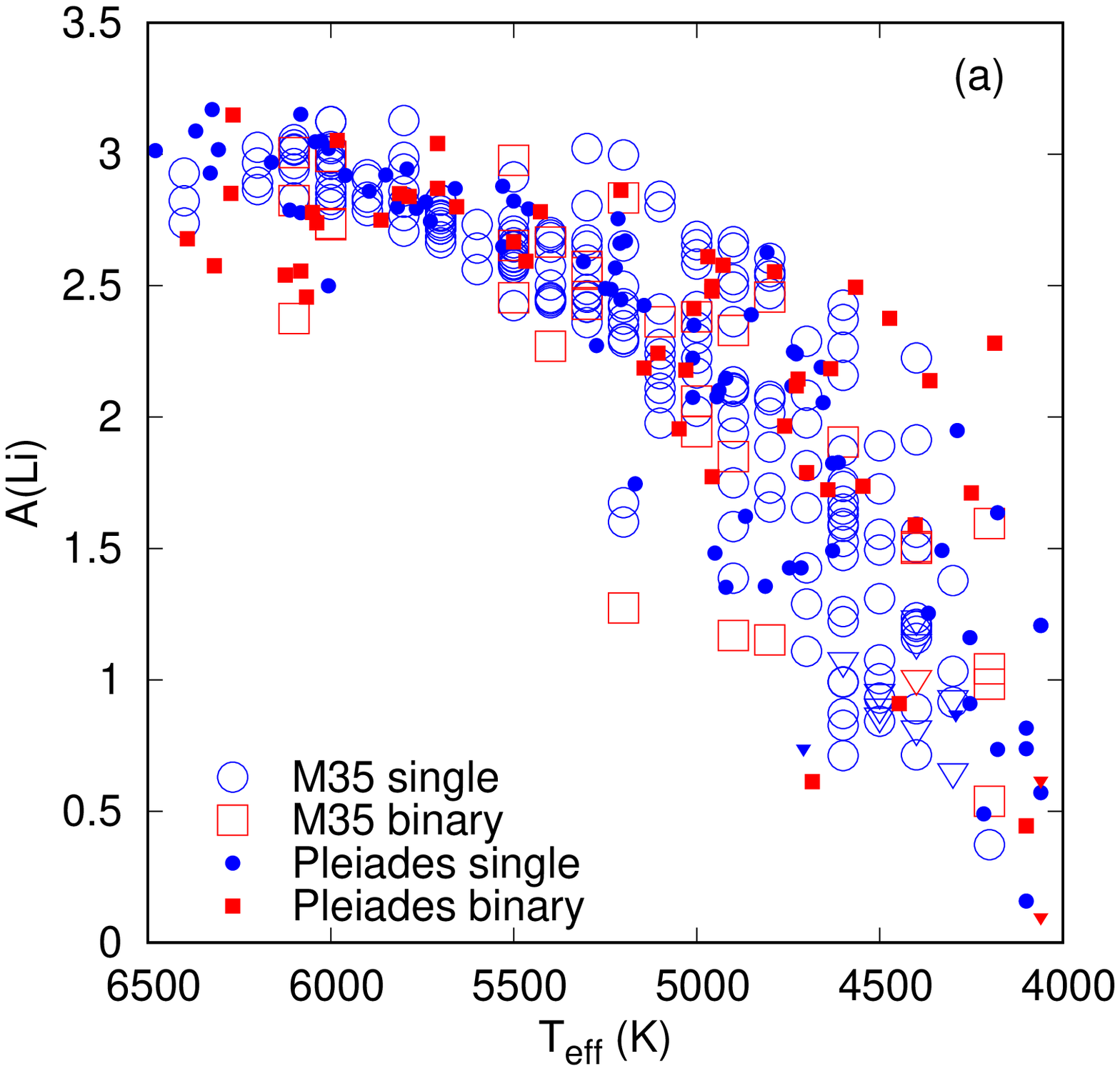}
  \end{minipage}
  \hfill
  \begin{minipage}[b]{0.32\textwidth}
    \includegraphics[width=\textwidth]{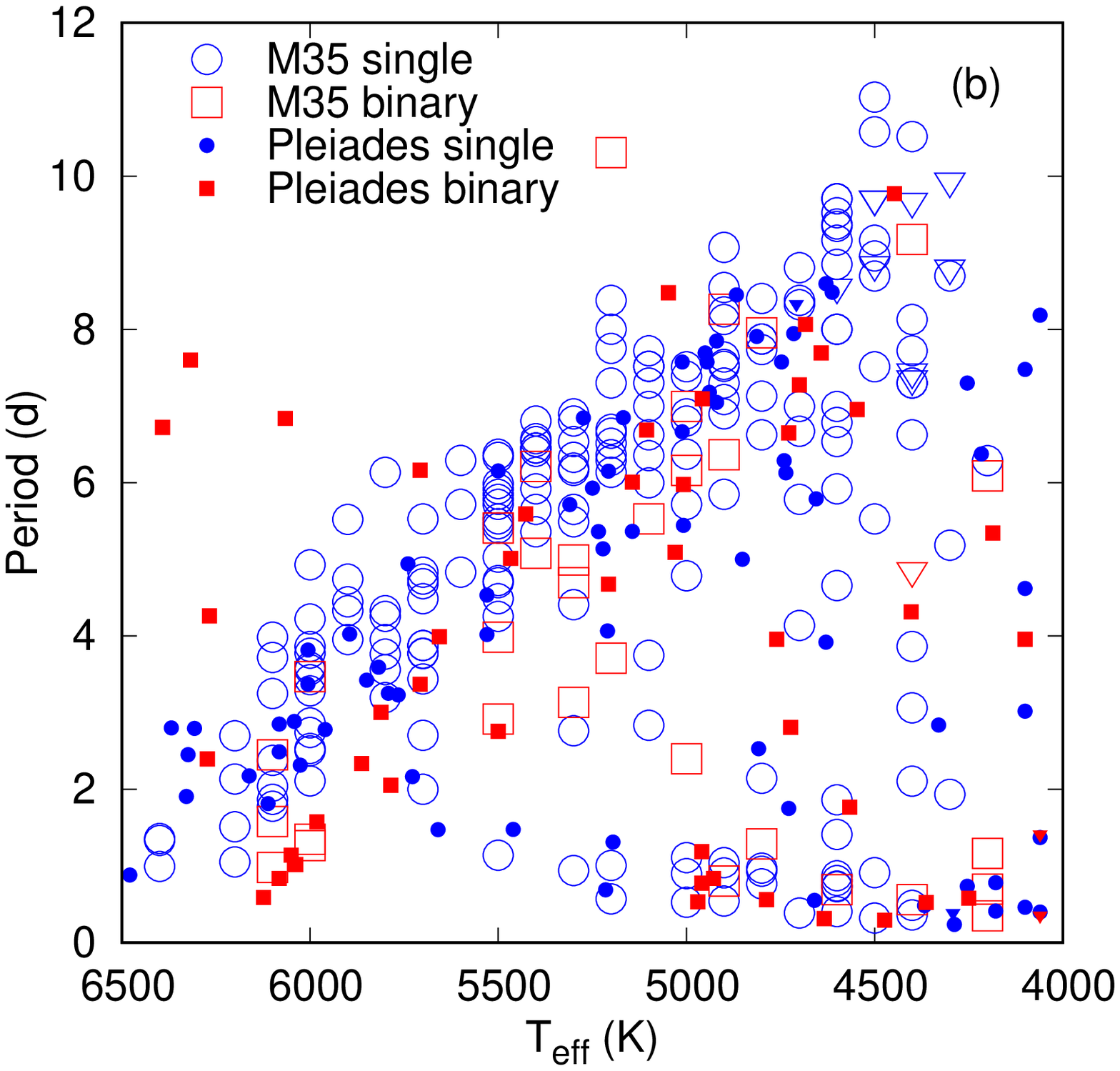}
  \end{minipage}
\begin{minipage}[b]{0.32\textwidth}
    \includegraphics[width=\textwidth]{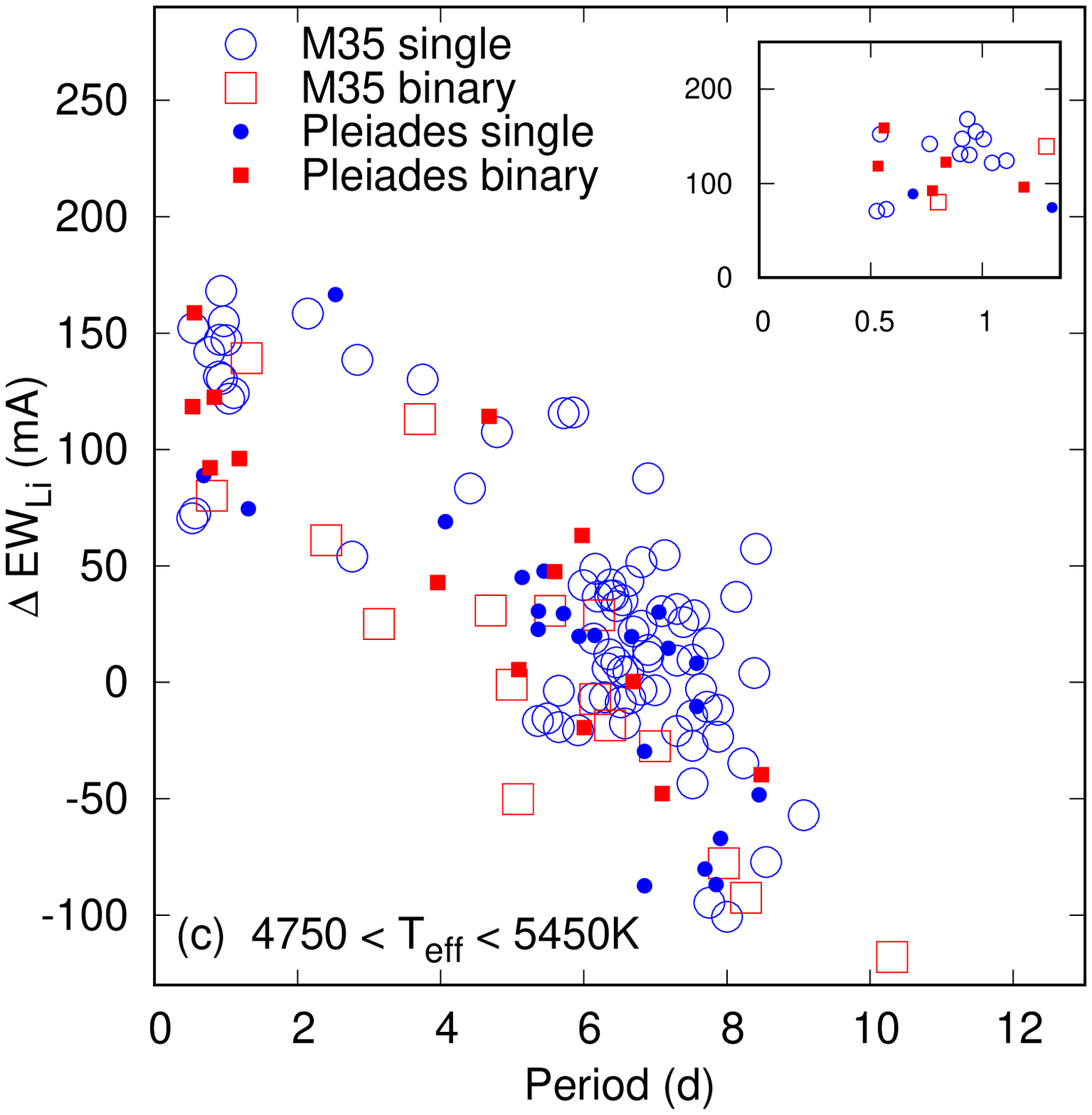}
  \end{minipage}

	\caption{A comparison of the lithium abundances, rotations periods and lithium-rotation connection for stars in M35 and the Pleiades. (a) Lithium abundances for the Pleiades, calculated from $T_{\rm eff}$ and EW(Li) presented by \protect\cite{Bouvier2018a}, compared with M35. (b) A comparison of rotation periods for the same stars. (c) The equivalent of Fig.~\ref{fig8}b but including data for the Pleiades.} 
\label{figpleiades}
\end{figure*} 

One of the primary motivations for this study was to probe the Li dispersion amongst G/K stars in a Pleiades-age cluster, but using a larger sample than available in the Pleiades itself. AT18 obtained Li measurements for about 80 G/K stars in M35 and made comparison with the Pleiades. They found that the mean level of Li abundance in M35 G/K stars was similar to, or slightly below, that in the Pleiades, but with significantly less dispersion.

Figure~\ref{figpleiades}a makes a comparison of our dataset with the same Pleiades sample considered by AT18, which comes from \cite[][B18]{Bouvier2018a}. The $T_{\rm eff}$ values for the Pleiades are taken directly from B18. The Li abundances were calculated by taking B18'S EW(Li) values and estimating abundances in exactly the same as described in Section~\ref{s3.3}. 


Our total sample of M35 members is 3 times larger than that of AT18 and contains many more stars at cooler temperatures. Figure~\ref{figpleiades} shows that the overall level and dispersion of Li abundance are very similar in M35 and the Pleiades, but these trends are also defined by about 2.5 times as many data points among the late G and K stars ($4500 < T_{\rm eff}<5500$\,K) of M35 than in the B18 Pleiades sample. There is some disagreement at the hottest $T_{\rm eff}$ values, where there is a hint that some Pleiades late-F (binary) stars are more Li-depleted and more rapidly rotating than in M35.

\begin{figure*}
	\centering
\begin{minipage}[b]{0.48\textwidth}
    \includegraphics[width=\textwidth]{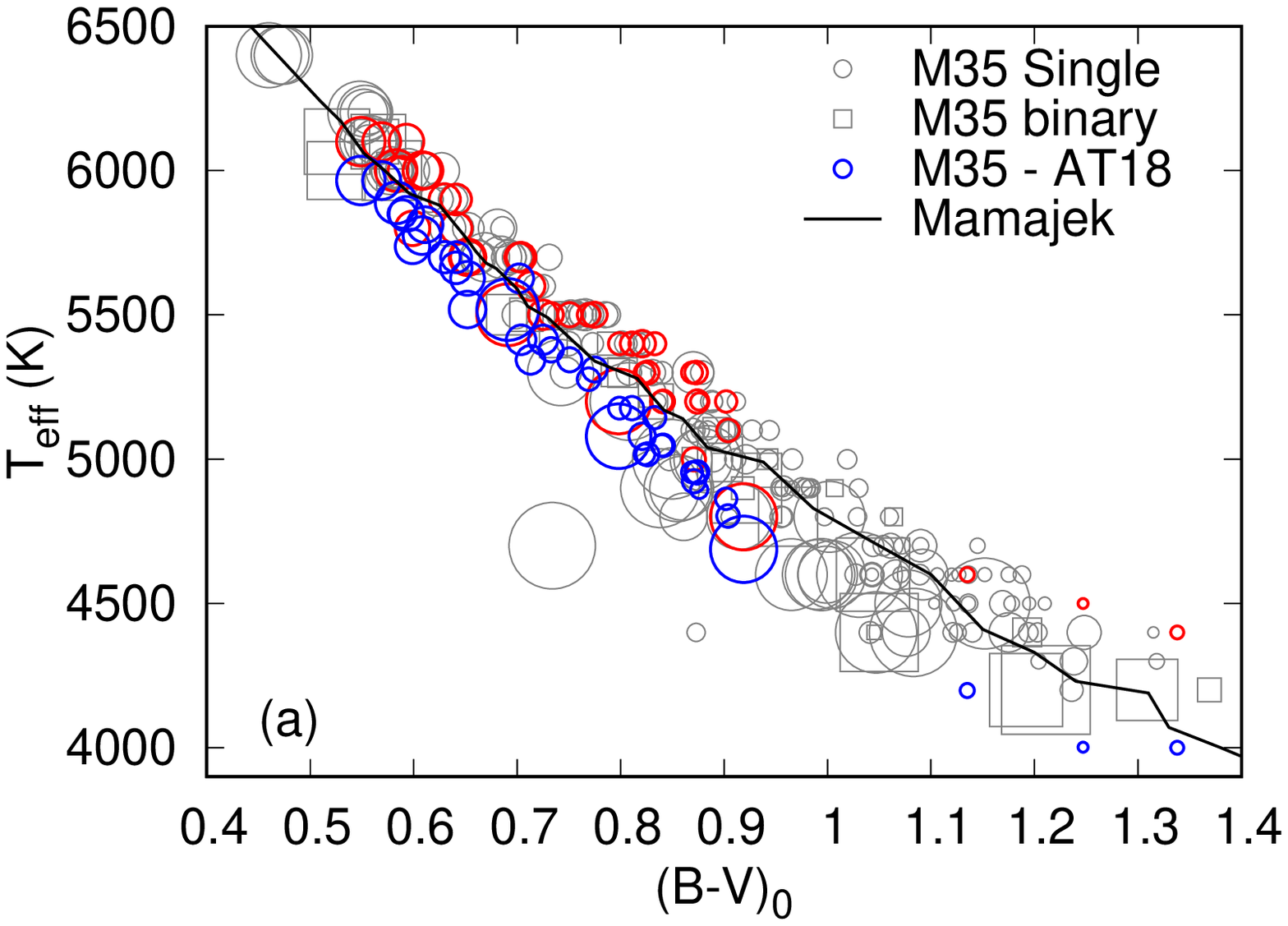}
  \end{minipage}
  \hfill
  \begin{minipage}[b]{0.48\textwidth}
    \includegraphics[width=\textwidth]{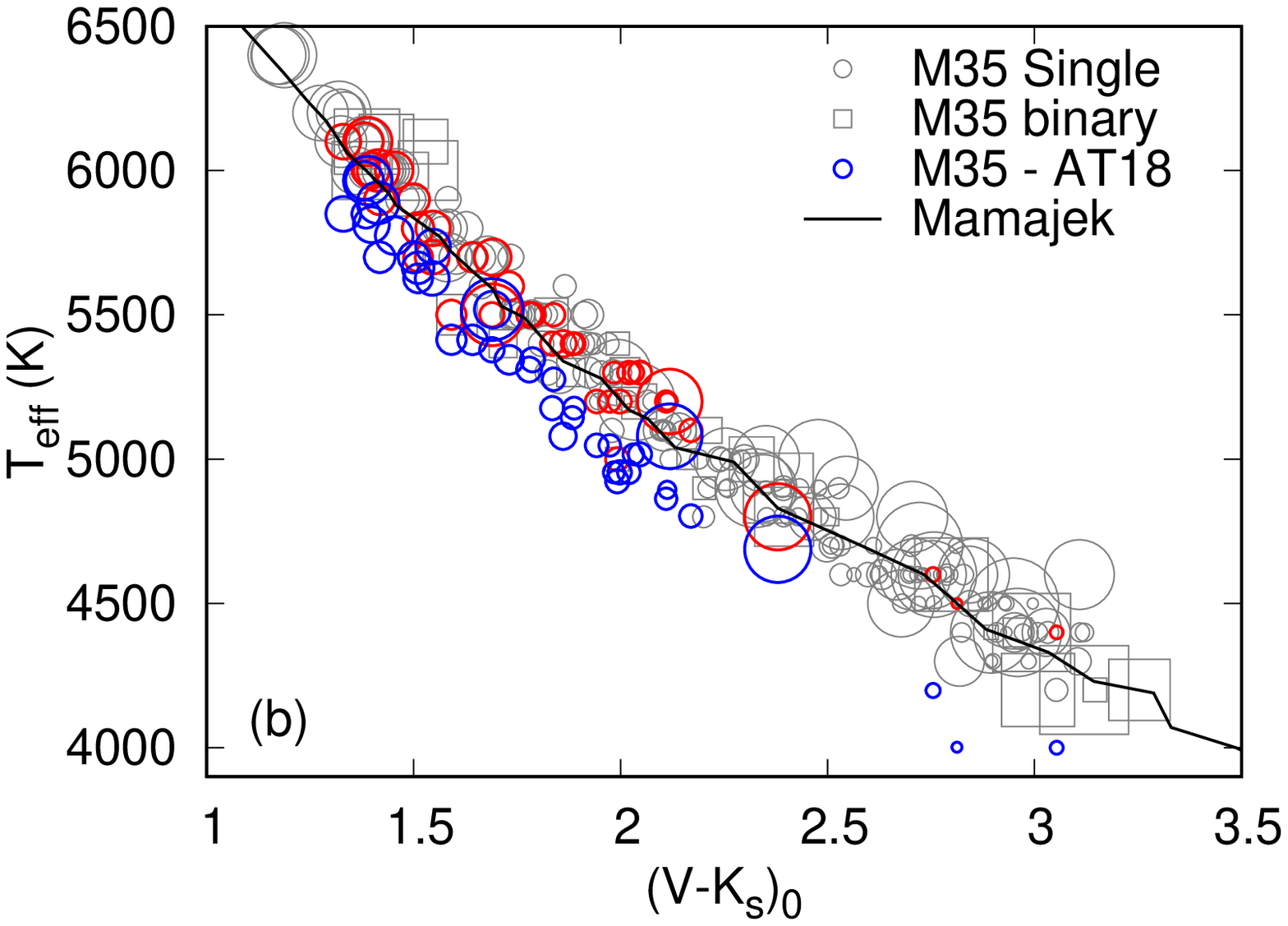}
  \end{minipage}
	\caption{Colour-$T_{\rm eff}$ relationships. (a) $T_{\rm eff}$ vs $(B-V)_0$; colours from Table~\ref{sedtable} and $E(B-V)=0.20$. (b) $T_{\rm eff}$ vs $(V-K_s)_0$; colours from Table~\ref{sedtable} and $E(V-K_s)=0.55$. In both plots, the grey  symbols are the M35 members from this paper with symbol size proportional to log angular velocity. The blue symbols are matched stars from \protect\cite{Anthony-Twarog2018a} plotted with $T_{\rm eff}$ values from that paper. The red symbols show the stars in our sample they are matched with. The solid line is a mean relationship for dwarf stars from \protect\cite{Pecaut2013a}, updated by Mamajek (2019).} 
\label{figtwarog}
\end{figure*} 
There are 44 stars in common between our sample and that of AT18, predominantly among the warmer stars. A comparison shows that any difference between our EW(Li) measurements and those of AT18 are small ($\langle \Delta {\rm EW(Li)} \rangle = -3$\,m\AA\ with $\sigma = 15$\,m\AA ) and consistent with the measurement uncertainties. However the AT18 temperatures for stars in common are approximately 200\,K cooler than used here.

Figure~\ref{figtwarog} plots two intrinsic colour-$T_{\rm eff}$ diagrams for M35. The plotted $T_{\rm eff}$ values for our sample are those in Table~\ref{resultstable}; the sample of stars in common with AT18 are identified and shown at the $T_{\rm eff}$ from AT18 (in blue) and the $T_{\rm eff}$ derived here (in red); the colours for both samples come from Table~\ref{sedtable}. Both datasets are compared with the mean 
colour-$T_{\rm eff}$ relationship for dwarf stars compiled by \cite{Pecaut2013a} and updated by Mamajek (2019)\footnote{http://www.pas.rochester.edu/\\$\sim$emamajek/EEM\_dwarf\_UBVIJHK\_colors\_Teff.dat}. These diagrams illustrate several points: (i) Whilst the $T_{\rm eff}$ values from this paper are reasonably consistent with Mamajek's mean relationship (although perhaps marginally hotter for a given $B-V$); the AT18 $T_{\rm eff}$ values are certainly much cooler. (ii) Most of the AT18 sample are slow rotators for their colour, and only one of the objects in common with this paper is classed as a binary here. (iii) The rapid rotators in our sample are bluer in $B-V$ for a given $T_{\rm eff}$ and redder in $V-K_s$ for a given $T_{\rm eff}$. This agrees with previous work on the Pleiades which arrived at similar conclusions on the rotation dependence of the colours \citep{Stauffer2003a, Kamai2014a, Covey2016a}. This has been attributed to spots or radius inflation (see Sections~\ref{s4.5} and~\ref{s5.2}).

Part of the difference between the $T_{\rm eff}$ values used here and those of AT18 may be due to the adoption of [M/H]$=-0.15$ in AT18 as opposed to solar metallicity here. The VOSA SED fitting tool does not provide a fine sampling of metallicity. We re-fitted the SEDs using the same models and reddening but with [M/H]$=-0.5$. Even with this large change, the mean $\Delta T_{\rm eff}$ only reduced from $(209 \pm 16)$ K to $(135 \pm 18)$ K. In summary, our $T_{\rm eff}$ values are metallicity-insensitive, might be too hot if M35 has a subsolar metallicity, but by $<100$\,K.

Small $T_{\rm eff}$ shifts or $T_{\rm eff}$ uncertainties will not systematically move the M35 G/K stars with respect to the overall Pleiades Li trend versus $T_{\rm eff}$ or lead to any difference in dispersion; lower $T_{\rm eff}$ also leads to lower $A$(Li) and the combined effect is to move points roughly parallel to the trend defined by the data (see Section~\ref{s4.4} for more detail). The cause of the limited dispersion in AT18's dataset appears to be due a lack of rapid rotators in their sample and the correlation between rapid rotation and high Li abundance. There are 65 stars with rotation periods \citep[from][]{Meibom2009a, Libralato2016a}  in AT18's sample; only 7 (11 per cent) have rotation periods less than 2 days, and these stars do follow the upper envelope of the Pleiades distribution in AT18. The rest of the AT18 sample are part of the "I-sequence" of slow rotators. In contrast (see Fig.~\ref{figpleiades}b), our sample of members contains 65 stars with rotation period $<2$ days (27 per cent) and another $\sim 36$ stars (14 per cent) in the transition region between these and the slow-rotating "I sequence". It is these fast and intermediate rotators that are responsible for much of the observed dispersion in Li abundances seen in Fig.~\ref{fig7} (see also Fig.~\ref{fig8}). The reason for the lack of rapid rotators in AT18's sample is discussed further in Section~\ref{s4.5}.

Figure~\ref{figpleiades}b shows that the distribution of rotation periods with $T_{\rm eff}$ is also very similar in the Pleiades and M35, as might be expected if they are roughly the same age. Like M35, the Pleiades sample of B18 also contains 35/125 (28 per cent) rapid rotators ($P<2$ d) in the range $4000<T_{\rm eff} < 6500$\,K and $\sim 19$ (15 per cent) of transition objects between these and the I-sequence. Again, we note the presence of several fast rotating late F-stars in the Pleiades that do not have counterparts in M35. 

The position of the slow-rotating I-sequence can be used as a "gyrochronological" age estimator in clusters \citep{Barnes2003a}. To define an approximate I-sequence locus, a straight line is fitted for $4000<T_{\rm eff}<6000$\,K to non-binary M35 members that are slower than the median rotation rate at a given $T_{\rm eff}$. Similar non-binary Pleiades objects are fitted by the same locus if it is shifted to shorter periods by a factor $1.07 \pm 0.03$. If these I-sequence stars obey a Skumanich-type spin-down law \citep{Skumanich1972a}, with $P \propto t^{1/2}$, then this would indicate that M35 is older by a factor $1.14 \pm 0.06$ than the Pleiades. For a Pleiades age of $125 \pm 8$ Myr \citep[from the lithium depletion boundary technique,][]{Stauffer1998a}, this would make the age of M35 $143 \pm 8$\,Myr on the same scale. However, this level of precision is spurious; aside from the difficulty of defining the I-sequence, there is the matter of uncertainties in the $T_{\rm eff}$ scale for M35, which in turn depend on the reddening and metallicity. A $\pm 100$ K shift in temperatures leads to a change in the M35 age estimate of $\mp 15$ Myr and if $T_{\rm eff}$s were $\sim 200$\,K cooler, as advocated by AT18, then M35 would be slightly younger than the Pleiades at 115 Myr. Our conclusion is that the rotation distributions of the two samples are quite similar and that the M35 gyrochronological age  is about $140 \pm 15$ Myr\footnote{A similar conclusion is reached if rotation period is plotted versus $(B-V)_0$, $(V-K_s)_0$ or $(G-K)_0$ \protect\citep[using a $G$-band extinction coefficient calibrated by][]{Casagrande2018a}.}. At the lower end of this range, the age of M35 may be consistent with the Pleiades. However, the upper end of the range is more consistent with the $\sim 50$ Myr  difference in the main-sequence turn-off ages between M35 and the Pleiades inferred by Deliyannis et al. (in preparation). 

Figure~\ref{figpleiades}c repeats Fig.~\ref{fig8}b, but with the Pleiades data added for comparison, using the same definition and baseline for $\Delta$EW$_{\rm Li}$. The Pleiades rotation periods are from Kepler K2 \citep{Rebull2016a}. The EW(Li) values from B18 generally have smaller uncertainties than those in M35, although there could be systematic differences in the EW(Li) measurements due to differences in continuum definition and the metallicity-dependent deblending corrections (see Section~\ref{s2.4}). In all respects the Pleiades data reinforce the features of the Li-rotation relationship seen in M35: there is a strong correlation with rotation period; there is a scatter around this correlation that is larger than the uncertainties, particularly at slow and intermediate rotation periods; and binaries follow the same relationship as single stars.

\subsection{A comparison with standard models}
\label{s4.4}

Figure~\ref{fig6} showed the inferred NLTE lithium abundances of M35 members along with the predictions of several "standard evolutionary models" at an age of 120\,Myr. This term refers to stellar evolutionary models that do not include non-convective mixing (e.g. diffusion or rotational mixing) or the structural influences of magnetic fields or rotation. Since in these models all the stars in the considered mass range have settled onto the ZAMS by 100 Myr and ceased Li depletion well before that, the exact choice of isochrone age does not affect the comparison.   

In order to use these models, which predict by how much lithium has been {\it depleted} from some initial value, an assumption needs to be made about the initial lithium abundance for the cluster, $A$(Li)$_0$. Here it is assumed that $A$(Li)$_0$ scales linearly with metallicity, so that $A$(Li)$_0$ is $3.26 +$ [Fe/H] \citep[e.g.][]{Cummings2011a}, where A(Li)$=3.26 \pm 0.05$ is the solar system meteoritic abundance \citep{Asplund2009a}.

The reader is cautioned that in addition to this possible source of systematic error, there are also systematic uncertainties in the $T_{\rm eff}$ scale to consider (see also Section~\ref{s4.3}, which affect both the plotted $T_{\rm eff}$ and $A$(Li) of the data in a correlated way. Two arrows on Fig.~\ref{fig6} show the effect of a $\pm 100$~K $T_{\rm eff}$ change for stars at $T_{\rm eff}=5500$\,K or 4500\,K, with a median $A$(Li) value at that $T_{\rm eff}$. The correlated uncertainty is kind, in the sense that points are moved roughly parallel to the evolutionary model isochrones.

The PROSECCO models, which use the Pisa version of the FRANEC code \citep{Tognelli2011a, Dellomodarme2012a} provide the closest match to the M35 data. The [Fe/H]=0.0 isochrone lies just below the median trend for $T_{\rm eff}<5400$\,K but over-predicts the Li abundance of hotter stars.
The [Fe/H]$=-0.15$ isochrone provides a good match to the upper envelope of the data for $T_{\rm eff} > 5000$\,K and moves towards, but still slightly above, the median abundance at lower temperatures. 

In contrast, the solar metallicity models of \cite{Baraffe2015a} over-predict the Li abundance for $T_{\rm eff}>5800$\,K, but under-predict the Li abundance, and follow the lower envelope of the M35 distribution, at $T_{\rm eff}<5500$\,K. Similarly, the Dartmouth solar metallicity evolution models of \cite{Dotter2008a} follow the lower envelope of the M35 distribution at $T_{\rm eff}<5500$\,K. Presumably, lower metallicity realisations of these models would provide a closer match to the median of $A$(Li) at lower $T_{\rm eff}$, moving in a similar metallicity-dependent way to that seen in the PROSECCO models.

None of the standard models provide any means for interpreting the two orders of magnitude spread of $A$(Li) at a given $T_{\rm eff}$ in the cooler stars. The discrepancies between their individual predictions can be attributed to differences in the treatment of convection (e.g. the adopted mixing length), the boundary conditions between the interior and photosphere, and the interior opacities. The latter may be the dominant factor, being dependent on the definition of "solar metallicity". The PROSECCO models use a solar heavy element mass fraction of $Z=0.013$, \cite{Baraffe2015a} uses $Z=0.0153$ and \cite{Dotter2008a} uses $Z=0.0189$. Increasing metallicity leads to more opacity, deeper convection zones on the PMS and more Li depletion at the photosphere \citep[e.g.][]{Chaboyer1995a, Piau2002a}.
These physical uncertainties in the models, together with uncertainty in the metallicity of M35, mean that it is difficult to say whether the spread of Li abundances at $T_{\rm eff}<5500$\,K  results from rapid rotators preserving more of their initial Li and being under-depleted with respect to standard models, or whether slow rotators have undergone more Li depletion than predicted by standard models. 

At $T_{\rm eff}>5800$\,K there is evidence that stars in M35 have depleted more Li than predicted by all the standard models. The likely culprit here is additional, rotation-driven mixing \citep[e.g.][]{Chaboyer1995a, Pinsonneault1997a, Eggenberger2012a, Somers2015a}. There is a potential signature of this seen in Fig.~\ref{fig8}d, in the form of decreasing $\Delta$EW$_{\rm Li}$ with increasing rotation rate; but as mentioned in Section~\ref{s4.2}, both the median EW(Li) and rotation period are changing rapidly with $T_{\rm eff}$ above 5800\,K. This, combined with the $T_{\rm eff}$ uncertainties and a relatively small dynamic range in EW(Li) and rotation period at a given $T_{\rm eff}$, means that the apparent correlation is suggestive, but not necessarily significant.

\subsection{Comparison with magnetic models}
\label{s4.5}
\begin{figure}
	\centering
	\includegraphics[width = 85mm]{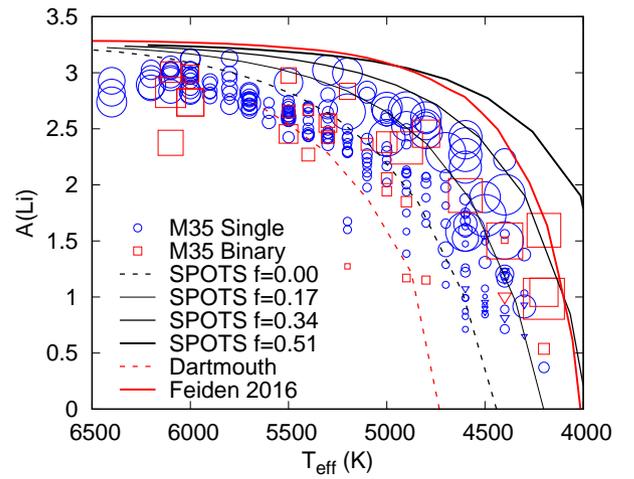}
	\caption{The NLTE lithium abundance of M35 members versus $T_{\rm eff}$ compared with "magnetic models". Symbols are as in Fig.~\ref{fig6}; the lines represent isochronal predictions of "magnetic" models at an age of 120 Myr, assuming an initial Li abundance of $3.26$. The isochrones are generated from the magnetic models of \protect\cite{Feiden2016a} and the {\sc spots} starspot models of \protect\cite{Somers2020a} for spot filling factors of 0.17 and 0.34. For comparison, the non-magnetic, counterparts of these isochrones (labelled "Dartmouth" and "SPOTS f=0.00" respectively) are also shown as dashed lines (see Section~\ref{s4.5}).} 
	\label{magmodels}
\end{figure}

\begin{figure*}
	\centering
\begin{minipage}[b]{0.48\textwidth}
    \includegraphics[width=\textwidth]{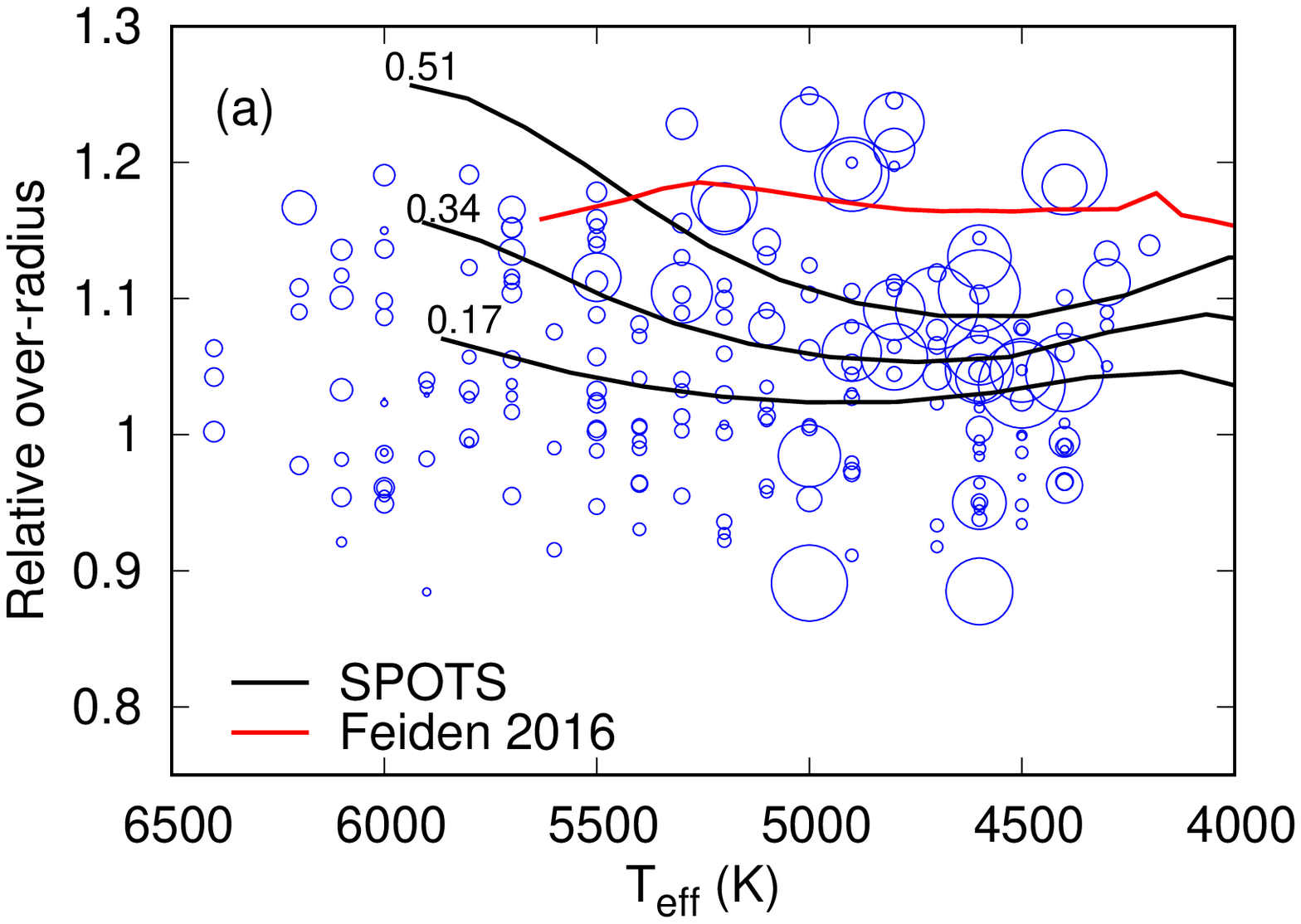}
  \end{minipage}
  \hfill
  \begin{minipage}[b]{0.48\textwidth}
    \includegraphics[width=\textwidth]{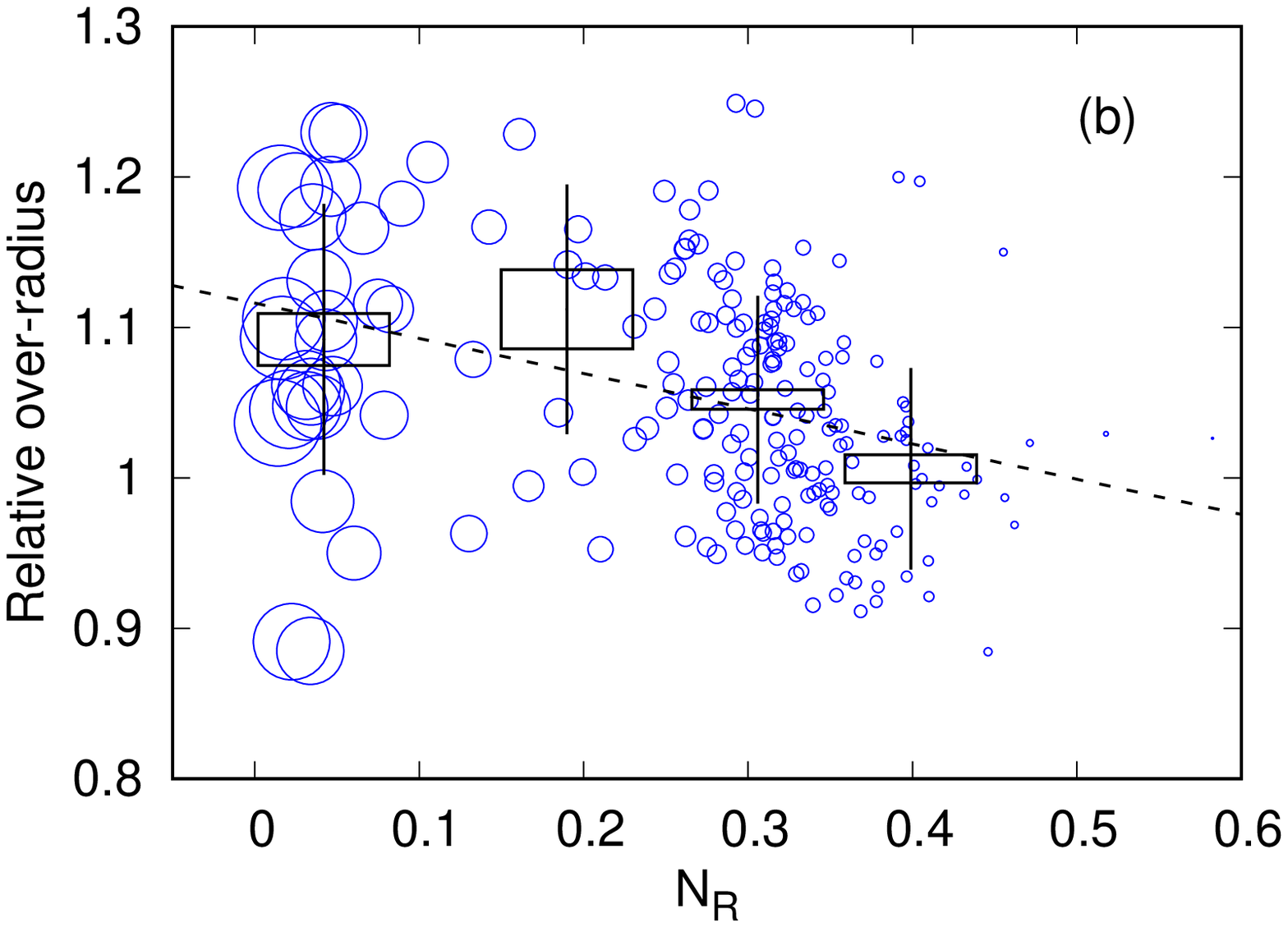}
  \end{minipage}
	\caption{The relationship between over-radius and magnetic activity for stars in M35. (a) Over-radius versus $T_{\rm eff}$, with symbol sizes proportional to the logarithm of the inverse Rossby number. The loci represent the predictions of the over-radius from magnetic models (see Section~\ref{s4.5}, where the labels are the spot filling factor in the {\sc spots} models), using their counterpart non-magnetic models as a baseline. (b) Over-radius versus Rossby number. The means in 4 bins of Rossby number are shown with boxes. The height of the box represents the standard error in the mean and the error bar is the standard deviation in that bin. The dashed line is a least squares fit to all the data.} 
\label{overradius}
\end{figure*} 
A hypothesis to explain the dispersion of Li abundances in the cooler stars and its connection with rotation is to invoke rotation-dependent levels of magnetic activity that affect the structure of a contracting PMS star. Proposed mechanisms are the magnetic inhibition of convective energy transport \citep{Ventura1998a, Macdonald2010a, Feiden2013a, Feiden2016a} or the blocking of flux at the photosphere by dark, magnetic starspots \citep{Jackson2014a, Somers2014a, Somers2015b}.
Both mechanisms lead to magnetically active cool stars having larger radii, cooler interior temperatures and consequently lower levels of photospheric Li depletion. The rotation dependence would emerge as a result of the well known relationship between faster rotation and higher levels of magnetic activity. Evidence has been accumulating that magnetically active stars are larger than inactive stars at the same $T_{\rm eff}$ \citep[e.g.][]{Jackson2018a}, that the properties of low-mass PMS eclipsing binaries and PMS Li depletion patterns are better explained if the stars are "inflated" \citep[e.g.][]{Lacy2016a, Jeffries2017a, Somers2020a, Murphy2020a} and a tripartite correlation between rapid rotation, radius inflation and lower Li depletion has been found in the Pleiades \citep{Somers2017a}.

Figure~\ref{magmodels} is the equivalent of Fig.~\ref{fig6} but now showing "magnetic isochrones" of Li depletion: (i) a model in which a dynamo-generated interior magnetic field suppresses convective flux, with a boundary condition of an equipartition magnetic field at the photosphere \citep[the magnetic Dartmouth models,][]{Feiden2016a}; (ii) models in which cool, surface starspots block flux at the surface, with spot coverage fractions of 0.17--0.51, and a spotted to unspotted photospheric temperature ratios of 0.8 \citep[{\sc spots},][]{Somers2020a}. All these isochrones are calculated at solar metallicity (assumed to be $Z= 0.0188$ and $Z=0.0165$ respectively), have ages of 120 Myr and are compared with their "non-magnetic" counterparts -- in the case of \cite{Feiden2016a} this is provided by the Dartmouth model, whilst \cite{Somers2020a} provide a variant of their model with zero spot coverage.

Both the magnetic models are capable of explaining the patterns of Li depletion in the cooler stars of M35 ($T_{\rm eff}<5500$\,K) if they have a range of magnetic activity that is correlated with their rotation rates. The magnetic Dartmouth isochrone and the {\sc spots} isochrone with a spot filling factor of 0.34 both follow the upper envelope of the M35 $A$(Li) distribution. However, all the magnetic isochrones under predict the amount of Li depletion in the stars with $T_{\rm eff}>5500$\,K, suggesting that if magnetic activity is important in these stars, then additional rotational mixing is even more important in explaining the additional Li depletion seen in early G- and F-type ZAMS stars. 
Note also that the differences between the magnetic and non-magnetic isochrones becomes much smaller at higher temperatures, perhaps explaining why a significant $A$(Li) dispersion is only seen for $T_{\rm eff}<5500$\,K.

If rotation-dependent stellar magnetism is responsible for the rotation-dependent dispersion in Li depletion, then we might expect to see that 
fast-rotating stars are "inflated" with respect to their slower rotating counterparts -- i.e. the amount of radius inflation seen should also be consistent with the predictions of the magnetic models and should be correlated with rotation. 

Figure.~\ref{overradius}a shows the over-radius $\rho$ (defined in Section~\ref{s3.4}) as a function of $T_{\rm eff}$, compared to the predicted over-radius of the magnetic models, using their non-magnetic counterparts as a baseline. 
In order to compare magnetic activity levels across a range of $T_{\rm eff}$ the symbol size is made proportional to the logarithm of the inverse of the Rossby number, $\log N_R^{-1}$, where $N_R$ is the ratio of rotation period to convective turnover time. Magnetic activity has been shown to be more tightly correlated with $N_R$ than with rotation period when aggregating data over a range of $T_{\rm eff}$, with magnetic activity increasing towards smaller Rossby numbers, with a flattening or "saturation" at $N_R<0.1$ \citep[e.g.][]{Pizzolato2003a, Jeffries2011a}. Convective turnover times were estimated from $(B-V)_0$ using the functional form proposed by \cite{Noyes1984a} and $N_R$ values are included in Table~\ref{resultstable}.

The largest values of $\rho$ would require spot coverage fractions $>50$ per cent or surface magnetic fields at around their equipartition value, the latter being reasonably consistent with the constraints from Li depletion in Fig.~\ref{magmodels}. The reader should note however, that the empirical baseline for $\rho$, although defined mostly by stars with slower rotation periods, may still be representative of a moderate level of magnetic activity, meaning that the true $\rho$ values with respect to magnetically inactive stars may be somewhat higher than shown. 

Figure.~\ref{overradius}b shows $\rho$ versus $N_R$. In both panels, only stars with $\rho<1.25$ are included, since we expect that most of the stars with $\rho>1.25$ are binaries, where the over-radius is overestimated due to the presence of a binary companion.
There is a general correlation between smaller $N_R$ and $\rho$, albeit with a large amount of scatter. The black symbols in the plot show the mean, standard deviation and standard error in the mean, for the data gathered into 4 broad bins of $N_R$. There is strong evidence for an increase in $\bar{\rho}$ for $0.1<N_R<0.5$ and then weaker evidence that the relationship flattens for smaller $N_R$, which would be reminiscent of how magnetic activity indicators behave in terms of "saturation" of activity for $N_R<0.1$. The dashed line in the plot is a simple least squares fit, which has a gradient of $-0.23 \pm 0.05$. The correlation between over-radius with Rossby number (and hence with rotation period) is probably the reason for the lack of many fast rotators in the sample of AT18 (see Section~\ref{s4.3}), since AT18 selected stars "close to the single-star fiducial sequence", which may have precluded the selection of many rapidly rotating and inflated stars. 

Some of the scatter in Fig.~\ref{overradius}b could be attributed to unidentified binary systems with moderate mass ratios. In particular, there are 5 clear outliers with large $\rho$ and large Rossby numbers that could fall into this category. However, there is no similar explanation for the several objects with small Rossby numbers and small over-radius values. 
The random measurement uncertainties in $\rho$ are approximately equal to the numerical uncertainties in $\log L/L_{\odot}$, plus uncertainties in the individual distances to cluster stars, and are $\leq 0.02$ dex; the propagated uncertainty in the radius baseline due to $T_{\rm eff}$ errors is only 0.02--0.03 dex. We conclude that the measurement uncertainties are much smaller than the standard deviations of 0.07--0.09 and that much of the dispersion in this plot is genuine, although this is revisited in Section~\ref{s5.3}. A similar conclusion can be drawn from a smaller sample of Pleiades rapid rotators \citep[see fig.~5 in][]{Somers2017a}, although slower rotators in the Pleiades are more tightly bunched around zero over-radius than seen here in M35. It is possible, since the census of binaries is more complete in the Pleiades, that unrecognised binaries in M35 are responsible for the larger scatter at slow rotation rates.

\subsection{A tripartite correlation between lithium, magnetic activity and over-radius}

\label{s4.6}

\begin{figure*}
	\centering
\begin{minipage}[b]{0.38\textwidth}
    \includegraphics[width=\textwidth]{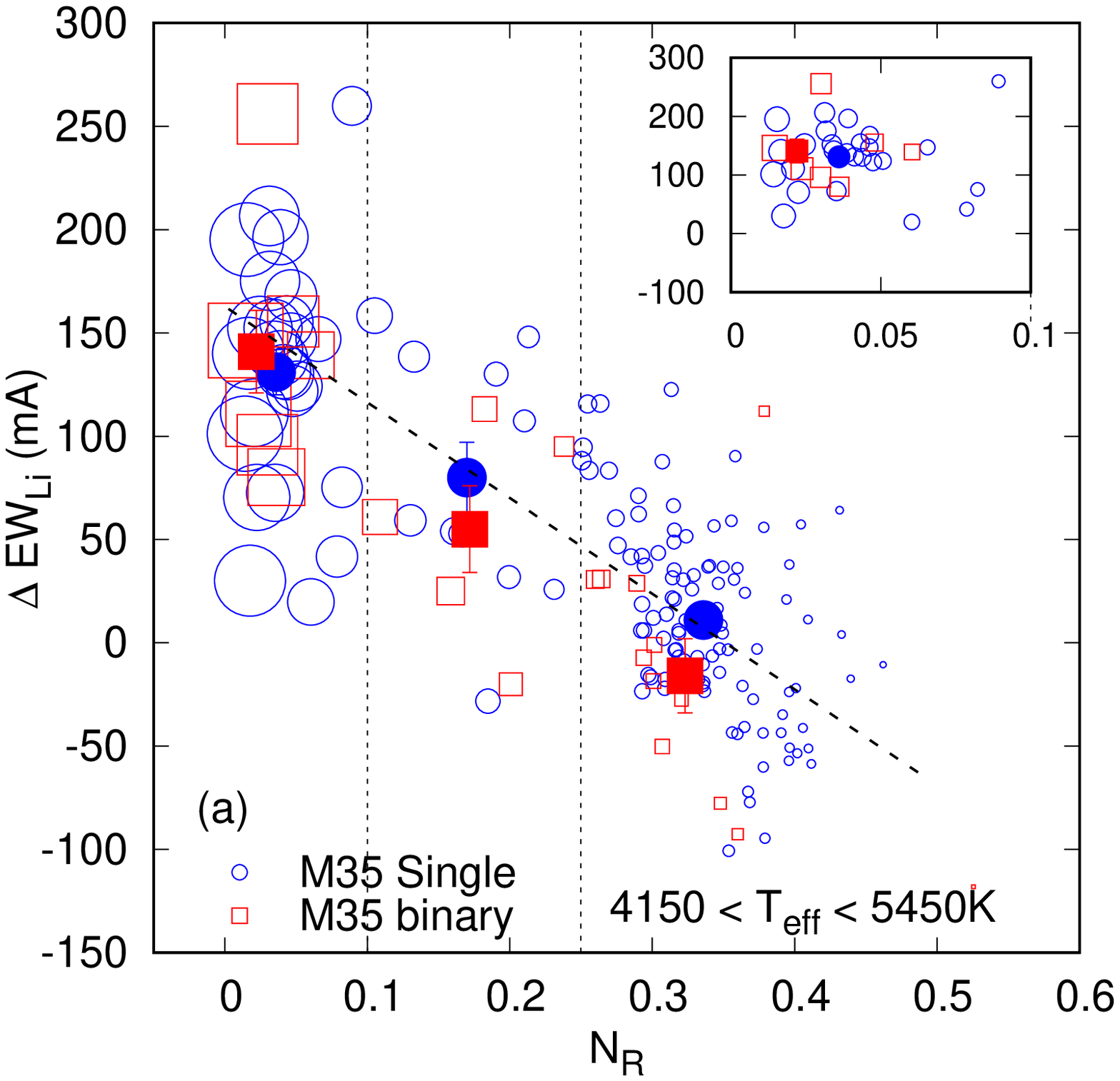}
  \end{minipage}
  \hfill
  \begin{minipage}[b]{0.58\textwidth}
    \includegraphics[width=\textwidth]{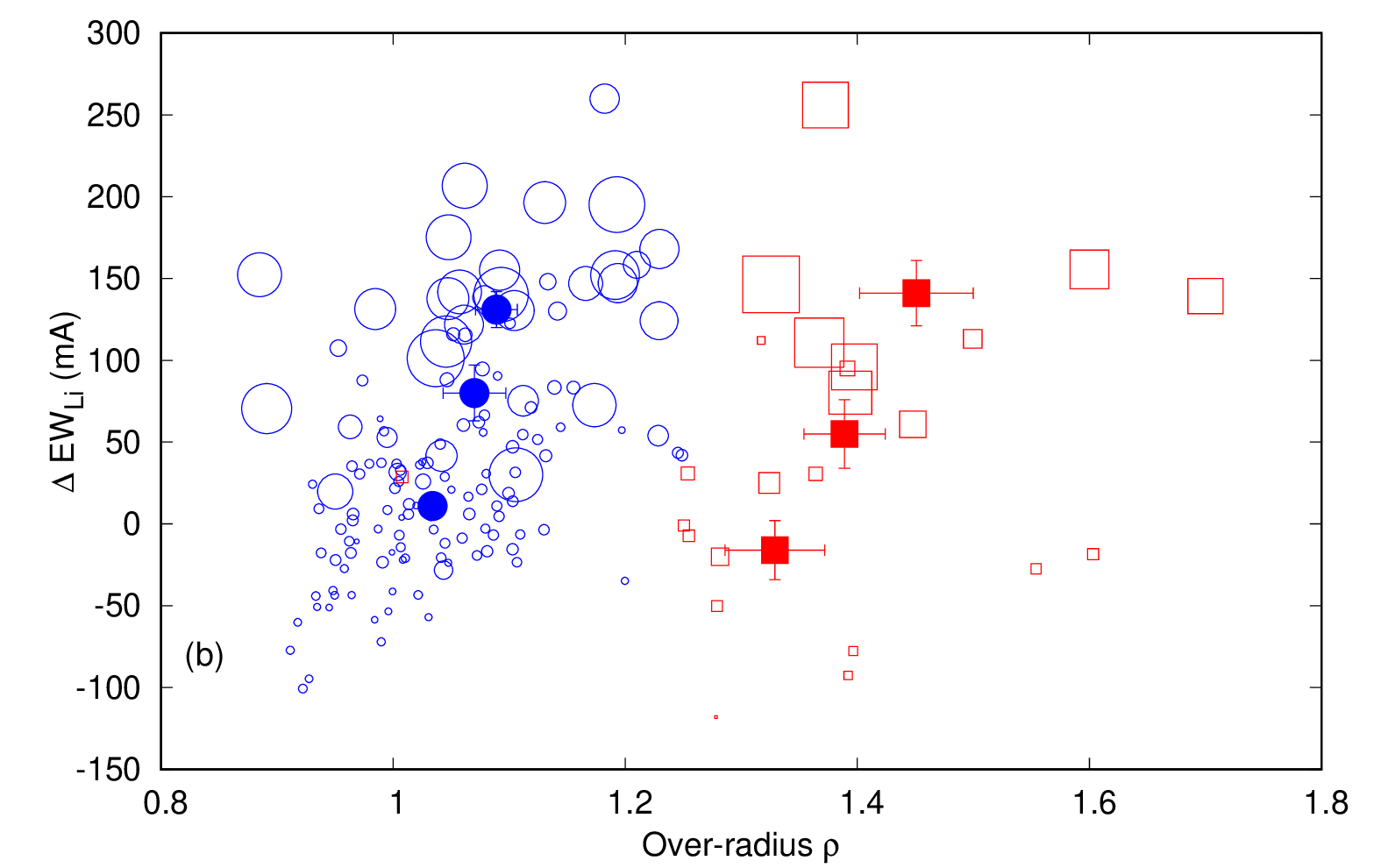}
  \end{minipage}
	\caption{Demonstrating the tripartite relationship between excess lithium ($\Delta$EW$_{\rm Li}$), Rossby Number ($N_R$) and over-radius ($\rho$). (a) $\Delta$EW$_{\rm Li}$ vs $N_R$ with symbol size proportional to  $\log (N_{R}^{-1})$. The filled symbols show the mean values in three $N_R$ ranges from Table~\ref{delEW}, with the error bars showing the standard error in the mean. The plotted mean $N_R$ value for binaries in the low $N_{R}$ bin has been decreased by 0.01 for clarity. The dashed line is a linear fit to the single star data. (b) $\Delta$EW$_{\rm Li}$ versus $\rho$. The symbol sizes and meanings are the same as in panel (a).}	 
\label{tripartite}
\end{figure*} 

\cite{Somers2017a} were able to establish a tripartite correlation between reduced Li depletion, rotation and over-radius in a sample of late-G and K-type Pleiades stars. In M35 there is the opportunity to explore this relationship with greater numbers.

Figure~\ref{tripartite} shows how $\Delta$EW$_{\rm Li}$ (see Section~\ref{s4.1}) depends on Rossby number and over-radius for stars with $4150< T_{\rm eff} < 5450$\,K -- the range where there is clear evidence for a dispersion in Li depletion at a given $T_{\rm eff}$ (see Fig.~\ref{fig8}). Table~\ref{delEW} shows the average level of $\Delta$EW$_{\rm Li}$ and $\rho$ for stars grouped into 3 bands of Rossby number.\footnote{Since convective turnover times only vary from $\sim 15$\,d to 24\,d over this range of $T_{\rm eff}$, whilst the rotation periods vary from 0.32--11.0\,d, then dividing the stars by rotation period rather than Rossby number would produce similar results.} These average values are shown as solid symbols in Fig~\ref{tripartite}.

Figure~\ref{tripartite}a shows that the strong correlation between $\Delta$EW$_{\rm Li}$ and rotation period is (unsurprisingly) repeated when using Rossby number as the independent variable. The behaviour of stars judged to be single or binary stars appears to be indistinguishable. The means and standard deviations of the single and binary stars in each of the $N_R$ bins of Table~\ref{delEW} are very similar. The scatter in the correlation, first noted for rotation period in Section~\ref{s4.2}, is still there. To demonstrate this, a straight line is fitted to the single star data in Fig.~\ref{tripartite}a. The reduced chi-squared of the fit (with 137 degrees of freedom) is $\chi^2_{\nu}= 5.89$.\footnote{Note that the additional uncertainties in $\Delta$EW$_{\rm Li}$ due to an assumed error in $T_{\rm eff}$ of $\pm 50$\,K have been included (see Section~\ref{s4.2}). If these were doubled then $\chi^{2}_{\nu}$ only decreases to 4.60.} This dispersion is apparent in all three $N_R$ ranges, despite the larger EW(Li) uncertainties for fast-rotating, cooler stars -- the rms dispersion around the fit for stars with $N_R<0.1$ is 57\,m\AA\, compared with their total rms uncertainties of  39\,m\AA. The scatter is of similar size, but is more significant in the $0.25 \geq N_R \geq 0.1$ and $N_R>0.25$ bins, where the uncertainties are smaller. The corresponding numbers for the rms dispersion (and rms uncertainties) are  55\,m\AA\ (21\,m\AA)  and 47\,m\AA\ (19\,m\AA) respectively. We recall that the robustness of the EW uncertainties were tested using the EW of the nearby Ca~{\sc i} 6717.7\AA\ line  (see Fig.~\ref{fig7}). The relative uncertainties in the Rossby numbers follow from the rotation period uncertainties discussed in Section~\ref{s4.2} and are unlikely to be important; any uncertainties due to photometry errors in calculating the turnover time are negligible.

\begin{table}
\caption{Average values of $\Delta$EW$_{\rm Li}$ and $\rho$ for  slow, medium and fast rotating cool stars ($4150 < T_{\rm eff}<5450$\,K) in probable single and binary stars.}
\begin{tabular}{lccc} 
\hline
		&	Slow rotators	&	Medium rotators	&	Fast rotators\\	
		& 	$N_R>$0.25	&	$0.25\geq N_R\geq0.1$	&	$N_R<0.1$
	\\\hline
  & \multicolumn{3}{c}{Single Stars} \\
    $n_{\rm star}$ & 102  & 11  & 26 \\			
		$\Delta$EW$_{\rm Li}$ (m\AA)	& $11 \pm 5$	&	$80 \pm 17$	&	$131 \pm 11$	\\
		$\rho$	&	$1.034 \pm 0.007$	&	$1.070 \pm 0.027$	&	$1.089 \pm 0.018$	\\\hline
 & \multicolumn{3}{c}{Binary stars} \\
    $n_{\rm star}$ & 12  & 5 & 7 \\			
		$\Delta$EW$_{\rm Li}$ (m\AA)	&	$-16 \pm 18$	&	$55 \pm 21$	&	$141 \pm 20$	\\
		$\rho$	&	$1.329 \pm 0.043$	&	$1.389 \pm 0.035$	&	$1.451 \pm 0.049$	\\\hline

    \end{tabular}
  \label{delEW}
\end{table}

Figure~\ref{tripartite}b shows the relationship between $\Delta$EW$_{\rm Li}$ and $\rho$.  The solid points here are the mean values of $\Delta$EW$_{\rm Li}$ and $\rho$ in the three $N_R$ bins listed in Table~\ref{delEW}. There are clearly general correlations between both higher $\Delta$EW$_{\rm Li}$ and higher $\rho$ for single and binary stars.  However, the scatter in these relationships is large. In particular, although there may be a few objects in the "single star" sample that are actually binaries and have an over-estimated $\rho$, note the presence of three objects with very low Rossby numbers that have $\Delta$EW$_{\rm Li}>60$\,m\AA\ but $\rho<1.0$ that cannot be explained in this way.  

The difference in $\rho$ between the smallest and largest  $N_R$ subsamples is $0.055 \pm 0.019$ for single stars and $0.122 \pm 0.065$ for binaries. Precision is hampered by the wide scatter in $\rho$ as a function of $N_R$, which has a standard deviation of about $\sigma_{\rho} \simeq 0.08$ for all the subsamples and which is several times larger than the measurement uncertainties in $\rho$. 

At first glance it may seem surprising that the binary stars follow similar correlations (but offset in $\rho$). However, if these are mainly wide binaries then the components probably behave like the sum of two independent single stars. In which case both the $\Delta$EW$_{\rm Li}$, $N_R$ and the $\Delta$EW$_{\rm Li}$, $\rho$ correlations will still be present, albeit with an offset and more scatter in $\rho$ because of the presence of binary companions with a range of mass ratios.

\section{Discussion}

\label{s5}

The large sample of stars that have been observed in M35, and the robust estimates of measurement uncertainties have allowed confirmation and a more detailed exploration of the connection between rotation and Li depletion, previously established for stars with $T_{\rm eff} < 5500$\,K in the Pleiades and other young clusters \citep[][B18]{Barrado2016a}. The new information established here is that, at the ZAMS, the relationship between $\Delta$EW$_{\rm Li}$ and rotation (or $N_R$) is not single-valued; there may not be a straightforwardly deterministic relationship between rotation rate and how much Li depletion is expected for a star at a given $T_{\rm eff}$. In addition, the presence of binary stars in the sample (albeit, not close, tidally-locked binary systems) is not responsible for any scatter, since such systems appear to follow the same relationship.

Explanations for the Li-rotation connection either suggest that fast rotators have had their Li depletion inhibited through some sort of magnetic inhibition of flux transport out of the star, that rapid rotation somehow inhibits internal mixing,  or that the stars have undergone additional, perhaps rotation-dependent mixing, such that their photospheres are more Li-depleted than expected by the time they reach the ZAMS.
Figures~\ref{fig6} and~\ref{magmodels} generally favour the latter class of explanation for hotter stars $T_{\rm eff}>5700$\,K. All of the standard models (those that feature only convective mixing and neglect the influences of rotation and magnetic fields) under-predict the Li depletion seen in M35 at these temperatures (if its initial Li abundance $A$(Li)$_0>3.1$). This has been noted before (and with higher quality data) in the hotter stars of M35 and has been attributed to slow rotational mixing, rather than diffusion \citep{Steinhauer2004a}, but other possibilities, including mixing by gravity waves, have been proposed to explain Li depletion beyond the PMS in solar-type stars \citep[e.g.][]{Garcia1991a, Schatzman1993a}. The data in Fig.~\ref{fig8}d suggest a weak relationship between faster rotation and increased Li depletion but the size of the uncertainties in the data compared with the range of EW(Li) and rotation rates makes this inconclusive.

In the cooler stars, the situation is less well-defined.
All the standard models in Fig.~\ref{fig6}, even those with the slightly sub-solar metallicity that may be appropriate for M35, over-predict the levels of Li depletion seen in the most rapidly rotating mid-G to K-type stars ($T_{\rm eff}<5500$\,K) and some predict as much depletion as seen in the slowest rotators. This suggests that rapid rotation inhibits PMS Li depletion. However, there is still sufficient uncertainty in (i) the microphysics in the models, especially the adopted solar metallicity and the assumed convective mixing length during PMS evolution; (ii) the metallicity of M35\footnote{Although note that there is no such uncertainty in the nearly solar-metallicity of the Pleiades, shown in Fig.~\ref{figpleiades}.}; and (iii) the initial Li abundance, that it is still possible that improved standard models, with perhaps lower interior opacities or smaller mixing lengths in PMS stars, may yet match the {\it upper} envelope of the M35 (and Pleiades) G- and K-type stellar Li distribution.. If so, then it cannot be ruled out that the dispersion seen is caused by {\it additional} depletion in the slower rotators, rather than inhibited depletion in the fastest rotators.

\subsection{The case for magnetic inflation}
\label{s5.1}

A number of authors have suggested that magnetic activity, either in the form of starspots or interior magnetic fields, suppresses the emergent radiative flux, resulting in a larger star with a lower core temperature and less Li depletion \citep{King2010a, Somers2015a, Somers2015b, Feiden2016a, Jeffries2017a}. The Li abundance would always be higher than that predicted by a standard model at a given $T_{\rm eff}$, as shown in Fig.~\ref{magmodels}, due to a combination of less Li depletion at a given mass and a reduction in $T_{\rm eff}$ at the ZAMS for a given mass. 

Figure~\ref{magmodels} suggests that matching the upper envelope of observed Li depletion in the cool stars of M35, requires some combination of equipartition-strength magnetic fields at the surface or dark spots covering $>30$ per cent of the photosphere. The Li-rotation connection would then emerge if there was a relationship between magnetic activity and rotation that produces significantly different degrees of radius inflation and consequent levels of photospheric Li depletion. Magnetic activity at these levels is expected in the fast-rotating cool stars of M35 -- the connection between magnetic activity measured by coronal and chromospheric emission and rotation is well known, but there is also evidence for: a correlation of global magnetic field strength with rotation deduced from spectropolarimetric observations \citep{Folsom2016a}; a large filling factor ($\sim 0.5$) of the surfaces of active young K-stars by equipartition  magnetic fields inferred from Zeeman broadening  \citep{Valenti2001a}; and rotation-dependent starspot filling factors of up to 0.5 for cool stars in the Pleiades, derived from the relative strengths of molecular bands \citep{Fang2016a}.  

Direct evidence in favour of this scenario comes from the differential levels of radius inflation between stars with fast and slow rotation rates, or between stars with small and large $N_R$ (Fig.~\ref{overradius}b), and the clear tripartite correlation between reduced levels of Li depletion, rapid rotation (or small $N_R$) and the over-radius (Fig.~\ref{tripartite}).
In relative terms, the $6 \pm 2$ per cent difference in $\rho$ between the stars with the smallest and largest $N_R$ is similar to the $\sim 10$ per cent difference inferred between fast and slow rotating K-stars in the Pleiades using similar techniques \citep{Somers2017a}, but is lower than the 14 per cent inflation estimated for fast-rotating M-stars in the Pleiades in comparison with radii predicted by standard models \citep{Jackson2018a}. Note though that the absolute over-radius could be larger than $\rho$, since $\rho$ is estimated in comparison to low luminosity stars in M35, which may themselves still be moderately magnetically active.  Taking into account observational uncertainties, the range of $\rho$ is also broadly consistent with the amount of inflation predicted by the same magnetic models that match the envelope of observed Li abundances (see Figs.~\ref{magmodels} and~\ref{overradius}a).

\subsection{Spots or magnetic inhibition of convection?}
\label{s5.2}

\begin{figure*}
	\centering
\begin{minipage}[b]{0.32\textwidth}
    \includegraphics[width=\textwidth]{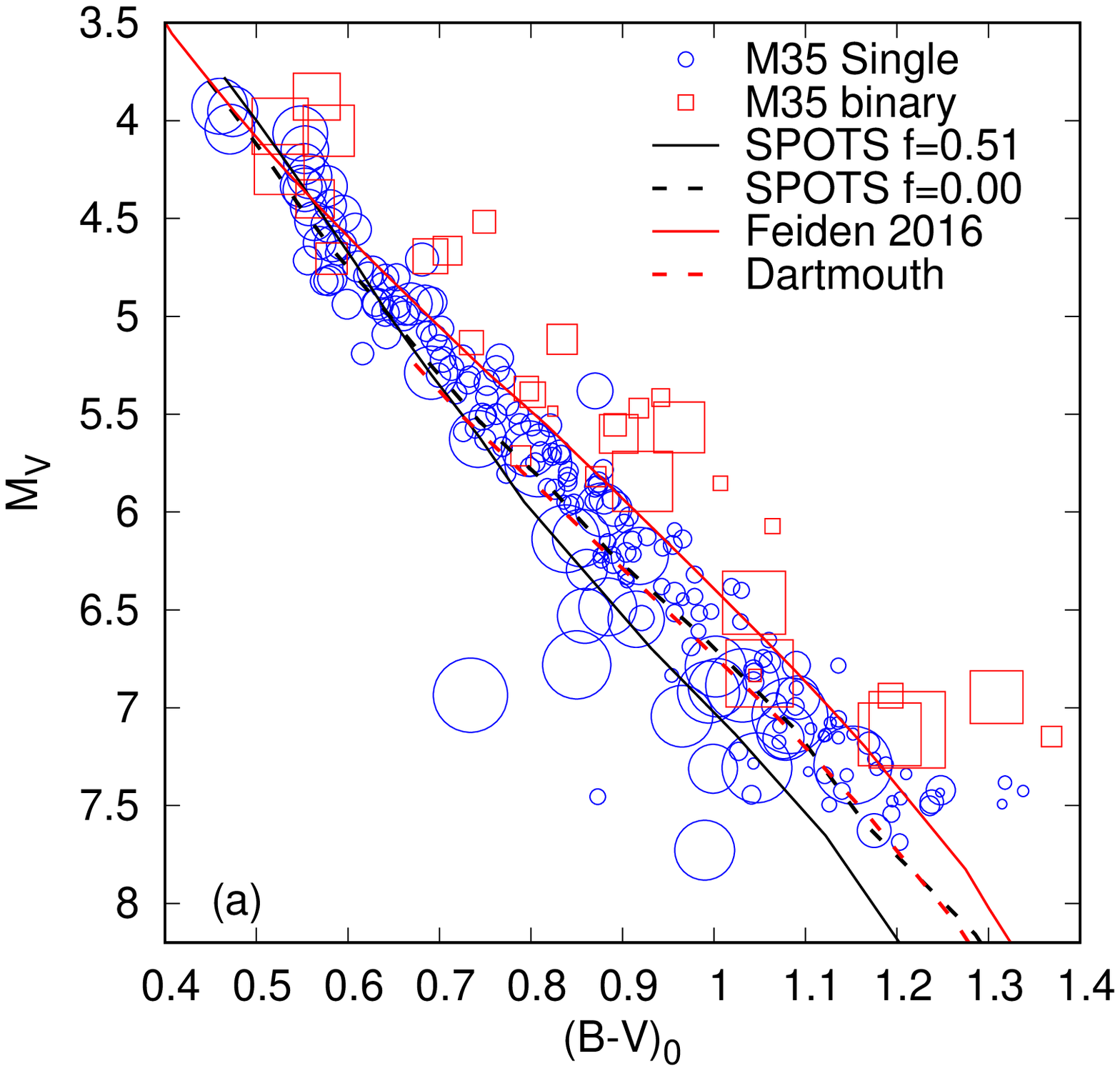}
  \end{minipage}
  \hfill
  \begin{minipage}[b]{0.32\textwidth}
    \includegraphics[width=\textwidth]{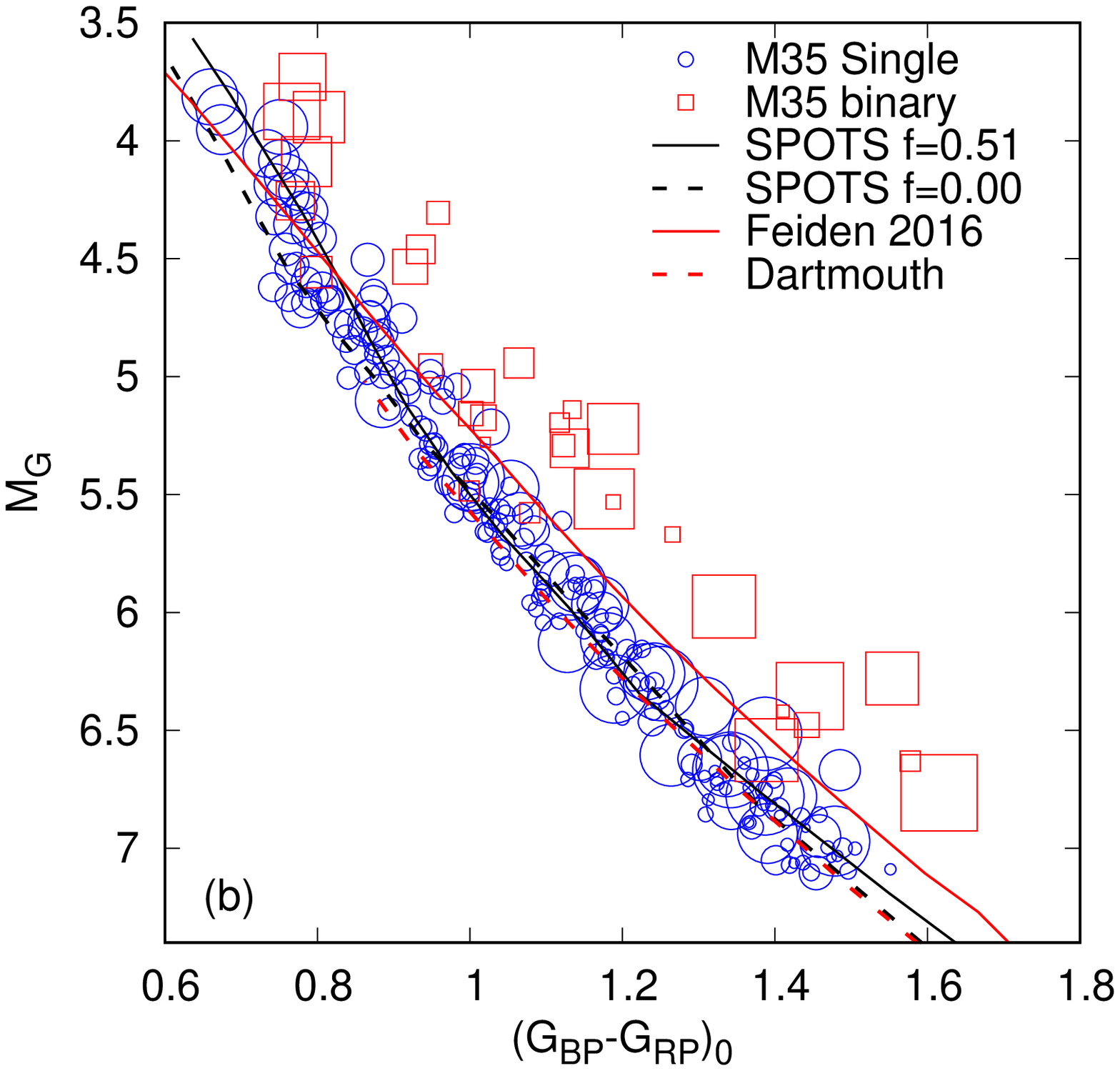}
  \end{minipage}
\begin{minipage}[b]{0.32\textwidth}
    \includegraphics[width=\textwidth]{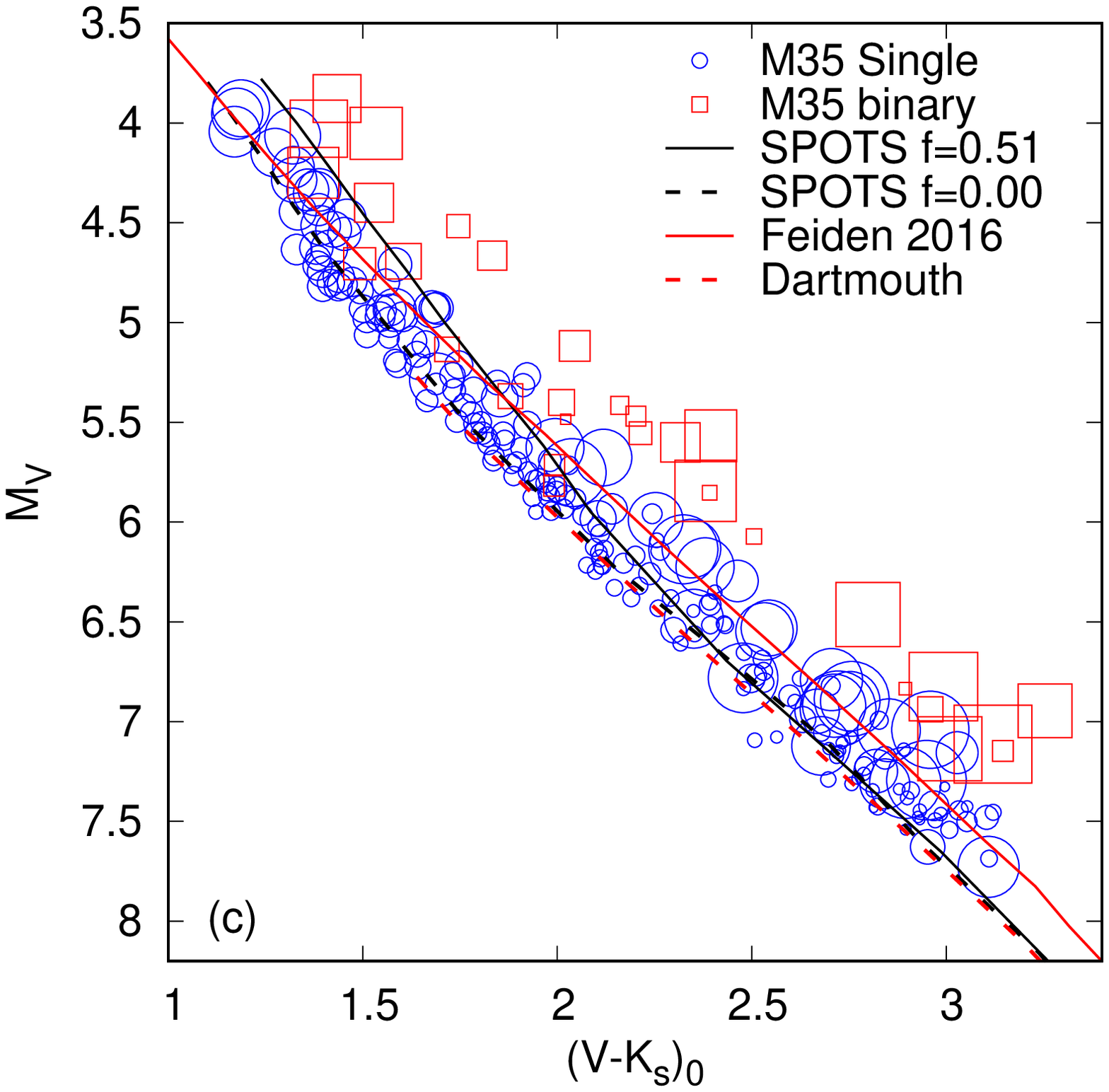}
  \end{minipage}

	\caption{Intrinsic colour vs absolute magnitude diagrams for M35. In each plot the symbol size is proportional to $\log (N_R^{-1})$. Isochrones at 120\,Myr are shown for both magnetic models (starspots with a filling factor of 0.51 and Feiden's magnetic inhibition of convection model, see Section~\ref{s4.5}) and these are compared with their non-magnetic counterparts (spot filling factor of zero and "Dartmouth" respectively).} 
\label{cmd}
\end{figure*} 

Whilst both starspots or the inhibition of convective flux could be responsible for inflating stars and reducing PMS Li depletion in a similar way, there are other observational consequences that can distinguish between the two mechanisms. Inflation of the star (by any mechanism) will decrease $T_{\rm eff}$ at a given luminosity, but
the inhomogeneous photospheres implied by spots have a different spectral energy distribution to a star of similar $T_{\rm eff}$ but uniform temperature. Unspotted regions on a spotted star, which dominate the observed flux, should be slightly hotter than in an unspotted star of the same mass or luminosity, and of course should be hotter than $T_{\rm eff}$ \citep{Spruit1986a, Jackson2009a, Jackson2014a, Somers2020a}.  This leads to colour anomalies whereby a spotted star is bluer in $B-V$ for a given luminosity but, depending on the temperature ratio of the spotted to unspotted photosphere, redder in colours like $V-K_s$ where the starspot flux makes a significant contribution. Such anomalies have been noted before in active stars \citep[e.g.][]{Stauffer2003a} and demonstrated to be rotation-dependent \citep{Kamai2014a, Covey2016a}.

Figure~\ref{cmd} shows three intrinsic colour vs absolute colour-magnitude diagrams (CMDs) for the M35 stars, using a distance of 885~pc, $E(B-V)=0.20$, $A_V=0.62$, $E(V-K_s)= 0.55$ (see Section~\ref{s2}) and values of $A_G=0.55$ and $E(G_{\rm BP} - G_{\rm RP})= 0.27$ using coefficients from \cite{Casagrande2018a}. Superposed are the magnetic model isochrones and their standard model counterparts discussed in Sections~\ref{s4.4} and~\ref{s4.5}. In order to remove any dependence on model atmospheres and bolometric correction calibrations from the comparisons between isochrones, the luminosity and $T_{\rm eff}$ of the non-magnetic Dartmouth and \cite{Feiden2016a} magnetic inhibition models have been converted to absolute magnitudes and colours using the same bolometric corrections used in the unspotted \cite{Somers2020a} isochrone.

The non-magnetic isochrones are very similar in each CMD, but the predictions of the magnetic models are quite different. The magnetic inhibition models of \cite{Feiden2016a} predict that magnetically active cool stars should be redder than inactive stars at a given absolute magnitude in all three CMDs, but with only small differences for the hotter stars. On the contrary, the {\sc spots} isochrones of \cite{Somers2020a} predict that heavily spotted cool stars should be bluer in $B-V$ than unspotted stars of the same absolute magnitude, but similar in $G_{\rm BP}-G_{\rm RP}$ and $V-K_s$, and that spotted hotter stars would have similar colours to unspotted stars in $B-V$, but would be redder in $G_{\rm BP}-G_{\rm RP}$ and $V-K_s$. In the data there is clear evidence that single stars with lower $N_R$, and presumably more magnetically active, are bluer in $B-V$, slightly redder in $V-K_s$ and very similar in $G_{\rm BP}-G_{\rm RP}$ to their less active siblings. In the hotter stars there is very little evidence for a displacement that depends on $N_R$, and the range in $N_R$ is smaller in any case. The binary stars are mostly found above and well separated from the single stars in most cases, as expected, though the separation is cleanest in the $G_{\rm BP}-G_{\rm RP}$ CMD.

These findings are qualitatively similar to that found in the Pleiades by \cite{Kamai2014a}. The rotation-dependent blueward displacement of the cooler K-stars in the $B-V$ CMD, which is the opposite of that expected by the simple inflation produced by globally inhibited convection, is strong evidence for photospheric temperature inhomogeneities. The size of the displacement appears compatible with the level of spot coverage that would be capable of explaining the Li results (see Fig.~\ref{magmodels}). The exact displacements will depend on both the spot filling factor and the ratio of spotted to unspotted photospheric temperatures (assumed to be 0.8 by Somers et al.). At redder colours, the lack of displacement in the $G_{\rm BP}-G_{\rm RP}$ CMD and the small rotation-dependent drift redward in $V-K_s$ CMD, is also qualitatively consistent with a heavily spotted model but probably needs a slightly larger spotted/unspotted temperature ratio to reach quantitative agreement. A caveat to these considerations is that none of these models include chromospheric emission or plages that may make a significant contribution to the $B$-band flux.

\subsection{Problems for magnetic inflation}
\label{s5.3}

Whilst the concordance of Li depletion, magnetic activity and the degree of radius inflation {\it on average} is encouraging, there are aspects of the M35 observations that are problematic for the "magnetic inflation" model. There is a wide dispersion in over-radius as a function of $N_R$ (see Fig.~\ref{overradius}b). Perhaps as a consequence, the relationships between $\Delta$EW$_{\rm Li}$ and $N_R$ and especially between $\Delta$EW$_{\rm Li}$ and $\rho$, also show significant scatter. This requires some explanation, because the $\Delta$EW$_{\rm Li}$, $\rho$ relationship should be fundamental to why there is a dispersion in Li at all. In particular it is a puzzle as to why there are a couple of examples of Li-rich stars with small $N_R$ but $\rho < 1$;  a few stars with small $N_R$ that are not very Li-rich; and some stars with large $N_R$ that are moderately Li-rich and with large $\rho$ (see Fig.~\ref{tripartite}). Whilst the last of these anomalies might be explained by unrecognised binarity, the other outliers are harder to understand.

These outlying stars and the wide dispersion ($\sigma_{\rho} \simeq 0.08$ and $\sigma_{\rm EW(Li)} \simeq 50$\,m\AA\ at a given $N_R$) are unlikely due to simple measurement uncertainties, but some part of the scatter may be caused by additional systematic errors associated with the stellar atmosphere. Photospheric inhomogeneities have both a short term and long term effect on the star. The modelling of \cite{Somers2020a} deals only with the long-term structural effects of spots and their {\it average} effect on the appearance of the star. It is well documented that active stars undergo large changes in spot coverage and brightness on timescales of days (associated with rotational modulation), weeks \citep[associated with the appearance and disappearance of spot groups, e.g.][]{Collier1995a, Jeffers2007a} and years \citep[possibly associated with activity cycles, e.g.][]{Innis1988a, Jarvinen2005a}. These changes are on timescales much shorter than the thermal timescale of the envelope and have no short-term effect on the radius of the star \citep[e.g][]{Spruit1986a}. However they do have short-term effects on the observed luminosity and hence on the derived $\rho$, depending on whether the photometry in all bands is cotemporal and how well the derived $T_{\rm eff}$ tracks the luminosity. In the visible and Kepler K2 bands these effects might add $\pm$2--5 per cent error to the average luminosity measured from single epoch photometry just due to rotational modulation \citep[e.g. see the Kepler K2 light curves of K-type Pleiades stars,][]{Rebull2016a}.  However, longer timescale variations may be more important. A number of field K-dwarfs with short rotation periods, comparable in age and activity to the cool stars of M35, have been monitored over years and decades. These exhibit long term variations that have full amplitudes of $\sim 0.2$ mag in $V$  \citep[e.g.][]{Messina2003b, Jarvinen2005a, Karmakar2016a}, which could lead to $\pm$5--10 per cent errors in estimated luminosity and feed through to an additional $\sim \pm 0.05$ scatter in $\rho$. This is possibly sufficient to explain the dispersion in the $\rho$, $N_R$ relation and perhaps even explain the scatter in $\Delta$EW$_{\rm Li}$ versus $\rho$. Indeed, Fig.~\ref{overradius}b is qualitatively reminiscent of fig.~11 in \cite{Fang2016a}, which shows a general correlation of increasing spot filling factor with decreasing $N_R$, but with a scatter that is significantly larger than the measurement uncertainties.

Spots and chromospheric activity in inhomogeneous atmospheres may also play a role in additional causing EW(Li) variations either through complex NLTE effects on the line formation or through the temperature inhomogenities contributed by starspots or chromospheric plages \citep[e.g.][]{Barrado2001b, King2004a, Xiong2005a, King2010a}. There is some empirical evidence for EW(Li) changes that correlate with the rotational modulation of starspots. The rms variations are of order 10--20\,m\AA\ \citep{Jeffries1994a, Hussain1997a}, but there was little evidence for any EW(Li) changes greater than a few m\AA\  in a sample of rapidly rotating Pleiades G/K-stars on timescales of a year \citep{Jeffries1999a}. It is possible that these additional sources of scatter may partially explain the additional dispersion in the $\Delta$EW$_{\rm Li}$ versus rotation or $N_R$ relationships.

A more interesting {\it physical} cause of dispersion in the $\Delta$EW$_{\rm Li}$ versus rotation or $N_R$ relation could be the rotational histories of stars. The rotation rates at the ZAMS do not necessarily reflect the rotation rates they had when they were depleting Li. According to the evolutionary tracks of \cite{Somers2020a} (with a spot filling factor of 0.34), stars with $5400\geq T_{\rm eff} \geq 4200$\,K in M35 have masses of $0.95 \geq M/M_{\odot} \geq 0.65$. Stars at the extreme ends of these ranges start and end their PMS Li depletion at ages of 3--15 Myr and 4--25 Myr respectively and it is their rotation and magnetic activity levels at these epochs that are crucial to the amount of Li depletion now seen in their photospheres. 

At both ends of this mass range, the stellar moment of inertia decreases by about a factor of 3.5 between the beginning and end of Li burning and then by a further factor of 3 by the time they reach the ZAMS. Towards the beginning of Li burning the stars may or may not be locked to an accretion disc that prevents their spin up and they may also have initial rotation rates that vary by a factor of 10. In the paradigm of early disc-locking, followed by PMS contraction and angular momentum loss through a magnetised wind  \citep[e.g.][]{Denissenkov2010a, Spada2011a}, there is a degeneracy between disc lifetime and initial rotation rate in determining the rotation rate at the ZAMS \citep{Gallet2013a, Gallet2015a}. Moderately slow rotators may have been born slow or have very long-lived discs, whereas moderately fast rotators may have been born fast or had short-lived discs. The importance of this is that stars with similar rotation rates on the ZAMS may have had different rotation rates at $\sim 10$ Myr when they were depleting their Li and this could lead to additional scatter in any relation between Li depletion and ZAMS rotation rate. If that were so, then perhaps the relationship should be tighter in younger clusters, during the epoch of Li destruction.

A further problem for the magnetic inflation idea is that a dispersion in Li depletion requires a dispersion in magnetic activity, internal magnetic fields or surface spot coverage that is correlated with rotation rate (or $N_R$), so that there is then a rotation-dependent degree of magnetic inflation that leads on to rotation-dependent Li depletion. There is plenty of evidence for a rotation-magnetic activity connection at slower rotation periods ($>3$ d) and larger Rossby numbers ($N_R>0.1$), but almost all indicators of magnetic activity exhibit a plateau or "saturation" at $N_R<0.1$, including chromospheric and coronal fluxes and the average surface magnetic flux \citep{Vilhu1984a, Pizzolato2003a, Reiners2009a, Marsden2009a, Wright2011a, Jeffries2011a}. 

This is a problem for a magnetic inflation explanation of the Li dispersion, since it is likely that all the M35 stars considered here had saturated levels of magnetic activity between ages of a few Myr and when their Li depletion phase was completed. Although their rotation periods were probably slightly slower on average than at the ZAMS \citep[$1<P_{\rm rot}/{\rm d}<10$ for subsolar mass stars in several star forming regions, e.g. see fig.~1 of][]{Gallet2015a}, the convective turnover times of PMS stars, which are fully or almost fully convective, are 3--20 times larger during the epoch of Li depletion than they are at the ZAMS \citep[again, using the spot models of][]{Somers2020a}. Even a $0.95M_{\odot}$ star with a rotation period of $\sim 10$ d would have $N_R \sim 0.15$ at the end of its Li depletion epoch, and $N_R$ would be smaller at shorter periods, younger ages or lower masses. If so, then it is hard to see how any rotation-dependence is injected into the Li depletion pattern, unless interior fields and surface spot filling factors do not saturate in the same way as other magnetic activity indicators. That spot filling factor may only saturate at much faster rotation rates and $N_{R}<0.02$ has been claimed by some authors, based on a continuing rise in light curve amplitudes at short periods \citep{Odell1995a, Messina2001b}. However, such studies only measure the asymmetric component of spot coverage and more sensitive spectroscopic studies that are sensitive to the total spot coverage do suggest
starspot saturation at $N_R \sim 0.1$, like other indicators, albeit with significant scatter \citep{Fang2016a}.

A plausible scenario that deserves consideration is that all the M35 stars had their PMS Li depletion inhibited by a similar amount at any given mass, regardless of their rotation rate, because of their saturated levels of magnetic activity. This would set the upper envelope of $A$(Li) at or above the upper envelope of rapid rotators in M35, with little dispersion. Subsequent to this, non-standard mixing mechanisms would act to provide additional Li depletion. If any additional mixing were more effective in {\it slower} rotators then the observed Li-rotation correlation in the cool stars might be reproduced. This cannot be the slow rotational mixing mechanism that may lead to more Li depletion in {\it faster} rotating F- and early G-stars, but possibilities include: mixing associated with rotational shear and core-envelope decoupling that develops as stars contract towards the ZAMS, with slow rotators experiencing more internal differential rotation \citep{Bouvier2008a, Gallet2015a}; or penetration of convective plumes into the radiative zone that becomes less effective in rapid rotators \citep{Montalban2000a}. So far, quantitative models of these processes have focused on stars at around a solar mass and predict a dispersion among ZAMS stars with $T_{\rm eff}>5500$\,K that isn't seen and too much depletion at lower $T_{\rm eff}$ \citep[e.g.][]{Eggenberger2012a, Baraffe2017a}. There is an urgent need to extend these models to lower $T_{\rm eff}$ where the most significant Li dispersion is observed.

\section{Summary}
\label{s6}

WIYN-Hydra observations of cool stars in the rich open cluster M35, along with Gaia DR2 astrometry, have allowed us to compile a database of 242 stars with secure membership, measurements of lithium, rotation periods from ground-based surveys and Kepler K2 observations, and SEDs based on multi-wavelength photometry. This is the largest sample from one open cluster with which to address the Li-rotation connection in young ZAMS stars. These measurements confirm earlier assertions from other clusters that the fast-rotating stars with $4100 < T_{\rm eff} <5500$ K are less depleted than slower rotating siblings at the same temperature by almost 2 orders of magnitude and less depleted than predicted by standard models of PMS stellar evolution. Instead, the upper envelope of Li abundance for cool stars in M35 is better represented by "magnetic models" which feature inhibition of convection by interior magnetic fields or the blocking of radiative flux at the photosphere by dark starspots. The magnetic models "inflate" the stars, making their interiors cooler, leading to lower levels of photospheric Li depletion.
Stars with $T_{\rm eff}>5500$\,K are more depleted than predicted by standard models; there is a hint that the faster rotating hot stars are more depleted, but any trend is masked by observational uncertainties and a relatively small range of Li abundance and rotation at the higher temperatures.

The Li depletion pattern and distribution of rotation periods with $T_{\rm eff}$ in M35 is very similar to that found in the Pleiades, but with a membership sample that is about 2.5 times larger. The slow rotating "I sequence" in M35 is a factor of $1.07\pm 0.03$ slower than in the Pleiades. This together with a consideration of uncertainties in the $T_{\rm eff}$ scale, reddening and metallicity yield an age of $140 \pm 15$\,Myr in comparison to an assumed age of 125\,Myr for the Pleiades.

Supporting evidence for the magnetic models is found in strong correlations between high EW(Li), rotation (or Rossby number, $N_R$) and relative stellar radius determined from SED modelling.  On average, the fastest rotating stars with the smallest $N_R$ are inflated by $6 \pm 2$ per cent with respect to the slow rotators and have higher EW(Li) at the same $T_{\rm eff}$. This corresponds well with what is expected from the same magnetic models that are capable of explaining the range of Li abundances; these require that convection is suppressed by global magnetic fields that reach equipartition levels at the surfaces of the fastest rotators, or that the fastest rotators have $>30$ per cent of their photospheres obscured by dark starspots. An examination of colour-magnitude diagrams betrays colour anomalies that get bigger for smaller $N_R$, which favours the starspot scenario. 

The EW(Li)-rotation (or $N_R$) correlation at $T_{\rm eff} < 5500$\,K has a dispersion larger than the measurement uncertainties. Since the photometric binaries in M35 follow a very similar relationship, unrecognised binarity is unlikely to play a role in this. The lack of a deterministic relationship between rotation, $T_{\rm eff}$ and EW(Li)  may be due to remaining uncertainties in the relationship between EW(Li) and Li abundance, caused by inhomogeneous photospheres and magnetic activity. Alternatively, it could be that the rotational history of the stars, and in particular the rotation rate at the epoch of Li destruction (3--30 Myr), which is not uniquely determined by their present rotation rates, may play the dominant role in determining the photospheric Li abundance at the ZAMS.

Any model where rotation-dependent magnetic activity leads to radius inflation and a rotation-dependent level of Li depletion faces an important challenge from the saturation of magnetic activity indicators observed to occur at $N_R < 0.1$. At the epoch of Li destruction we expect almost all the M35 stars to have had $N_R < 0.1$. Unless starspot coverage or interior magnetic fields saturate at significantly lower $N_R$, then it is difficult to see how a rotation-dependence is imprinted on the Li depletion pattern. Instead, the data are also consistent with the idea that magnetic inflation reduces the PMS Li depletion of all stars by a similar amount, regardless of rotation rate, and that subsequent, rotation-dependent mixing causes the slow rotators to deplete more of their photospheric Li depletion by the time they reach the age of M35.   Resolving these issues requires more modelling efforts and would benefit from similar observational studies to track both the extent and rotation-dependence of photospheric Li depletion for clusters both during and immediately after the main epoch of PMS Li destruction at 3--30\,Myr.

\section*{Acknowledgments}

Data presented herein were obtained at the WIYN 3.5m Observatory from
telescope time allocated to NN-EXPLORE through (a) the scientific
partnership of the National Aeronautics and Space Administration, the
National Science Foundation, and the National Optical Astronomy
Observatory, and (b) Indiana University's share of time on the WIYN
3.5-m. This work was supported by a NASA WIYN PI Data Award,
administered by the NASA Exoplanet Science Institute, though JPL RSA \#
1560105.   This publication makes use of VOSA, developed under the 
Spanish Virtual Observatory project supported from the Spanish 
MINECO through grant AyA2017-84089.
RDJ and RJJ also wish to thank the UK Science and Technology
Facilities Council for financial support.

\section*{Data Availability Statement}
This work is based on spectra obtained on the WIYN-3.5m telescope. The raw data are available in the NOAO archive (http://archive1.dm.noao.edu/) by searching for Program Number 2017B-0281.
All other data are either included in the online Tables or are obtained from published work and catalogues as referenced in the text.
\appendix

\section{Calculation of the probability of cluster membership of individual targets}
\label{appa}

\begin{figure*}

	\begin{minipage}[t]{0.9\textwidth}
	\includegraphics[width = 160mm]{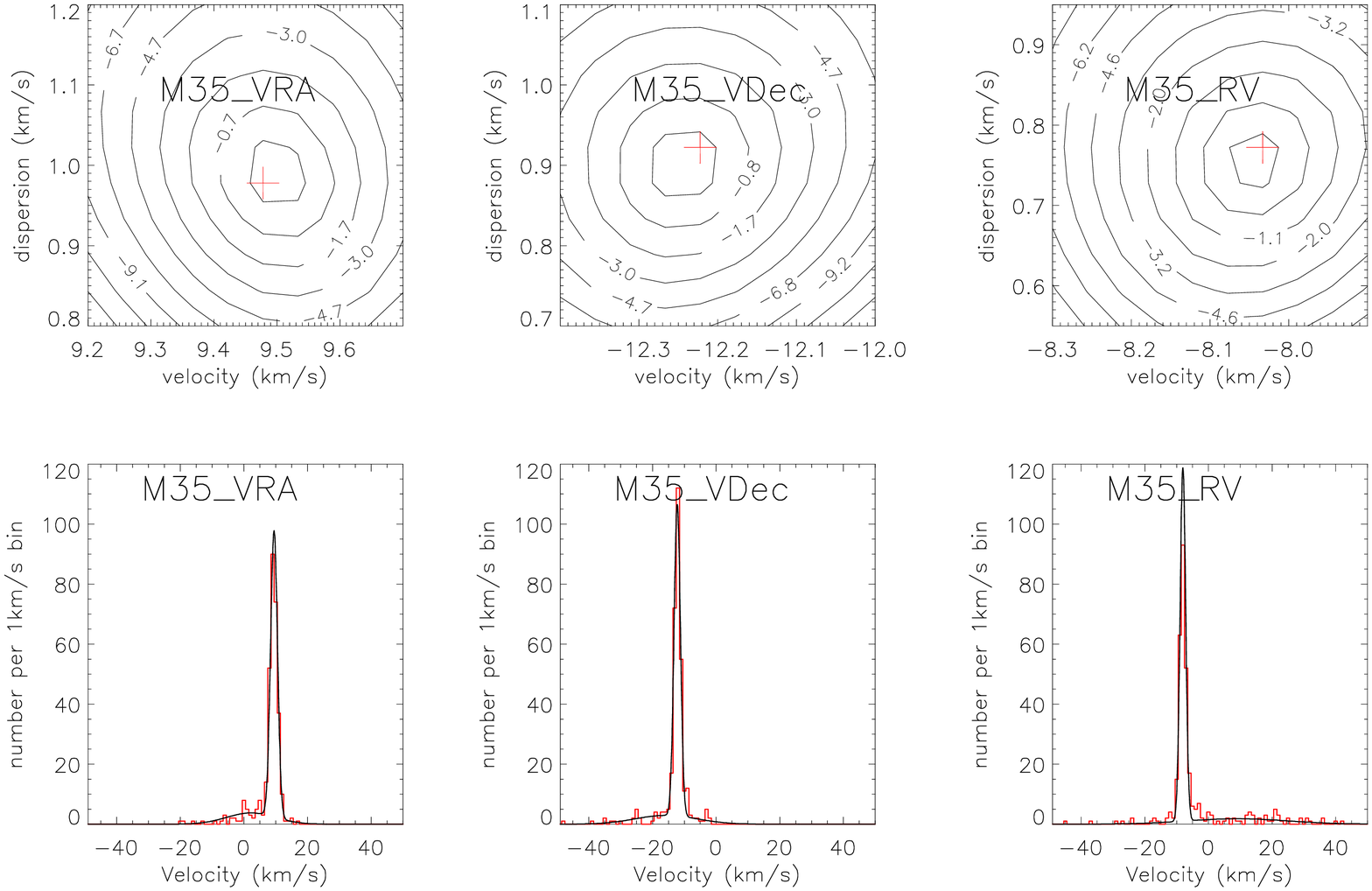}
	\end{minipage}
	\caption{Results of the 3D maximum likelihood analysis used to determine the cluster properties and target membership probabilities. The upper plots shows contours of log likelihood of cluster intrinsic velocity and dispersion relative to the maximum likelihood value which is located at the cross on each plot. The lower plots show histograms of measured velocities together with a predicted model distribution evaluated at the maximum likelihood values of cluster velocity, dispersion and fraction that are cluster members (0.78). }
	\label{figA1}
	
\end{figure*}
A maximum likelihood technique was used to determine the cluster membership probability for each target, using their their radial velocity (RV) and proper motion velocities ($V_{\rm RA}$ and $V_{\rm DEC}$). The intrinsic probability density of the targets in velocity space was modelled as the sum of two  3-dimensional (3D) Gaussian distributions. The first is a relatively narrow distribution representing cluster members and the second a broader distribution representing a background population of field stars. This intrinsic distribution was broadened by measurement uncertainties in velocity and, in the case of the $RV$ component, by the effects of binary motion on the measured RV of binary stars, to give a model distribution of target velocities.

A maximum likelihood method was used to determine the properties of the intrinsic Gaussians and the fraction $f_{\rm C}$ of targets belonging to the cluster. Full details of the modelling procedure are presented in \citet{Jackson2020a}. To model the effects of binarity on the RV distribution, a binary fraction of 0.4 was assumed and the binary period and flat mass ratio distribution found by \citet{Raghavan2010a} for field stars. As shown in \citet{Jackson2020a}, the assumed parameters of the binary distribution hardly affect the membership probability estimates but do have small effects on the derived intrinsic velocity dispersion. The effect of binarity on the proper motion measurements are ignored, as is any dispersion in the distance to the cluster members. The former effect was shown by \citet{Jackson2020a} to be much smaller than that of binarity on the RV distribution because of the averaging effect of taking Gaia DR2 measurements over 22 months. The latter is negligible at the 885\,pc distance to M35. Uncertainties in the average distance lead to uncertainties in the tangential velocity dispersions but have no effect on membership probabilities.

The results of the analysis are given in Fig.~\ref{figA1} and Table~\ref{cluster3D}. The upper plots in Fig.~\ref{figA1} show contours of log likelihood for different combinations of model parameters along each dimension of the velocity space. The lower plots compare the data and best-fit model distribution (using the median uncertainties). Membership probabilities are computed for each target and are listed in Table~\ref{targets}. Note that the cluster parameters listed in Table~\ref{cluster3D} are a first approximation to the true velocity dispersions. There are a number of systematic effects (rotation, expansion, asymmetry etc.) that have not been considered that could affect the velocity dispersions but are unlikely to change the membership probabilities significantly \citep[see][for a discussion]{Jackson2020a}.

\begin{table}
\caption{Results of the 3D maximum likelihood analysis used to determine the intrinsic velocities, velocity dispersions and fraction of cluster members for the model Gaussian distribution of representing the cluster population.}
\begin{tabular}{l@{\hspace{0.5\tabcolsep}}ccccc} 
\hline
             	             &	V$_{\rm RA}$ (km\,s$^{-1}$)&	V$_{\rm Dec}$ (km\,s$^{-1}$)& RV (km\,s$^{-1}$)\\\hline
Cluster velocity 	         &	9.50$\pm$0.08	&	-12.24$\pm$0.07	&	-8.10$\pm$0.07	\\
Cluster dispersion	         &	 1.00$\pm$0.07	&	0.90$\pm$0.07	&	0.74$\pm$0.07	\\
$f_{\rm C}$ 	 &	 \multicolumn{3}{c}{$0.78\pm0.02$}	\\\hline
\end{tabular}
  \label{cluster3D}
\end{table}

\section{Photometry used in the SED fitting}
\label{appb}

Table~\ref{sedtable} lists the photometry used in the SED fitting (Section~\ref{s3.2}). The photometric data (with uncertainties) were gathered from various catalogues  (cross-matching co-ordinates within 2 arcsecond);
\begin{enumerate}
	\item $G$, $G_{\rm BP}$ and $G_{\rm RP}$ magnitudes from the Gaia DR2 catalogue \citep{Gaia2018a}.
	\item $J$, $H$ and $K_s$ from the 2MASS catalogue \citep{Skrutskie2006a}.
	\item Near infrared $W1$ and $W2$ magnitudes from the ALLWISE catalogue \citep{Cutri2013a}. 
	\item $U$, $B$, $V$, $R_{\rm C}$ and $I_{\rm C}$ magnitudes were taken where possible from a recent homogeneous photometric survey of M35 (see AT18). $U$ magnitudes were available for only 33 per cent of members.
\end{enumerate}

\begin{table*}
\caption{Photometry and SED-fitting results for members of M35 with rotation periods (Section~\ref{s3.2}).  The object names correspond to those in Table~\ref{targets}. There are 242 rows and the the columns are listed here; the full table is only available electronically.}
\begin{tabular}{lccccc} 
\hline
Object &	 $U$ & $B$ & $V$ & $R_{\rm C}$ & $I_{\rm C}$ \\
       &	$G$ & $G_{\rm BP}$ 	& $G_{\rm RP}$ && \\
       &    $J\ \ $	& $H\ \  $ &	$K_s\ \ $	& $W_1$  & $W_2$ 	\\
  & \multicolumn{5}{c}{mag} \\ \hline
J06070616+2402101 &  $99.999\pm   9.999$&      $99.999\pm   9.999$&    $99.999\pm   9.999$&    $99.999\pm   9.999$&      $99.999\pm   9.999$\\      & $16.6331\pm   0.0032$&     $17.2495\pm   0.0117$&    $15.7899\pm   0.0086$\\
 & $14.681\pm   0.028$&    $14.106\pm   0.033$&    $14.072\pm   0.042$&    $14.048\pm   0.031$&     $14.239\pm   0.053$\\    
  J06072249+2421401 &  $99.999\pm   9.999$&      $17.537\pm   0.024$&    $16.482\pm   0.01$&     $15.884\pm   0.026$&      $15.238\pm   0.010$\\
  & $16.1573\pm   0.0044$&     $16.7887\pm   0.017$&     $15.3801\pm   0.0101$& & \\
  & $14.264\pm   0.027$&    $13.741\pm   0.026$&    $13.577\pm   0.033$&    $13.560\pm    0.026$&     $13.654\pm   0.038$ \\    
  J06072843+2416426 &  $99.999\pm   9.999$&      $16.155\pm   0.017$&    $15.286\pm   0.015$&    $14.788\pm   0.007$&      $14.304\pm   0.013$\\    &  $15.0479\pm   0.001$&      $15.5359\pm   0.0034$&    $14.4004\pm   0.0032$& &\\
  & $13.648\pm   0.027$&    $13.199\pm   0.030$&     $13.144\pm   0.028$&    $12.985\pm   0.025$&     $12.997\pm   0.030$\\
\hline
  \end{tabular}
  \label{sedtable}
\end{table*}

\bibliographystyle{mn2e.bst} 
\bibliography{references}

\begin{thebibliography}{}

\bibitem[\protect\citeauthoryear{{Allard}, {Homeier} \& {Freytag}}{{Allard}
  et~al.}{2012}]{Allard2012a}
{Allard} F.,  {Homeier} D.,    {Freytag} B.,  2012, Philosophical Transactions
  of the Royal Society of London Series A, 370, 2765

\bibitem[\protect\citeauthoryear{{Anthony-Twarog}, {Deliyannis}, {Harmer},
  {Lee-Brown}, {Steinhauer}, {Sun} \& {Twarog}}{{Anthony-Twarog}
  et~al.}{2018}]{Anthony-Twarog2018a}
{Anthony-Twarog} B.~J.,  {Deliyannis} C.~P.,  {Harmer} D.,  {Lee-Brown} D.~B.,
  {Steinhauer} A.,  {Sun} Q.,    {Twarog} B.~A.,  2018, \aj, 156, 37

\bibitem[\protect\citeauthoryear{{Asplund}, {Grevesse}, {Sauval} \&
  {Scott}}{{Asplund} et~al.}{2009}]{Asplund2009a}
{Asplund} M.,  {Grevesse} N.,  {Sauval} A.~J.,    {Scott} P.,  2009, \araa, 47,
  481

\bibitem[\protect\citeauthoryear{{Balachandran}, {Lambert} \&
  {Stauffer}}{{Balachandran} et~al.}{1988}]{Balachandran1988a}
{Balachandran} S.,  {Lambert} D.~L.,    {Stauffer} J.~R.,  1988, \apj, 333, 267

\bibitem[\protect\citeauthoryear{{Baraffe}, {Chabrier}, {Allard} \&
  {Hauschildt}}{{Baraffe} et~al.}{1998}]{Baraffe1998a}
{Baraffe} I.,  {Chabrier} G.,  {Allard} F.,    {Hauschildt} P.~H.,  1998, \aap,
  337, 403

\bibitem[\protect\citeauthoryear{{Baraffe}, {Homeier}, {Allard} \&
  {Chabrier}}{{Baraffe} et~al.}{2015}]{Baraffe2015a}
{Baraffe} I.,  {Homeier} D.,  {Allard} F.,    {Chabrier} G.,  2015, \aap, 577,
  A42

\bibitem[\protect\citeauthoryear{{Baraffe}, {Pratt}, {Goffrey}, {Constantino},
  {Folini}, {Popov}, {Walder} \& {Viallet}}{{Baraffe}
  et~al.}{2017}]{Baraffe2017a}
{Baraffe} I.,  {Pratt} J.,  {Goffrey} T.,  {Constantino} T.,  {Folini} D.,
  {Popov} M.~V.,  {Walder} R.,    {Viallet} M.,  2017, \apj, 845, L6

\bibitem[\protect\citeauthoryear{{Barnes}}{{Barnes}}{2003}]{Barnes2003a}
{Barnes} S.~A.,  2003, \apj, 586, 464

\bibitem[\protect\citeauthoryear{{Barrado}, {Bouy}, {Bouvier}, {Moraux},
  {Sarro}, {Bertin}, {Cuillandre}, {Stauffer}, {Lillo-Box} \&
  {Pollock}}{{Barrado} et~al.}{2016}]{Barrado2016a}
{Barrado} D.,  {Bouy} H.,  {Bouvier} J.,  {Moraux} E.,  {Sarro} L.~M.,
  {Bertin} E.,  {Cuillandre} J.-C.,  {Stauffer} J.~R.,  {Lillo-Box} J.,
  {Pollock} A.,  2016, \aap, 596, A113

\bibitem[\protect\citeauthoryear{{Barrado y Navascu{\'e}s}, {Deliyannis} \&
  {Stauffer}}{{Barrado y Navascu{\'e}s} et~al.}{2001}]{Barrado2001a}
{Barrado y Navascu{\'e}s} D.,  {Deliyannis} C.~P.,    {Stauffer} J.~R.,  2001,
  \apj, 549, 452

\bibitem[\protect\citeauthoryear{{Barrado y Navascu{\'e}s}, {Garc{\'\i}a
  L{\'o}pez}, {Severino} \& {Gomez}}{{Barrado y Navascu{\'e}s}
  et~al.}{2001}]{Barrado2001b}
{Barrado y Navascu{\'e}s} D.,  {Garc{\'\i}a L{\'o}pez} R.~J.,  {Severino} G.,
   {Gomez} M.~T.,  2001, \aap, 371, 652

\bibitem[\protect\citeauthoryear{{Bayo}, {Rodrigo}, {Barrado Y Navascu{\'e}s},
  {Solano}, {Guti{\'e}rrez}, {Morales-Calder{\'o}n} \& {Allard}}{{Bayo}
  et~al.}{2008}]{Bayo2008a}
{Bayo} A.,  {Rodrigo} C.,  {Barrado Y Navascu{\'e}s} D.,  {Solano} E.,
  {Guti{\'e}rrez} R.,  {Morales-Calder{\'o}n} M.,    {Allard} F.,  2008, \aap,
  492, 277

\bibitem[\protect\citeauthoryear{{Bershady}, {Barden}, {Blanche}, {Blanco},
  {Corson}, {Crawford}, {Glaspey}, {Habraken}, {Jacoby}, {Keyes}, {Knezek},
  {Lemaire}, {Liang}, {McDougall}, {Poczulp}, {Sawyer}, {Westfall} \&
  {Willmarth}}{{Bershady} et~al.}{2008}]{Bershady2008a}
{Bershady} M.,  {Barden} S.,  {Blanche} P.-A.,  {Blanco} D.,  {Corson} C.,
  {Crawford} S.,  {Glaspey} J.,  {Habraken} S.,  {Jacoby} G.,  {Keyes} J.,
  {Knezek} P.,  {Lemaire} P.,  {Liang} M.,  {McDougall} E.,  {Poczulp} G.,
  {Sawyer} D.,  {Westfall} K.,    {Willmarth} D.,  2008, SPIE, 7014E, 15B

\bibitem[\protect\citeauthoryear{{Bildsten}, {Brown}, {Matzner} \&
  {Ushomirsky}}{{Bildsten} et~al.}{1997}]{Bildsten1997a}
{Bildsten} L.,  {Brown} E.~F.,  {Matzner} C.~D.,    {Ushomirsky} G.,  1997,
  \apj, 482, 442

\bibitem[\protect\citeauthoryear{{Bodenheimer}}{{Bodenheimer}}{1965}]{Bodenheimer1965a}
{Bodenheimer} P.,  1965, \apj, 142, 451

\bibitem[\protect\citeauthoryear{{Bouvier}}{{Bouvier}}{2008}]{Bouvier2008a}
{Bouvier} J.,  2008, \aap, 489, L53

\bibitem[\protect\citeauthoryear{{Bouvier}, {Barrado}, {Moraux}, {Stauffer},
  {Rebull}, {Hillenbrand}, {Bayo}, {Boisse}, {Bouy}, {DiFolco}, {Lillo-Box} \&
  {Calder{\'o}n}}{{Bouvier} et~al.}{2018}]{Bouvier2018a}
{Bouvier} J.,  {Barrado} D.,  {Moraux} E.,  {Stauffer} J.,  {Rebull} L.,
  {Hillenbrand} L.,  {Bayo} A.,  {Boisse} I.,  {Bouy} H.,  {DiFolco} E.,
  {Lillo-Box} J.,    {Calder{\'o}n} M.~M.,  2018, \aap, 613, A63

\bibitem[\protect\citeauthoryear{{Bouvier}, {Lanzafame}, {Venuti} \& {et
  al.}}{{Bouvier} et~al.}{2016}]{Bouvier2016a}
{Bouvier} J.,  {Lanzafame} A.~C.,  {Venuti} L.,    {et al.} 2016, \aap, 590,
  A78

\bibitem[\protect\citeauthoryear{{Bouy}, {Bertin}, {Barrado}, {Sarro},
  {Olivares}, {Moraux}, {Bouvier}, {Cuilland re}, {Ribas} \& {Beletsky}}{{Bouy}
  et~al.}{2015}]{Bouy2015a}
{Bouy} H.,  {Bertin} E.,  {Barrado} D.,  {Sarro} L.~M.,  {Olivares} J.,
  {Moraux} E.,  {Bouvier} J.,  {Cuilland re} J.~C.,  {Ribas} {\'A}.,
  {Beletsky} Y.,  2015, \aap, 575, A120

\bibitem[\protect\citeauthoryear{{Butler}, {Cohen}, {Duncan} \&
  {Marcy}}{{Butler} et~al.}{1987}]{Butler1987a}
{Butler} R.~P.,  {Cohen} R.~D.,  {Duncan} D.~K.,    {Marcy} G.~W.,  1987,
  \apjl, 319, L19

\bibitem[\protect\citeauthoryear{{Casagrande} \& {VandenBerg}}{{Casagrande} \&
  {VandenBerg}}{2018}]{Casagrande2018a}
{Casagrande} L.,  {VandenBerg} D.~A.,  2018, \mnras, 479, L102

\bibitem[\protect\citeauthoryear{{Chaboyer}, {Demarque} \&
  {Pinsonneault}}{{Chaboyer} et~al.}{1995}]{Chaboyer1995a}
{Chaboyer} B.,  {Demarque} P.,    {Pinsonneault} M.~H.,  1995, \apj, 441, 876

\bibitem[\protect\citeauthoryear{{Coelho}, {Barbuy}, {Melendez}, {Schiavon} \&
  {Castilho}}{{Coelho} et~al.}{2005}]{Coelho2005a}
{Coelho} P.,  {Barbuy} B.,  {Melendez} J.,  {Schiavon} R.,    {Castilho} B.,
  2005, VizieR Online Data Catalog, 6120, 0

\bibitem[\protect\citeauthoryear{{Collier Cameron}}{{Collier
  Cameron}}{1995}]{Collier1995a}
{Collier Cameron} A.,  1995, \mnras, 275, 534

\bibitem[\protect\citeauthoryear{{Covey}, {Ag{\"u}eros}, {Law}, {Liu},
  {Ahmadi}, {Laher}, {Levitan}, {Sesar} \& {Surace}}{{Covey}
  et~al.}{2016}]{Covey2016a}
{Covey} K.~R.,  {Ag{\"u}eros} M.~A.,  {Law} N.~M.,  {Liu} J.,  {Ahmadi} A.,
  {Laher} R.,  {Levitan} D.,  {Sesar} B.,    {Surace} J.,  2016, \apj, 822, 81

\bibitem[\protect\citeauthoryear{{Cummings}}{{Cummings}}{2011}]{Cummings2011a}
{Cummings} J.,  2011, PhD thesis, Indiana University

\bibitem[\protect\citeauthoryear{{Cutri}, {Wright}, {Conrow}, {Fowler},
  {Eisenhardt}, {Grillmair}, {Kirkpatrick} \& {et al.}}{{Cutri}
  et~al.}{2013}]{Cutri2013a}
{Cutri} R.~M.,  {Wright} E.~L.,  {Conrow} T.,  {Fowler} J.~W.,  {Eisenhardt}
  P.~R.~M.,  {Grillmair} C.,  {Kirkpatrick} J.~D.,    {et al.} 2013, Technical
  report, {Explanatory Supplement to the AllWISE Data Release Products}

\bibitem[\protect\citeauthoryear{{D'Antona} \& {Mazzitelli}}{{D'Antona} \&
  {Mazzitelli}}{1997}]{Dantona1997a}
{D'Antona} F.,  {Mazzitelli} I.,  1997, Memorie della Societa Astronomica
  Italiana, 68, 807

\bibitem[\protect\citeauthoryear{{Deliyannis}, {Demarque} \&
  {Kawaler}}{{Deliyannis} et~al.}{1990}]{Deliyannis1990a}
{Deliyannis} C.~P.,  {Demarque} P.,    {Kawaler} S.~D.,  1990, \apjs, 73, 21

\bibitem[\protect\citeauthoryear{{Dell'Omodarme}, {Valle}, {Degl'Innocenti} \&
  {Prada Moroni}}{{Dell'Omodarme} et~al.}{2012}]{Dellomodarme2012a}
{Dell'Omodarme} M.,  {Valle} G.,  {Degl'Innocenti} S.,    {Prada Moroni} P.~G.,
   2012, \aap, 540, A26

\bibitem[\protect\citeauthoryear{{Denissenkov}, {Pinsonneault}, {Terndrup} \&
  {Newsham}}{{Denissenkov} et~al.}{2010}]{Denissenkov2010a}
{Denissenkov} P.~A.,  {Pinsonneault} M.,  {Terndrup} D.~M.,    {Newsham} G.,
  2010, \apj, 716, 1269

\bibitem[\protect\citeauthoryear{{Dotter}, {Chaboyer}, {Jevremovi{\'c}},
  {Kostov}, {Baron} \& {Ferguson}}{{Dotter} et~al.}{2008}]{Dotter2008a}
{Dotter} A.,  {Chaboyer} B.,  {Jevremovi{\'c}} D.,  {Kostov} V.,  {Baron} E.,
   {Ferguson} J.~W.,  2008, \apjs, 178, 89

\bibitem[\protect\citeauthoryear{{Duncan} \& {Jones}}{{Duncan} \&
  {Jones}}{1983}]{Duncan1983a}
{Duncan} D.~K.,  {Jones} B.~F.,  1983, \apj, 271, 663

\bibitem[\protect\citeauthoryear{{Eggenberger}, {Haemmerl{\'e}}, {Meynet} \&
  {Maeder}}{{Eggenberger} et~al.}{2012}]{Eggenberger2012a}
{Eggenberger} P.,  {Haemmerl{\'e}} L.,  {Meynet} G.,    {Maeder} A.,  2012,
  \aap, 539, A70

\bibitem[\protect\citeauthoryear{{Fang}, {Zhao}, {Zhao}, {Chen} \& {Bharat
  Kumar}}{{Fang} et~al.}{2016}]{Fang2016a}
{Fang} X.-S.,  {Zhao} G.,  {Zhao} J.-K.,  {Chen} Y.-Q.,    {Bharat Kumar} Y.,
  2016, \mnras, 463, 2494

\bibitem[\protect\citeauthoryear{{Feiden}}{{Feiden}}{2016}]{Feiden2016a}
{Feiden} G.~A.,  2016, \aap, 593, A99

\bibitem[\protect\citeauthoryear{{Feiden} \& {Chaboyer}}{{Feiden} \&
  {Chaboyer}}{2013}]{Feiden2013a}
{Feiden} G.~A.,  {Chaboyer} B.,  2013, \apj, 779, 183

\bibitem[\protect\citeauthoryear{{Folsom}, {Petit}, {Bouvier}, {L{\`e}bre},
  {Amard}, {Palacios}, {Morin}, {Donati}, {Jeffers}, {Marsden} \&
  {Vidotto}}{{Folsom} et~al.}{2016}]{Folsom2016a}
{Folsom} C.~P.,  {Petit} P.,  {Bouvier} J.,  {L{\`e}bre} A.,  {Amard} L.,
  {Palacios} A.,  {Morin} J.,  {Donati} J.~F.,  {Jeffers} S.~V.,  {Marsden}
  S.~C.,    {Vidotto} A.~A.,  2016, \mnras, 457, 580

\bibitem[\protect\citeauthoryear{{Gaia Collaboration}, {Babusiaux}, van Leeuwen
  \& {et al.}}{{Gaia Collaboration} et~al.}{2018}]{Gaia2018b}
{Gaia Collaboration} {Babusiaux} C.,  van Leeuwen F.,    {et al.} 2018, \aap,
  616, A10

\bibitem[\protect\citeauthoryear{{Gaia Collaboration}, {Brown}, {Vallenari} \&
  {et al.}}{{Gaia Collaboration} et~al.}{2018}]{Gaia2018a}
{Gaia Collaboration} {Brown} A.~G.~A.,  {Vallenari} A.,    {et al.} 2018, \aap,
  616, A1

\bibitem[\protect\citeauthoryear{{Gallet} \& {Bouvier}}{{Gallet} \&
  {Bouvier}}{2013}]{Gallet2013a}
{Gallet} F.,  {Bouvier} J.,  2013, \aap, 556, A36

\bibitem[\protect\citeauthoryear{{Gallet} \& {Bouvier}}{{Gallet} \&
  {Bouvier}}{2015}]{Gallet2015a}
{Gallet} F.,  {Bouvier} J.,  2015, \aap, 577, A98

\bibitem[\protect\citeauthoryear{{Garcia Lopez} \& {Spruit}}{{Garcia Lopez} \&
  {Spruit}}{1991}]{Garcia1991a}
{Garcia Lopez} R.~J.,  {Spruit} H.~C.,  1991, \apj, 377, 268

\bibitem[\protect\citeauthoryear{{Geller}, {Mathieu}, {Braden}, {Meibom},
  {Platais} \& {Dolan}}{{Geller} et~al.}{2010}]{Geller2010a}
{Geller} A.~M.,  {Mathieu} R.~D.,  {Braden} E.~K.,  {Meibom} S.,  {Platais} I.,
     {Dolan} C.~J.,  2010, \aj, 139, 1383

\bibitem[\protect\citeauthoryear{{Horne}}{{Horne}}{1986}]{Horne1986a}
{Horne} K.,  1986, \pasp, 98, 609

\bibitem[\protect\citeauthoryear{{Hussain}, {Unruh} \& {Collier
  Cameron}}{{Hussain} et~al.}{1997}]{Hussain1997a}
{Hussain} G.~A.~J.,  {Unruh} Y.~C.,    {Collier Cameron} A.,  1997, \mnras,
  288, 343

\bibitem[\protect\citeauthoryear{{Innis}, {Thompson}, {Coates} \&
  {Evans}}{{Innis} et~al.}{1988}]{Innis1988a}
{Innis} J.~L.,  {Thompson} K.,  {Coates} D.~W.,    {Evans} T.~L.,  1988,
  \mnras, 235, 1411

\bibitem[\protect\citeauthoryear{{Jackson}, {Deliyannis} \&
  {Jeffries}}{{Jackson} et~al.}{2018}]{Jackson2018a}
{Jackson} R.~J.,  {Deliyannis} C.~P.,    {Jeffries} R.~D.,  2018, \mnras, 476,
  3245

\bibitem[\protect\citeauthoryear{{Jackson} \& {Jeffries}}{{Jackson} \&
  {Jeffries}}{2014}]{Jackson2014a}
{Jackson} R.~J.,  {Jeffries} R.~D.,  2014, \mnras, 441, 2111

\bibitem[\protect\citeauthoryear{{Jackson}, {Jeffries} \& {Maxted}}{{Jackson}
  et~al.}{2009}]{Jackson2009a}
{Jackson} R.~J.,  {Jeffries} R.~D.,    {Maxted} P.~F.~L.,  2009, \mnras, 399,
  L89

\bibitem[\protect\citeauthoryear{{Jackson}, {Jeffries}, {Wright} \& {et
  al.}}{{Jackson} et~al.}{2020}]{Jackson2020a}
{Jackson} R.~J.,  {Jeffries} R.~D.,  {Wright} N.~J.,    {et al.} 2020, \mnras,
  496, 4701

\bibitem[\protect\citeauthoryear{{J{\"a}rvinen}, {Berdyugina}, {Tuominen},
  {Cutispoto} \& {Bos}}{{J{\"a}rvinen} et~al.}{2005}]{Jarvinen2005a}
{J{\"a}rvinen} S.~P.,  {Berdyugina} S.~V.,  {Tuominen} I.,  {Cutispoto} G.,
  {Bos} M.,  2005, \aap, 432, 657

\bibitem[\protect\citeauthoryear{{Jeffers}, {Donati} \& {Collier
  Cameron}}{{Jeffers} et~al.}{2007}]{Jeffers2007a}
{Jeffers} S.~V.,  {Donati} J.-F.,    {Collier Cameron} A.,  2007, \mnras, 375,
  567

\bibitem[\protect\citeauthoryear{{Jeffries}}{{Jeffries}}{1999}]{Jeffries1999a}
{Jeffries} R.~D.,  1999, \mnras, 309, 189

\bibitem[\protect\citeauthoryear{{Jeffries}}{{Jeffries}}{2006}]{Jeffries2006a}
{Jeffries} R.~D.,  2006, in {Randich} S.,  {Pasquini} L.,  eds, Chemical
  Abundances and Mixing in Stars in the Milky Way and its Satellites
  {Pre-Main-Sequence Lithium Depletion}.
p.~163

\bibitem[\protect\citeauthoryear{{Jeffries}, {Byrne}, {Doyle}, {Anders},
  {James} \& {Lanzafame}}{{Jeffries} et~al.}{1994}]{Jeffries1994a}
{Jeffries} R.~D.,  {Byrne} P.~B.,  {Doyle} J.~G.,  {Anders} G.~J.,  {James}
  D.~J.,    {Lanzafame} A.~C.,  1994, \mnras, 270, 153

\bibitem[\protect\citeauthoryear{{Jeffries}, {Jackson}, {Briggs}, {Evans} \&
  {Pye}}{{Jeffries} et~al.}{2011}]{Jeffries2011a}
{Jeffries} R.~D.,  {Jackson} R.~J.,  {Briggs} K.~R.,  {Evans} P.~A.,    {Pye}
  J.~P.,  2011, \mnras, 411, 2099

\bibitem[\protect\citeauthoryear{{Jeffries}, {Jackson}, {Franciosini} \& {et
  al.}}{{Jeffries} et~al.}{2017}]{Jeffries2017a}
{Jeffries} R.~D.,  {Jackson} R.~J.,  {Franciosini} E.,    {et al.} 2017,
  \mnras, 464, 1456

\bibitem[\protect\citeauthoryear{{Jeffries}, {James} \& {Thurston}}{{Jeffries}
  et~al.}{1998}]{Jeffries1998a}
{Jeffries} R.~D.,  {James} D.~J.,    {Thurston} M.~R.,  1998, \mnras, 300, 550

\bibitem[\protect\citeauthoryear{{Jeffries}, {Totten} \& {James}}{{Jeffries}
  et~al.}{2000}]{Jeffries2000a}
{Jeffries} R.~D.,  {Totten} E.~J.,    {James} D.~J.,  2000, \mnras, 316, 950

\bibitem[\protect\citeauthoryear{{Kamai}, {Vrba}, {Stauffer} \&
  {Stassun}}{{Kamai} et~al.}{2014}]{Kamai2014a}
{Kamai} B.~L.,  {Vrba} F.~J.,  {Stauffer} J.~R.,    {Stassun} K.~G.,  2014,
  \aj, 148, 30

\bibitem[\protect\citeauthoryear{{Karmakar}, {Pandey}, {Savanov}, {Tas},
  {Pandey}, {Misra}, {Joshi}, {Dmitrienko}, {Sakamoto}, {Gehrels} \&
  {Okajima}}{{Karmakar} et~al.}{2016}]{Karmakar2016a}
{Karmakar} S.,  {Pandey} J.~C.,  {Savanov} I.~S.,  {Tas} G.,  {Pandey} S.~B.,
  {Misra} K.,  {Joshi} S.,  {Dmitrienko} E.~S.,  {Sakamoto} T.,  {Gehrels} N.,
    {Okajima} T.,  2016, \mnras, 459, 3112

\bibitem[\protect\citeauthoryear{{King} \& {Schuler}}{{King} \&
  {Schuler}}{2004}]{King2004a}
{King} J.~R.,  {Schuler} S.~C.,  2004, \aj, 128, 2898

\bibitem[\protect\citeauthoryear{{King}, {Schuler}, {Hobbs} \&
  {Pinsonneault}}{{King} et~al.}{2010}]{King2010a}
{King} J.~R.,  {Schuler} S.~C.,  {Hobbs} L.~M.,    {Pinsonneault} M.~H.,  2010,
  \apj, 710, 1610

\bibitem[\protect\citeauthoryear{{Kurucz}}{{Kurucz}}{1992}]{Kurucz1992a}
{Kurucz} R.~L.,  1992, in {Barbuy} B.,  {Renzini} A.,  eds, The Stellar
  Populations of Galaxies Vol.~149 of IAU Symposium, {Model Atmospheres for
  Population Synthesis}.
p.~225

\bibitem[\protect\citeauthoryear{{Lacy}, {Fekel}, {Pavlovski}, {Torres} \&
  {Muterspaugh}}{{Lacy} et~al.}{2016}]{Lacy2016a}
{Lacy} C. H.~S.,  {Fekel} F.~C.,  {Pavlovski} K.,  {Torres} G.,
  {Muterspaugh} M.~W.,  2016, \aj, 152, 2

\bibitem[\protect\citeauthoryear{{Leone}}{{Leone}}{2007}]{Leone2007a}
{Leone} F.,  2007, \apjl, 667, L175

\bibitem[\protect\citeauthoryear{{Libralato}, {Bedin}, {Nardiello} \&
  {Piotto}}{{Libralato} et~al.}{2016}]{Libralato2016a}
{Libralato} M.,  {Bedin} L.~R.,  {Nardiello} D.,    {Piotto} G.,  2016, \mnras,
  456, 1137

\bibitem[\protect\citeauthoryear{{Lim}, {Sung}, {Kim}, {Bessell}, {Hwang} \&
  {Park}}{{Lim} et~al.}{2016}]{Lim2016a}
{Lim} B.,  {Sung} H.,  {Kim} J.~S.,  {Bessell} M.~S.,  {Hwang} N.,    {Park}
  B.-G.,  2016, \apj, 831, 116

\bibitem[\protect\citeauthoryear{{Lindegren} \& {et al.}}{{Lindegren} \& {et
  al.}}{2018}]{Lindegren2018a}
{Lindegren} L.,  {et al.} 2018, \aap, 616, A2

\bibitem[\protect\citeauthoryear{{Lindegren}, {Lammers}, Bastian \& {et.
  al.}}{{Lindegren} et~al.}{2016}]{Lindegren2016a}
{Lindegren} L.,  {Lammers} U.,  Bastian U.,    {et. al.} 2016, \aap, 595, A4

\bibitem[\protect\citeauthoryear{{Macdonald} \& {Mullan}}{{Macdonald} \&
  {Mullan}}{2010}]{Macdonald2010a}
{Macdonald} J.,  {Mullan} D.~J.,  2010, \apj, 723, 1599

\bibitem[\protect\citeauthoryear{{MacDonald} \& {Mullan}}{{MacDonald} \&
  {Mullan}}{2013}]{MacDonald2013a}
{MacDonald} J.,  {Mullan} D.~J.,  2013, \apj, 765, 126

\bibitem[\protect\citeauthoryear{{Magic}, {Collet}, {Asplund}, {Trampedach},
  {Hayek}, {Chiavassa}, {Stein} \& {Nordlund}}{{Magic}
  et~al.}{2013}]{Magic2013a}
{Magic} Z.,  {Collet} R.,  {Asplund} M.,  {Trampedach} R.,  {Hayek} W.,
  {Chiavassa} A.,  {Stein} R.~F.,    {Nordlund} {\r{A}}.,  2013, \aap, 557, A26

\bibitem[\protect\citeauthoryear{{Marsden}, {Carter} \& {Donati}}{{Marsden}
  et~al.}{2009}]{Marsden2009a}
{Marsden} S.~C.,  {Carter} B.~D.,    {Donati} J.,  2009, \mnras, 399, 888

\bibitem[\protect\citeauthoryear{{Meibom}, {Mathieu} \& {Stassun}}{{Meibom}
  et~al.}{2009}]{Meibom2009a}
{Meibom} S.,  {Mathieu} R.~D.,    {Stassun} K.~G.,  2009, \apj, 695, 679

\bibitem[\protect\citeauthoryear{{Messina} \& {Guinan}}{{Messina} \&
  {Guinan}}{2003}]{Messina2003b}
{Messina} S.,  {Guinan} E.~F.,  2003, \aap, 409, 1017

\bibitem[\protect\citeauthoryear{{Messina}, {Lanzafame}, {Feiden}, {Millward},
  {Desidera}, {Buccino}, {Curtis}, {Jofr{\'e}}, {Kehusmaa}, {Medhi}, {Monard}
  \& {Petrucci}}{{Messina} et~al.}{2016}]{Messina2016a}
{Messina} S.,  {Lanzafame} A.~C.,  {Feiden} G.~A.,  {Millward} M.,  {Desidera}
  S.,  {Buccino} A.,  {Curtis} I.,  {Jofr{\'e}} E.,  {Kehusmaa} P.,  {Medhi}
  B.~J.,  {Monard} B.,    {Petrucci} R.,  2016, \aap, 596, A29

\bibitem[\protect\citeauthoryear{{Messina}, {Rodon{\`o}} \& {Guinan}}{{Messina}
  et~al.}{2001}]{Messina2001b}
{Messina} S.,  {Rodon{\`o}} M.,    {Guinan} E.~F.,  2001, \aap, 366, 215

\bibitem[\protect\citeauthoryear{{Montalb{\'a}n} \&
  {Schatzman}}{{Montalb{\'a}n} \& {Schatzman}}{2000}]{Montalban2000a}
{Montalb{\'a}n} J.,  {Schatzman} E.,  2000, \aap, 354, 943

\bibitem[\protect\citeauthoryear{{Morales}, {Ribas}, {Jordi}, {Torres},
  {Gallardo}, {Guinan}, {Charbonneau}, {Wolf}, {Latham}, {Anglada-Escud{\'e}},
  {Bradstreet}, {Everett}, {O'Donovan}, {Mandushev} \& {Mathieu}}{{Morales}
  et~al.}{2009}]{Morales2009a}
{Morales} J.~C.,  {Ribas} I.,  {Jordi} C.,  {Torres} G.,  {Gallardo} J.,
  {Guinan} E.~F.,  {Charbonneau} D.,  {Wolf} M.,  {Latham} D.~W.,
  {Anglada-Escud{\'e}} G.,  {Bradstreet} D.~H.,  {Everett} M.~E.,  {O'Donovan}
  F.~T.,  {Mandushev} G.,    {Mathieu} R.~D.,  2009, \apj, 691, 1400

\bibitem[\protect\citeauthoryear{{Murphy}, {Lawson}, {Onken}, {Yong}, {Da
  Costa}, {Zhou}, {Mamajek}, {Bell}, {Bessell} \& {Feinstein}}{{Murphy}
  et~al.}{2020}]{Murphy2020a}
{Murphy} S.~J.,  {Lawson} W.~A.,  {Onken} C.~A.,  {Yong} D.,  {Da Costa} G.~S.,
   {Zhou} G.,  {Mamajek} E.~E.,  {Bell} C. P.~M.,  {Bessell} M.~S.,
  {Feinstein} A.~D.,  2020, \mnras, 491, 4902

\bibitem[\protect\citeauthoryear{{Nardiello}, {Bedin}, {Nascimbeni} \& et
  al.}{{Nardiello} et~al.}{2015}]{Nardiello2015a}
{Nardiello} D.,  {Bedin} L.~R.,  {Nascimbeni} V.,    et al. 2015, \mnras, 447,
  3536

\bibitem[\protect\citeauthoryear{{Noyes}, {Weiss} \& {Vaughan}}{{Noyes}
  et~al.}{1984}]{Noyes1984a}
{Noyes} R.~W.,  {Weiss} N.~O.,    {Vaughan} A.~H.,  1984, \apj, 287, 769

\bibitem[\protect\citeauthoryear{{O'Dell}, {Panagi}, {Hendry} \& {Collier
  Cameron}}{{O'Dell} et~al.}{1995}]{Odell1995a}
{O'Dell} M.~A.,  {Panagi} P.,  {Hendry} M.~A.,    {Collier Cameron} A.,  1995,
  \aap, 294, 715

\bibitem[\protect\citeauthoryear{{Pecaut} \& {Mamajek}}{{Pecaut} \&
  {Mamajek}}{2013}]{Pecaut2013a}
{Pecaut} M.~J.,  {Mamajek} E.~E.,  2013, \apjs, 208, 9

\bibitem[\protect\citeauthoryear{{Piau} \& {Turck-Chi{\`e}ze}}{{Piau} \&
  {Turck-Chi{\`e}ze}}{2002}]{Piau2002a}
{Piau} L.,  {Turck-Chi{\`e}ze} S.,  2002, \apj, 566, 419

\bibitem[\protect\citeauthoryear{{Pinsonneault}}{{Pinsonneault}}{1997}]{Pinsonneault1997a}
{Pinsonneault} M.,  1997, \araa, 35, 557

\bibitem[\protect\citeauthoryear{{Pizzolato}, {Maggio}, {Micela}, {Sciortino}
  \& {Ventura}}{{Pizzolato} et~al.}{2003}]{Pizzolato2003a}
{Pizzolato} N.,  {Maggio} A.,  {Micela} G.,  {Sciortino} S.,    {Ventura} P.,
  2003, \aap, 397, 147

\bibitem[\protect\citeauthoryear{{Raghavan}, {McAlister}, {Henry}, {Latham},
  {Marcy}, {Mason}, {Gies}, {White} \& {ten Brummelaar}}{{Raghavan}
  et~al.}{2010}]{Raghavan2010a}
{Raghavan} D.,  {McAlister} H.~A.,  {Henry} T.~J.,  {Latham} D.~W.,  {Marcy}
  G.~W.,  {Mason} B.~D.,  {Gies} D.~R.,  {White} R.~J.,    {ten Brummelaar}
  T.~A.,  2010, \apjs, 190, 1

\bibitem[\protect\citeauthoryear{{Randich}}{{Randich}}{2009}]{Randich2009a}
{Randich} S.,  2009, in {Mamajek} E.~E.,  {Soderblom} D.~R.,   {Wyse} R. F.~G.,
   eds, The Ages of Stars Vol.~258 of IAU Symposium, {On the use of lithium to
  derive the ages of stars like our Sun}.
pp 133--140

\bibitem[\protect\citeauthoryear{{Randich}, {Martin}, {Garcia Lopez} \&
  {Pallavicini}}{{Randich} et~al.}{1998}]{Randich1998a}
{Randich} S.,  {Martin} E.~L.,  {Garcia Lopez} R.~J.,    {Pallavicini} R.,
  1998, \aap, 333, 591

\bibitem[\protect\citeauthoryear{{Randich}, {Pallavicini}, {Meola}, {Stauffer}
  \& {Balachandran}}{{Randich} et~al.}{2001}]{Randich2001a}
{Randich} S.,  {Pallavicini} R.,  {Meola} G.,  {Stauffer} J.~R.,
  {Balachandran} S.~C.,  2001, \aap, 372, 862

\bibitem[\protect\citeauthoryear{{Rebull}, {Stauffer}, {Bouvier}, {Cody},
  {Hillenbrand}, {Soderblom}, {Valenti}, {Barrado}, {Bouy}, {Ciardi} \&
  {Pinsonneault}}{{Rebull} et~al.}{2016}]{Rebull2016a}
{Rebull} L.~M.,  {Stauffer} J.~R.,  {Bouvier} J.,  {Cody} A.~M.,  {Hillenbrand}
  L.~A.,  {Soderblom} D.~R.,  {Valenti} J.,  {Barrado} D.,  {Bouy} H.,
  {Ciardi} D.,    {Pinsonneault} M.,  2016, \aj, 152, 113

\bibitem[\protect\citeauthoryear{{Reiners}, {Basri} \& {Browning}}{{Reiners}
  et~al.}{2009}]{Reiners2009a}
{Reiners} A.,  {Basri} G.,    {Browning} M.,  2009, \apj, 692, 538

\bibitem[\protect\citeauthoryear{{Rieke} \& {Lebofsky}}{{Rieke} \&
  {Lebofsky}}{1985}]{Rieke1985a}
{Rieke} G.~H.,  {Lebofsky} M.~J.,  1985, ApJ, 288, 618

\bibitem[\protect\citeauthoryear{{Schatzman}}{{Schatzman}}{1993}]{Schatzman1993a}
{Schatzman} E.,  1993, \aap, 279, 431

\bibitem[\protect\citeauthoryear{{Siess}, {Dufour} \& {Forestini}}{{Siess}
  et~al.}{2000}]{Siess2000a}
{Siess} L.,  {Dufour} E.,    {Forestini} M.,  2000, \aap, 358, 593

\bibitem[\protect\citeauthoryear{{Skrutskie}, {Cutri}, {Stiening}, {Weinberg},
  {Schneider}, {Carpenter}, {Beichman}, {Capps}, {Chester}, {Elias} \&
  {Huchra}}{{Skrutskie} et~al.}{2006}]{Skrutskie2006a}
{Skrutskie} M.~F.,  {Cutri} R.~M.,  {Stiening} R.,  {Weinberg} M.~D.,
  {Schneider} S.,  {Carpenter} J.~M.,  {Beichman} C.,  {Capps} R.,  {Chester}
  T.,  {Elias} J.,    {Huchra} J.,  2006, \aj, 131, 1163

\bibitem[\protect\citeauthoryear{{Skumanich}}{{Skumanich}}{1972}]{Skumanich1972a}
{Skumanich} A.,  1972, \apj, 171, 565

\bibitem[\protect\citeauthoryear{{Sneden}, {Bean}, {Ivans}, {Lucatello} \&
  {Sobeck}}{{Sneden} et~al.}{2012}]{Sneden2012a}
{Sneden} C.,  {Bean} J.,  {Ivans} I.,  {Lucatello} S.,    {Sobeck} J., , 2012,
  {MOOG: LTE line analysis and spectrum synthesis}

\bibitem[\protect\citeauthoryear{{Soderblom}}{{Soderblom}}{2010}]{Soderblom2010a}
{Soderblom} D.~R.,  2010, \araa, 48, 581

\bibitem[\protect\citeauthoryear{{Soderblom}, {Stauffer}, {Hudon} \&
  {Jones}}{{Soderblom} et~al.}{1993}]{Soderblom1993a}
{Soderblom} D.~R.,  {Stauffer} J.~R.,  {Hudon} J.~D.,    {Jones} B.~F.,  1993,
  ApJs, 85, 315

\bibitem[\protect\citeauthoryear{{Somers}, {Cao} \& {Pinsonneault}}{{Somers}
  et~al.}{2020}]{Somers2020a}
{Somers} G.,  {Cao} L.,    {Pinsonneault} M.~H.,  2020, \apj, 891, 29

\bibitem[\protect\citeauthoryear{{Somers} \& {Pinsonneault}}{{Somers} \&
  {Pinsonneault}}{2014}]{Somers2014a}
{Somers} G.,  {Pinsonneault} M.~H.,  2014, \apj, 790, 72

\bibitem[\protect\citeauthoryear{{Somers} \& {Pinsonneault}}{{Somers} \&
  {Pinsonneault}}{2015a}]{Somers2015a}
{Somers} G.,  {Pinsonneault} M.~H.,  2015a, \apj, 807, 174

\bibitem[\protect\citeauthoryear{{Somers} \& {Pinsonneault}}{{Somers} \&
  {Pinsonneault}}{2015b}]{Somers2015b}
{Somers} G.,  {Pinsonneault} M.~H.,  2015b, \mnras, 449, 4131

\bibitem[\protect\citeauthoryear{{Somers} \& {Stassun}}{{Somers} \&
  {Stassun}}{2017}]{Somers2017a}
{Somers} G.,  {Stassun} K.~G.,  2017, \aj, 153, 101

\bibitem[\protect\citeauthoryear{{Spada}, {Lanzafame}, {Lanza}, {Messina} \&
  {Collier Cameron}}{{Spada} et~al.}{2011}]{Spada2011a}
{Spada} F.,  {Lanzafame} A.~C.,  {Lanza} A.~F.,  {Messina} S.,    {Collier
  Cameron} A.,  2011, \mnras, 416, 447

\bibitem[\protect\citeauthoryear{{Spruit} \& {Weiss}}{{Spruit} \&
  {Weiss}}{1986}]{Spruit1986a}
{Spruit} H.~C.,  {Weiss} A.,  1986, \aap, 166, 167

\bibitem[\protect\citeauthoryear{{Stauffer}, {Jones}, {Backman}, {Hartmann},
  {Barrado y Navascu{\'e}s}, {Pinsonneault}, {Terndrup} \& {Muench}}{{Stauffer}
  et~al.}{2003}]{Stauffer2003a}
{Stauffer} J.~R.,  {Jones} B.~F.,  {Backman} D.,  {Hartmann} L.~W.,  {Barrado y
  Navascu{\'e}s} D.,  {Pinsonneault} M.~H.,  {Terndrup} D.~M.,    {Muench}
  A.~A.,  2003, \aj, 126, 833

\bibitem[\protect\citeauthoryear{{Stauffer}, {Schultz} \&
  {Kirkpatrick}}{{Stauffer} et~al.}{1998}]{Stauffer1998a}
{Stauffer} J.~R.,  {Schultz} G.,    {Kirkpatrick} J.~D.,  1998, \apjl, 499,
  L199+

\bibitem[\protect\citeauthoryear{{Steinhauer} \& {Deliyannis}}{{Steinhauer} \&
  {Deliyannis}}{2004}]{Steinhauer2004a}
{Steinhauer} A.,  {Deliyannis} C.~P.,  2004, \apjl, 614, L65

\bibitem[\protect\citeauthoryear{{Stuik}, {Bruls} \& {Rutten}}{{Stuik}
  et~al.}{1997}]{Stuik1997a}
{Stuik} R.,  {Bruls} J.~H.~M.~J.,    {Rutten} R.~J.,  1997, \aap, 322, 911

\bibitem[\protect\citeauthoryear{{Sung} \& {Bessell}}{{Sung} \&
  {Bessell}}{1999}]{Sung1999a}
{Sung} H.,  {Bessell} M.~S.,  1999, \mnras, 306, 361

\bibitem[\protect\citeauthoryear{{Thorburn}, {Hobbs}, {Deliyannis} \&
  {Pinsonneault}}{{Thorburn} et~al.}{1993}]{Thorburn1993a}
{Thorburn} J.~A.,  {Hobbs} L.~M.,  {Deliyannis} C.~P.,    {Pinsonneault} M.~H.,
   1993, \apj, 415, 150

\bibitem[\protect\citeauthoryear{{Tognelli}, {Prada Moroni} \&
  {Degl'Innocenti}}{{Tognelli} et~al.}{2011}]{Tognelli2011a}
{Tognelli} E.,  {Prada Moroni} P.~G.,    {Degl'Innocenti} S.,  2011, \aap, 533,
  A109

\bibitem[\protect\citeauthoryear{{Torres}}{{Torres}}{2013}]{Torres2013a}
{Torres} G.,  2013, Astronomische Nachrichten, 334, 4

\bibitem[\protect\citeauthoryear{{Valenti} \& {Johns-Krull}}{{Valenti} \&
  {Johns-Krull}}{2001}]{Valenti2001a}
{Valenti} J.~A.,  {Johns-Krull} C.,  2001, in {Mathys} G.,  {Solanki} S.~K.,
  {Wickramasinghe} D.~T.,  eds, Magnetic Fields Across the Hertzsprung-Russell
  Diagram Vol.~248 of Astronomical Society of the Pacific Conference Series,
  {Magnetic Field Measurements for Cool Stars}.
p.~179

\bibitem[\protect\citeauthoryear{{Ventura}, {Zeppieri}, {Mazzitelli} \&
  {D'Antona}}{{Ventura} et~al.}{1998}]{Ventura1998a}
{Ventura} P.,  {Zeppieri} A.,  {Mazzitelli} I.,    {D'Antona} F.,  1998, \aap,
  331, 1011

\bibitem[\protect\citeauthoryear{{Vilhu}}{{Vilhu}}{1984}]{Vilhu1984a}
{Vilhu} O.,  1984, \aap, 133, 117

\bibitem[\protect\citeauthoryear{{von Hippel}, {Steinhauer}, {Sarajedini} \&
  {Deliyannis}}{{von Hippel} et~al.}{2002}]{vonHippel2002a}
{von Hippel} T.,  {Steinhauer} A.,  {Sarajedini} A.,    {Deliyannis} C.~P.,
  2002, \aj, 124, 1555

\bibitem[\protect\citeauthoryear{{Wright}, {Drake}, {Mamajek} \&
  {Henry}}{{Wright} et~al.}{2011}]{Wright2011a}
{Wright} N.~J.,  {Drake} J.~J.,  {Mamajek} E.~E.,    {Henry} G.~W.,  2011,
  \apj, 743, 48

\bibitem[\protect\citeauthoryear{{Xiong} \& {Deng}}{{Xiong} \&
  {Deng}}{2005}]{Xiong2005a}
{Xiong} D.~R.,  {Deng} L.,  2005, \apj, 622, 620

\end{thebibliography}



\bsp 
\label{lastpage}
\end{document}